\documentclass[12pt,a4paper,twoside]{kkd}
\usepackage{epsfig}
\usepackage{amssymb}
\usepackage{setspace}
\usepackage[round]{natbib}

\usepackage{amsmath}
\usepackage{bm,psfrag}
\usepackage{color}

\usepackage{ragged2e} %Added 29/06/2009
\makeatletter
\begin{document}

\pagenumbering{roman}
\addcontentsline{toc}{chapter}{Title page}
 
%\newpage
%\begin{singlespacing}
\setlength{\textheight}{285.9 mm}
\oddsidemargin 1.2 cm 

\thispagestyle{empty}
\def\maketitle{%
  \null
  \thispagestyle{empty}%
%  \vfill
  \begin{center}%\leavevmode
   % \normalfont
    {\Large {\bf \@title\par}}%
    \vskip 1.0 cm
    %{\bf {\sl Thesis submitted to}}\\
    {\bf   {\em Thesis submitted to the}}\\
    %\vskip .2 cm
    {\textbf{\em Indian Institute of Technology, Kharagpur}}\\
    %\vskip .2 cm
    {\bf{\em For the award of the degree}}\\
    \vspace{0.75cm}
    {\bf {\em of}}\\
    \vspace{0.35cm}
    {\Large {\bf Doctor of Philosophy }}\\
    \vspace{.25cm}   
    {\bf {\em by}}\\
    %\vspace{.25cm}
    {\Large {\bf\@author\par}}%
    \vspace{0.7cm}
    {\bf {\em Under the guidance of}}\\
    \vspace{.35cm}
    {\Large {\bf Prof. Somnath Bharadwaj}}\\
    \vspace*{0.2cm} 
    \end{center}%
  %\vfill
  \null
}
\setlength{\topmargin}{-15. mm}

\begin{singlespacing}
\title{PROBING TURBULENCE IN THE
INTERSTELLAR MEDIUM USING RADIO-INTERFEROMETRIC OBSERVATIONS OF
NEUTRAL HYDROGEN}
\author{Prasun Dutta}
\maketitle
\vspace{.1cm}
\begin{figure}[h]
\centerline{\psfig{figure=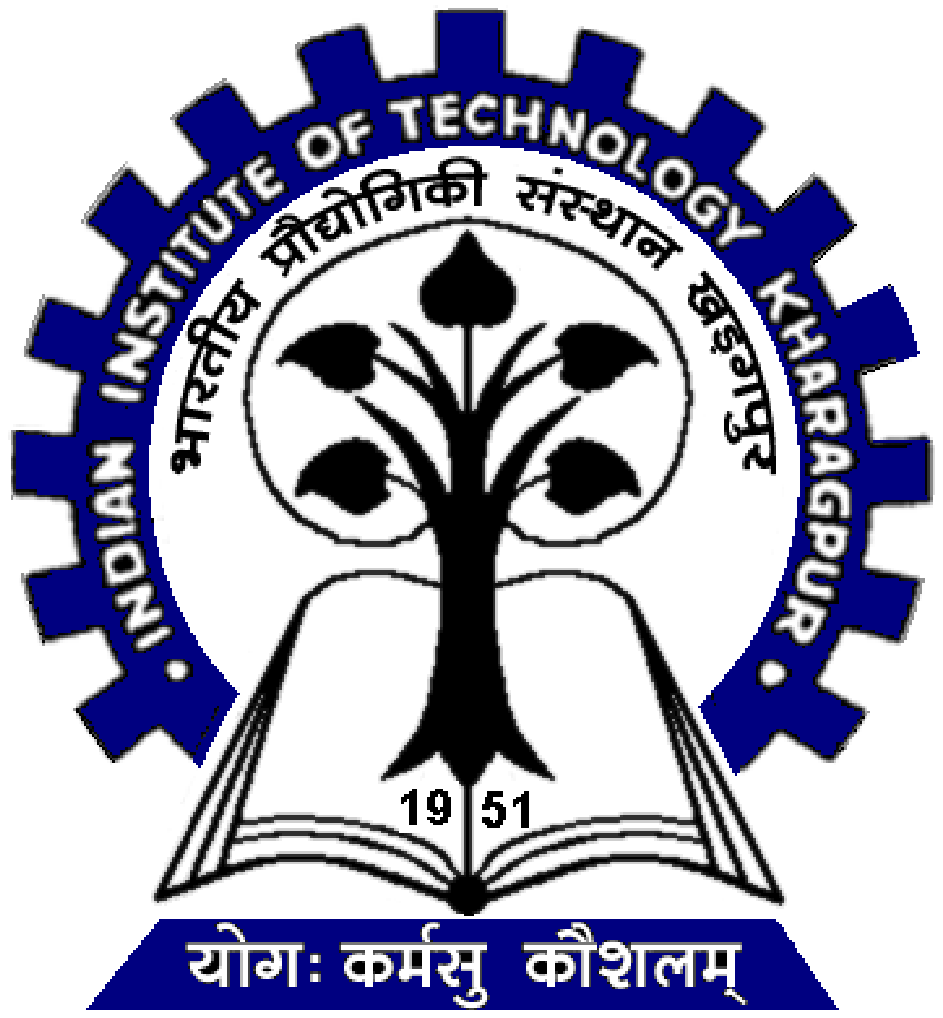,width=4.5cm,height=4.5cm}}
\end{figure}
\begin{center}
%\vspace{.3cm}
{\large {\bf DEPARTMENT OF PHYSICS AND METEOROLOGY}}\\
\vspace{.15cm}
{\large {\bf INDIAN INSTITUTE OF TECHNOLOGY, KHARAGPUR}}\\
\vspace{.15cm}
{\large {\bf FEBRUARY 2011}}\\
\vspace{.25cm}
{\bf $\copyright$ 2011 Prasun Dutta. All rights reserved.}

\end{center}

\end{singlespacing}
\pagebreak

\setlength{\topmargin}{0.05cm} %\Added by Partha

\textheight 21.1 cm 
\textwidth 16.0 cm 
\pagestyle{headings}

\newcommand{\degr}{\ensuremath{^{\circ}}}

\renewcommand{\labelenumi}{(\roman{enumi})}
\renewcommand{\thefootnote}{\fnsymbol{footnote}}

\setlength{\headheight}{10.0 mm}
\setlength{\headsep}{10.0 mm}
\setlength{\topskip}{0.00 mm}
\setlength{\textheight}{215.9 mm}
\setlength{\textwidth}{152.4 mm}

\setlength{\parskip}{.75 mm}
\setlength{\parindent}{6.00 mm}
\setlength{\floatsep}{4 mm}
\setlength{\textfloatsep}{4 mm} \setlength{\intextsep}{4 mm}
\setcounter{secnumdepth}{3} \setcounter{tocdepth}{2}
\newtheorem{theorem}{Theorem}[section]
\newtheorem{lemma}{Lemma}[chapter]
\newtheorem{Remark}{Remark}[chapter]
\newtheorem{corollary}{Corollary}
\newtheorem{definition}[theorem]{Definition}
\newtheorem{observation}[theorem]{Observation}
\newtheorem{fact}{Fact}[theorem]
\newtheorem{proposition}[theorem]{Proposition}
\newtheorem{rule-def}[theorem]{Rule}
\newcommand{\ccnt}[1]{\multicolumn{1}{|c|}{#1}}
\newcommand{\fns}[1]{\footnotesize{#1}}

%\singlespacing
%\doublespacing
% defineing short form------

\newcommand{\sce}{\setcounter{equation}}
\newcommand{\nn}{\nonumber}
\newcommand{\lb}{\label}
\newcounter{saveeqn}
\newcommand{\alpheqn}{\setcounter{saveeqn}{\value{equation}}
\stepcounter{saveeqn}\setcounter{equation}{0}%
\renewcommand{\theequation}{\mbox{\arabic{chapter}.\arabic{saveeqn}\alph{equation}}}}
\newcommand{\reseteqn}{\setcounter{equation}{\value{saveeqn}}%
\renewcommand{\theequation}{\arabic{chapter}.\arabic{equation}}}
\newcounter{savecite}
\newcommand{\alphcite}{\setcounter{savecite}{\value{cite}}
\stepcounter{savecite}\setcounter{cite}{0}%
\renewcommand{\thecite}{\mbox{\arabic{savecite}\alph{cite}}}}
\newcommand{\resetcite}{\setcounter{cite}{\value{savecite}}%
\renewcommand{\thecite}{\arabic{cite}}}

\newcommand{\be}{\begin{equation}}
\newcommand{\e}{\end{equation}}
\newcommand{\bear}{\begin{eqnarray}}
\newcommand{\ear}{\end{eqnarray}}
\newcommand{\nline}{\nonumber \\}
\newcommand{\f}{\frac}
\newcommand{\de}{{\rm d}}
\newcommand{\del}{\partial}

%%%%%%%%%%%%%%%%%%%%%%%%%%%%%
%%%% journals %%%%%%%%%%%%%%%
%%%%%%%%%%%%%%%%%%%%%%%%%%%%%
\def\apj{Astrophysical Journal}
\def\apjl{Astrophysical Journal Letters}
\def\apjs{Astrophysical Journal Supplement Series}
\def\aap{Astronomy and Astrophysics}
\def\aapr{Astronomy and Astrophysics Review}
\def\aaps{Astronomy and Astrophysics Suppliment}
\def\mnras{Monthly Notices of Royal Astronomical Society}
\def\mnrasl{Monthly Notices of Royal Astronomical Society: Letters}
\def\araa{Annual Review of Astronomy \& Astrophysics}
\def\aj{Astronomical Journal}
\def\prd{Physical Review D}
\def\pre{Physical Review E}
\def\prl{Physical Review Letters}
\def\phyrep{Physics Report}
\def\japa{Journal of Astrophysics and Astronomy}
\def\jcap{Journal of Cosmology and Astroparticle Physics}
\def\pasp{Astronomical Society of the Pacific, Publications}
\def\skytel{Sky and Telescope}
\def\tifr{Tata Institute of Fundamental Rrsearch: deemed University}
\def\memras{Member of Royal Astronomical Society}
\def\AA{$\AA$}
%%%%%%%%%%%%%%%%%%%%%%%%%%%%%%%%%%%%
\clearpage{\pagestyle{empty}\cleardoublepage}
\thispagestyle{empty}
\newcommand{\dead}[2]{ & & & & & & & {{\hspace{5cm}\bf #1}} &  {{\it #2}} \\}
\newcommand{\deads}{& & & & & & & & \\}
%\topskip 18cm
%\begin{center}
\begin{tabular}{lcccccccl}
\deads
\deads
\deads
\deads
\deads
\deads
\deads
\deads
\deads
\deads
\deads
\deads
\deads
\deads
\deads
\hline
\dead{\large {To}}{My parents, Chotomama,}
\dead{}{Rupai, Chutki, Babi, Didi}
\dead{}{and all my friends}
\end{tabular}

\vspace{4.75cm}
{\bf ...in   the   infinite   beauty,   we   all   join   in   one...}
%\Large { ...In the infinite beauty we all join in one...}
%\topskip 0cm
\newpage

\clearpage{\pagestyle{empty}\cleardoublepage}
\addcontentsline{toc}{chapter}{Certificate of Approval}
%MAKE IT TOTALLY BLANK
%\thispagestyle{empty}
%\vspace{1cm}
 %\draftstring{}
%\watermarkgraphic{water.png} \watermark
\begin{center}
%{\Large \underline{\bf CERTIFICATE OF APPROVAL}}
{\Large \underline{\bf APPROVAL OF THE VIVA-VOCE BOARD}}
\end{center}

%\vspace{.5cm}
\hspace{12cm} Date: $21/2/2011$ 

\justifying

\hspace{-.55cm}

\noindent
Certified that the thesis entitled {\bf Probing Turbulence In
the Interstellar Medium using Radio-Interferometric Observations of
Neutral Hydrogen}
submitted by {\bf PRASUN DUTTA} to 
the Indian Institute of Technology, Kharagpur, for the award of the
degree of Doctor of Philosophy has been accepted by the external
examiners and that the student has successfully defended the thesis in
the viva-voce examination held today.

\vspace{2.95cm}

\noindent
Signature\hspace{4.25cm}Signature\hspace{4.25cm}Signature

\vspace{0.1cm}
\noindent
(Member of the DSC)\hspace{2.1cm}(Member of the
DSC)\hspace{2.2cm}(Member of the DSC)

\vspace{2.95cm}

\noindent
Signature\hspace{4.25cm}Signature\hspace{4.25cm}Signature

\vspace{0.1cm}
\noindent
(Supervisor)\hspace{3.65cm}(External Examiner)\hspace{2.2cm}(Chairman)

\clearpage{\pagestyle{empty}\cleardoublepage}
\addcontentsline{toc}{chapter}{Certificate by the Supervisor}
%MAKE IT TOTALLY BLANK
%\thispagestyle{empty}
\setcounter{page}{7}
\vspace{3cm}
 %\draftstring{}
%\watermarkgraphic{water.png} \watermark
\begin{center}
{\Large \underline{\bf CERTIFICATE}}
\end{center}

\vspace{2.5cm}

\noindent
This is to certify that the thesis entitled {\bf Probing Turbulence In
the Interstellar Medium using Radio-Interferometric Observations of
Neutral Hydrogen},
submitted by {\bf Prasun Dutta} to  
the Indian Institute of Technology, Kharagpur, is a record of bona
fide research work under my supervision and is of
consideration for the award of the degree of Doctor of Philosophy of
the Institute.

\vspace{5.0cm}
\noindent

\hspace{9.5cm}\underline{\hspace{1.cm} Supervisor \hspace{1.cm}}

\noindent
\hspace{9.75cm}Prof. Somnath Bharadwaj

\vspace{0.5cm}
\noindent
\hspace{10cm}Date:

\clearpage{\pagestyle{empty}\cleardoublepage} 
\addcontentsline{toc}{chapter}{Declaration}
\oddsidemargin 0.9 cm 
\evensidemargin 1.4 cm
\setlength{\textwidth}{152.4 mm}
 \vspace*{2cm}
 %\draftstring{}
%\watermarkgraphic{water.png} \watermark
\begin{singlespacing}

{\Large {\bf DECLARATION}}

\vspace{1cm} I certify that 

\begin{description}

\item 
\hspace{0.1cm} a. The work contained in the thesis is original and has 
been done by myself under the general supervision of my supervisor.

\item 
\hspace{0.1cm} b. The work has not been submitted to any other 
Institute for any degree or diploma.

\item
\hspace{0.1cm} c. I have followed the guidelines provided by the 
Institute in writing the thesis.

\item
\hspace{0.1cm} d. I have conformed to the norms and guidelines given 
in the Ethical Code of Conduct of the Institute.

\item
\hspace{0.1cm} e. Whenever I have used materials (data, theoretical 
analysis, and text) from other sources, I have given due credit to
them by citing them in the text of the thesis and giving their details
in the references.

\item
\hspace{0.1cm} f. Whenever I have quoted written materials from other
sources, I have put them under quotation marks and given due credit to
the sources by citing them and giving required details in the
references.

\end{description}

%\end{enumerate}

\vspace{2.5cm}
\hspace{9.5cm} Signature of the Student

%%\vspace{5.65cm}
%%\hspace{10.5cm} iv

\end{singlespacing}

%\clearpage{\pagestyle{empty}\cleardoublepage} %%%%%%%%%%%%%%%%%%%%
%\chaptermark{cv}
\clearpage{\pagestyle{empty}\cleardoublepage} 
\addcontentsline{toc}{chapter}{Acknowledgements}
\vspace{-0.75cm}

\oddsidemargin 0.9 cm 
\evensidemargin -.4 cm
\setlength{\textwidth}{152.4 mm}

\vspace{-2.5cm}
\begin{center}
{\Large \bf Acknowledgments}
\end{center}
 It is 
the collective effort of many individuals, my parents and relatives,
teachers, friends,  that have poured into a nice A4 size book only
to become my thesis. Though this section bears the
name "Acknowledgements", I have no words to acknowledge or thank. I will
only present here a few of my  memorable experiences.

My parents always encouraged me to learn as much as I can and then
share the knowledge with everybody. Chotomama taught me photography,
but more than that he seeded in me the quest for asking how things
work. My teachers Anjan babu, Debasis babu, Sajal babu, Biswanath
babu, Ashesh babu, Saibal babu and Tarasankar babu have always encouraged
me to pursue higher education. They prepared the base over which I
have only tried adding things. 

Back home, I shared numerous  moments of true joy   with my
cousins, Didi, Babi, Buronda, Rajada, Tunku di, Sonali di.  Many
of my time that I have spent with my sisters Chutki and Rupai, 
are the times that I will always like to go back to. They, being very
good students have helped me solve some of my problems too. 

If I were ever to do another Ph. D, I would prefer working with
Prof. Somnath Bharadwaj again. If I were ever to guide a student, I will
just try to emulate how he prepared us. He showed us the art
of doing research in a professional way yet with passion and
love. He has always clarified all my doubts and then
blessed me by showing his magical power of reasoning every aspect
starting from the very basics. He has been a true teacher, a friend, 
philosopher and guide to me always. Working in Collaboration with
Jayaram, Ayesha, SPK, Tapomoy and Nirupam has also been encouraging
and enjoyable. I will be very happy to continue working with them. 

CTS was never felt far from home. Friends here, Prakash, Biswambhar
(Bisu), Somnath (Bhaipo), Suman (Mahajan), Tatan, Soma di and Partha
da, Sanjit (Dos), Suman (Choto), Abhik (Jethu), Ganesh, Subhasis (Panda),
Sudipta, Tapamoy (Mukhanov), all were the part of a big happy family. 
Supratik da, Ratna di, Anupam da, Biswajit da, Saijad da and Kanan da
were like elder brothers and sisters and helped me to
overcome my initial imbalances. The occasional chats with all
the members here over a glass of tea at Kalida have been truly
refreshing in between work. I have learnt a lot of physics from
Prof. Sayan Kar, Prof Sugata Pratik Khastgir, Prof. Arghya Taraphdar,
Prof. Anirvan Dasgupta, Prof Krishna Kumar and Prof. Anushree Roy.  At
times I have interacted with the M Sc students here, Niladri, Anishur,
Kamakshya and Upasana. They also had to offer me there share of
knowledge.  Members of CTS, Ujal da, Venu da, Gopal da and
Subhabrata da were always available whenever I needed them.

I will miss the time with SPK. Be it a very scientific discussion
amongst us, or a pure``adda" at his residence, the scene was always
inspirational. Added to the delightful snacks, tea and coffee from
boudi those are the evenings I will always return to.

I am one of the privileged few to have great friends around. Bidesh
has been  there since my early school days and Prakash shares every
moments of 
mine here at IIT. Sanjit, Suman, Sourav, Surja, Samaresh, Goutam,
Tapomoy have never let me 
walk alone and Subhasis, Somnath (Nagakasi), Kamal, Somnath (bhaipo),
Ganesh have been 
more like my brothers. Nirupam, Sambit, Chandryee di at NCRA, Tapan da,
Saranya da, Tuhin at IUCAA and Wasim, Ruta, Yogesh, Kshitij at RRI
made my visits over there very pleasant. Starting from our RAS days
Wasim and I have discussed and quarreled probably over thousands of
hours in understanding different aspects of radio astronomy. However,
our friendship surely extends well out of our profession. I must not
forget mentioning here of Subhasis, Somnath, Kamal, Tapomoy, Shrobana and
Upasana again for wasting their time  in proof-reading this thesis.  

Finally, I will like to thank IIT Kharagpur, Department of Physics and
Meteorology, AJCB hall and all the  staff members for  making my stay 
of seven years comfortable and enjoyable.

\vspace{.5cm}
\hspace{10cm}Signature

%{\setlength{\baselineskip}{14pt} \setlength{\parskip}{2pt}

%\clearpage

%\newpage

%\clearpage{\pagestyle{empty}\cleardoublepage} %Added by PSD
\addcontentsline{toc}{chapter}{List of Symbols}
%\thispagestyle{empty}
%\newpage

\newcommand{\sline}[2]{ {{\bf #1}} & & & &  {#2} \\}
\newcommand{\sskip}{& & & & \\}
%%%%%%%%%%%%%%%%% definitions %%%%%%%%%%%%%%%%%%%%%%%%%%%
\def\HI{H~{\sc i}\, }
\def\V{{\mathcal{V}}}
\def\P{{\mathcal{P}}}
\def\A{\mbox{\boldmath$A$}}
\def\U{\mbox{\boldmath$U$}}
\def\n{\mbox{\boldmath$n$}}
\def\k{\mbox{\boldmath$k$}}
\def\d{\mbox{\boldmath$d$}}
\def\vr{\mbox{\boldmath$r$}}
\def\vx{\mbox{\boldmath$x$}}
\def\vt{\mbox{\boldmath$\theta$}}
\def\Nt{{\bf N_{2}}}
\def\S{\mathcal{S}}
\def\N{\mathcal{N}}
\def\Wt{\tilde{W}}

\def\E{\mathcal{P}}
\def\B{\mathcal{B}}
\newcommand{\wt}{\widetilde}

\begin{singlespacing}
\vspace{-3.5cm}
%%%%%%%%%%%%%%%%%%%%%%%%%%%%%%%%%%%%%%%%%%%%%%%%%%%%%%%%%%
\chapter*{List of Symbols}
\vspace{-1.cm}
%\centering
\begin{center}
{\bf \Large{  Acronyms}}
\end{center}
\begin{center}
\begin{tabular}{lcccl}
\hline
\sskip
\sline{\large {\bf Acronym  }}{\large {\bf Full form }} 
%\sskip
\hline \hline
\sskip
\sline{H~{\sc i}\, }{Neutral hydrogen}
\sskip
\sline{ISM}{Interstellar Medium}
\sskip
\sline{SFR}{Star Formation Rate}
%\sskip
%\sline{TRGB}{Tip of the Red Giant Branch}
\sskip
\sline{NGC}{New General Catalogue}
\sskip
\sline{IC}{Index Catalogue}
\sskip
\sline{SMC}{Small Magellanic Cloud}
\sskip
\sline{LMC}{Large Magellanic Cloud}
\sskip
\sline{CMBR}{Cosmic Microwave Background Radiation}
\sskip
\sline{AIPS}{Astrophysical Image Processing System}
%\sskip
%\sline{IRAS}{Infrared Astronomical Satellite}
\sskip
\sline{VLA}{Very Large Array}
\sskip
\sline{GMRT}{Giant Meterwave Radio Telescope}
%\sskip
%\sline{ATCA}{Australian Telescope Compact Array}
\sskip
\sline{WSRT}{Westerbork Synthesis Radio Telescope}
\sskip
\sline{FIGGS}{Faint Irregular Galaxy GMRT Survey}
\sskip
\sline{THINGS}{The HI Nearby Galaxy Survey}
\hline
\end{tabular}
\end{center}
\newpage

\begin{center}
{\bf \Large{Symbols}}
\end{center}

\begin{center}
\begin{tabular}{lcccl}
\hline
\sskip
\sline{\large { \bf Symbols  }}{\large {\bf Definitions }} 
%\sskip
\hline \hline
\sskip
\sline{$\U$}{Baseline in kilo wavelength}
\sskip
\sline{$\nu$}{Observing frequency in MHz}
\sskip
\sline{$\vt$}{Angle in the sky measured}
\sline{}{from the centre of the galaxy}
\sskip
\sline{$\V(\U, \nu)$ or $\V(\U)$}{Visibility}
\sskip
\sline{$I(\vt, \nu)$}{Specific Intensity}
\sskip
\sline{$\tilde{I}_{\nu}$}{Smooth part of $I(\vt, \nu)$}
\sskip
\sline{$\delta I(\vt, \nu)$}{Fluctuating part of $I(\vt, \nu)$}
\sskip
\sline{$W(\vt, \nu)$ or $W(\vt)$}{Window function}
\sskip
\sline{$\theta_{0}$}{Galaxy angular radius}
\sskip
\sline{$A(\vt, \nu)$}{Antenna primary beam}
\sskip
\sline{$\N(\U, \nu)$}{Noise at baseline $\U$ and frequency $\nu$}
\sskip
\sline{$\tilde{\delta I}(\U, \nu)$}{Fourier transform of $\delta
  I(\vt,   \nu)$}
\sskip
\sline{$\delta^{2}_{D}(\U)$}{Two dimensional Dirac delta function}
\sskip
\sline{$P_{HI}(\U, \nu)$}{Power spectrum}
\sskip
\sline{$\tilde{W}(\U, \nu)$ or $\tilde{W}(\U)$}{Fourier transform of
  $W(\vt, \nu)$ or $W(\vt)$} 
\sskip
\sline{$\P(\U)$}{Power spectrum estimator}
\sskip
\sline{$U_{m}$}{Lowest baseline above which}
\sline{}{power spectrum is estimated}
\sskip
\sline{$1 M_{\odot}$}{Solar mass: $1 M_{\odot}$  = $1.99 \times 10^{30}$ kg}
\hline
\end{tabular}
\end{center}
\newpage
\begin{center}
\begin{tabular}{lcccl}
\hline
\sskip
\sline{\large { \bf Symbols  }}{\large {\bf Definitions }} 
%\sskip
\hline \hline
\sskip
\sline{$U_{min}$}{Minimum value of the baseline above}
\sline{}{which power spectrum is a power law}
\sskip
\sline{$U_{max}$}{Maximum value of the baseline below}
\sline{}{which power spectrum is a power law}
\sskip
\sline{$R_{min}$}{Minimum length-scale above which}
\sline{}{power spectrum is a power law}
\sskip
\sline{$R_{max}$}{Maximum length-scale below which}
\sline{}{power spectrum is a power law}
\sskip
\sline{$A$}{Amplitude of the power law}
\sskip
\sline{$\alpha$}{Power law index}
\sskip
\sline{$\sigma_{\P}$}{Variance in the power spectrum estimator}
\sskip
\sline{$N_{2}$}{Noise covariance}
\sskip
\sline{$N_{g}$}{Number of independent estimates of $\P(U)$}
\sline{}{at a given $U$ bin}
\sskip
\sline{$N_{b}$}{Number of visibility pairs}
\sline{}{at a given $U$ bin}
\sskip
\sline{$\hat{h}(\vr)$}{Homogeneous and isotropic}
\sline{}{Gaussian random field}
\sskip
\sline{$P(k)$}{Three dimensional power spectrum}
\sskip
\sline{$\gamma$}{Power law index of the}
\sline{}{three dimensional power spectrum}
\sskip
\sline{$G(\vr)$}{Three dimensional structure of the galaxy}
\sskip
\sline{$z_{h}$}{Scale height of the galaxy}
\sskip
\sline{1 pc}{parsec: 1 parsec = $3 \times 10^{16}$ m}
\hline
\end{tabular}
\end{center}
\end{singlespacing}
\newpage
\thispagestyle{empty}

%\clearpage{\pagestyle{empty}\cleardoublepage} %Added by PSD
\addcontentsline{toc}{chapter}{List of Tables}
%\newpage
\begin{singlespacing}
\listoftables
\end{singlespacing}

%\clearpage
%\newpage

%\setcounter{page}{17}

%\thispagestyle{empty}   %Due to this line there is problem in contents

%\newpage

%Just add the line \setcounter{page}{17} complile then remove
%problem may be solved temporarily 

%\chaptermark{$\empty$}
%\clearpage{\pagestyle{empty}\cleardoublepage} %Added by PSD
\addcontentsline{toc}{chapter}{List of Figures}
\begin{singlespacing}
\listoffigures
\end{singlespacing}

%\newpage

%\thispagestyle{empty}

%\newpage

%\clearpage
\clearpage{\pagestyle{empty}\cleardoublepage} 
\chaptermark{$\empty$}
\addcontentsline{toc}{chapter}{Abstract}

%\thispagestyle{empty}

%\pagenumbering{roman}

%\vspace{-0.5cm}
%\vspace{2cm}    %%%Partha

\begin{center}
{\Large \bf Abstract}
\end{center}
%\vspace{.2cm}

\begin{singlespacing}
In this thesis we use radio-interferometric observations to probe the
intensity fluctuation power spectrum of the neutral hydrogen (\HI) 21-cm
emission from the interstellar medium (ISM) of the external 
galaxies. We develop a visibility based power spectrum estimator to
probe the power spectrum directly from the interferometric
observations and also estimate the errors in it. A numerical
simulation of synthetic observations is also performed to access the
efficacy and limitations of this estimator. We use this estimator to
evaluate the power spectrum of three individual  spiral galaxies, a
dwarf galaxy sample and THINGS  \footnote{THINGS: The HI Nearby Galaxy
  Survey} spiral galaxies. In
each case, the power spectrum is found to follow a power law $P_{\rm
  HI}(U) = A U^{\alpha}$ over a specific length scale range. The
estimated  value for the slope $\alpha$ ranges from $\sim -1.5$  to
$\sim -2.6$ for the sample of dwarf galaxies. We interpret this bi-modality 
as arising due to 2D turbulence on length scales
much larger than the scale-height of the galaxy disk and 3D otherwise.
The power law slope also show a weak correlation with the surface
density of star formation rate for these galaxies. We found for the
external spiral galaxies the power spectrum is a power law up to a
length scale of $10$ to $16$ kpc, indicating turbulence to be operational
at these large length scales. We measure the scale-height of the
external nearly face-on spiral galaxy NGC~1058 to be $\sim 500$ pc
interpreting  a change in the slope observed in it's power
spectrum. Power spectrum of  the  harassed galaxy NGC~4254 is found to
be affected 
by galaxy harassment. For 18 spiral galaxies derived from the THINGS
sample we found no correlation between the power law slope and the
different dynamical parameters of these galaxies. For most of the
dwarf and spiral galaxies, the intensity fluctuation power spectrum is
found to be  a direct probe of the density fluctuation and not
affected by the velocity fluctuations in ISM. We would attempt to
understand  these new observations in terms of physical models of the
ISM in future.

\vspace{3cm}

{\bf Keywords}: Interstellar Medium, Turbulence, Galaxies,
Scale-height, Power Spectrum, visibility correlation.

\end{singlespacing}

%\newpage
\clearpage

\thispagestyle{empty}

\newpage
%\newpage

%\clearpage {\pagestyle{empty}\cleardoublepage}
\addcontentsline{toc}{chapter}{Contents}
\tableofcontents
\newpage
\thispagestyle{empty}

\setcounter{chapter}{0}

\setcounter{section}{0}
\setcounter{subsection}{0}
\setcounter{subsubsection}{2}
\setcounter{equation}{0}
%%\pagenumbering{arabic}
\pagenumbering{arabic} 
\setcounter{page}{1}    %%%%%KKD used \setcounter{page}{0}  
\oddsidemargin 0.9 cm 
\evensidemargin -.4 cm
\setlength{\textwidth}{152.4 mm}

\chapter{Introduction}
\label{chap:intro}
\vspace{-.5cm}
It is now well established \citep{1998gaas.book.....B} that
all galaxies have an interstellar medium (ISM) irrespective of their
morphological class. This medium carries $10-15 \%$ of the baryonic
mass of a galaxy and  is  known to contain rarefied gas (mostly
hydrogen and helium), dust (silicate and graphite),  
relativistic electrons and large scale magnetic fields. The
structure of the ISM is different in the  spiral and irregular
galaxies compared to that in ellipticals. Moreover, for most of the
observed galaxies, the gaseous ISM extends well beyond the stellar
disk. Most of this gaseous medium  is hydrogen
($70\, \%$) and helium  ($28\, \%$), whereas there is a trace of the
heavier elements ($1.5\, \%$) injected to the ISM during
supernovae. Interstellar matter is distributed in discrete clouds of
molecular and atomic gas with no preferred cloud scale. The molecular
clouds have a temperature of about $100-200$ K compared to the cold
atomic gas with temperature $50-100$ K. These denser clouds occupy only
$1-2 \, \%$ of the interstellar space and constitutes half of its mass. The
rest of the ISM in between these clouds consists of warm neutral and
ionized gas. These various forms of interstellar matter are highly coupled
and is regarded as a single dynamical entity. In the process of star
formation, clumps of ISM driven by gravity collapses to give birth to
stars, whereas, dying stars in supernovae, ejects a huge 
amount of gas, dust and metal in the ISM making it highly coupled with
the stellar disk of the galaxy.

\section{A short overview on the ISM turbulence}

Over the last few
decades the structure and dynamics of the ISM has been  extensively studied
by observing it in both emission and absorption. The ISM is
found to have rich scale invariant structures arranged in clouds of
different shapes and sizes. Based on various observations and
theoretical understandings it is now believed that the ISM is
turbulent.  
 \citet{2004ARA&A..42..211E} give a
review of the progress in observations 
of ISM turbulence over the last few decades
whereas \citet{2004ARA&A..42..275S} review the theoretical implications
and effects of these observations.

Pioneering work in the theory of ISM was done as early as 1951
by \citet{1951ApJ...114..165V}. He outlined a model for the ISM with a
hierarchy of  
structures formed by the supersonic turbulence stirred at the large
length-scales and dissipated at small scales by atomic viscosity.  In
the following year one of the first statistical models of a continuous
and correlated gas distribution was proposed by \citet{1952ApJ...115..103C},
who successfully applied it to the extinction fluctuations of Milky
Way surface brightness. Observationally, absorption of starlight by
the Na and Ca$^{+}$ ions revealed clumpy structure of the
ISM \citep{1998gaas.book.....B}. 

Recent
observations and numerical modellings, suggest that 
supersonic turbulent flows play a large role in the processes of star
formation \citep{2003ApJ...592..975L}.  \citet{2003ASPC..287...81V}
have shown, that first the transient clumpy molecular
cloud structures are generated through turbulent fragmentation and
then the individual clumps collapse to form proto-stars.

Turbulence generates scale-invariant stochastic fluctuations in the
velocity and density field of the ISM. These fluctuations can be
traced observing the ISM in emission and absorption lines of gas and
metals. Observationally, statistical estimators are used to quantify
the nature of these fluctuations as well as the turbulence that generated
by them. Velocity fluctuations are quantified using structure function of
order two whereas auto-correlation function or the power spectrum is
used to determine the density fluctuations. Most of the knowledge of
ISM turbulence is acquired using these statistical measures. Here we
briefly discuss a few statistical tools which are widely used to
measure and quantify the ISM turbulence.

\subsection*{Auto-correlation function and power spectrum}
The stochastic 
density field generated by fully developed turbulence is statistically
homogeneous and isotropic and is believed to follow the Gaussian statistics.
The statistical nature of such  fluctuations can be completely
specified by either the  
auto-correlation function or the power spectrum.
The auto-correlation function
$\xi_{A}(\mid \delta \vr \mid)$ of any  homogeneous and isotropic 
scalar field $A$ is defined as  
\begin{equation}
\xi_{A}(\mid \delta \vr \mid) \ =\ \left < A (\vr)\, A(\vr
+ \delta \vr) \right >. 
\end{equation}
Fourier transform of the auto-correlation function is known as
the power spectrum, i.e,  
\begin{equation}
P_{A}(k) = \int  d \vr
e^{i \k \cdot \vr} \xi_{A}(\mid \delta \vr \mid) \, ,
\end{equation}
where $k = \mid \k \mid$. For both the above
definitions, the angular brackets denote ensemble averages.  For a
scale-invariant fluctuation both the auto-correlation function as well
as the power spectrum are expected to follow power laws, i.e,
$P_{A}(k) \sim k^{\alpha}$. The slope of these power law $\alpha$ is
believed to be a good quantifier of the nature of the turbulence
generating these structures. 

\subsection*{Structure function}
The structure function of order $p$ for an observable $A$ is defined as 
\begin{equation}
S_{p}(\mid \delta \vr \mid) = \left < \mid A(\vr) - A(\vr
+ \delta \vr) \mid ^{p}\right >. 
\end{equation}
For a  scale-invariant field, 
$S_{p}(\delta r)$ are also expected to be power laws, i.e, $S_{p}(\delta r)
 \sim \delta r^{\zeta_{p}}$. It is to be noted that the structure
 function of order two  is directly related to the auto-correlation
 function by $S_{2}(\mid \delta \vr \mid) = 2 ( <A^{2}(\vr)> - \xi(\mid \delta \vr \mid) )$. 

\subsection*{Delta variance}
Another common  measure of a stochastic field is the  delta
variance defined as
\begin{equation}
\sigma^{2}_{\Delta} (L) \ =\ \left < \int_{0}^{3L/2}\, d\vx \left
[ \left ( A(\vr + \vx ) - <A> \right ) \Theta (\vx) \right
]^{2} \right >,
\end{equation}
where the cutoff function $\Theta$ is given by
\begin{equation}
\Theta(\vx) = \pi \, \left ( \frac{2}{L} \right ) ^{2} \, \times \,   
\begin{cases} 1 & \text{if $x<L/2$,}
\\
\frac{1}{8} &\text{if $L/2 < x < 3L/2$.}
\end{cases}
\end{equation}
Power spectra measurements are typically limited by sample variance at
large length-scales. Delta variance avoids this
problem by using the cutoff function $\Theta(\vx)$.

It is to keep in mind that the estimators discussed above do not
provide complete information if the stochastic field they probe has
non-Gaussianity. Probes for non Gaussian fluctuations include structure
function of higher order, higher order correlation functions,
bispectrum, trispectrum etc. However, measuring these from the present
observations is very difficult and is not usually addressed. 

Most of the known facts about ISM turbulence is inferred by estimating
the above quantifiers from mostly spectroscopic observations and then
comparing to existing models of turbulence. Here we discuss how some
important aspects of ISM turbulence are inferred from various
observations. 

\subsection{Observations an inferences}

\citet{1951ZA.....30...17V} found that the structure function of velocity
fluctuations in the direction of the Orion nebulae is a power law. The
slope  was found to be $\sim 0.25 - 0.5$ which
indicated a Kolmogorov energy spectrum. However, using a better data
\citet{1959ApJS....4..199W} got a steeper power law with slope $\sim0.6$ and
inferred in favour of compressible turbulence. 

Apart from being scale-invariant, the density field generated by
turbulence is also intermittent.  Radio scintillation observations
showed the clumpy nature of ISM at very small scales down to $10^7$ m or
lower \citep{1973MNRAS.162..329L}.  These could be understood using a
model of magnetic turbulence generated by streaming instabilities
\citep{1968ApJ...152..987W}. However, the relation of turbulence at
these small length-scales and that observed at parsec length-scales was
not very well understood.

During 1980s, a series of observation of spectral line-width at
different line of sight revealed power law
correlations. \citet{1990ApJ...365L..27G} introduced wavelet
transformation for the analysis of $^{13}$CO centroid velocities in
L~1551, and the measured velocity-size slope was
$0.7$.  \citet{1992MNRAS.256..457H} used a 
clump finding algorithm and various  other correlation techniques for
HCO$^{+}$ and HCN along M~17SW. They found scale-invariant fluctuations
for scales $<1$ pc. \citet{1994ApJ...432..622M} used the structure
function to determine a slope of $0.43\pm0.15$ for the velocity-size 
relation.

Comparing delta variance observations with models with and
without gravity \citet{2001A&A...379.1005O} concluded that self gravity
increases power at small scales. 
The observations of the Polaries flare by \citet{2002A&A...390..307O}
is compared with the models of delta variance for isothermal MHD
turbulence. They found that the turbulence is driven from outside
and the dissipation scale is smaller than their resolution
scale. 

Observing the extinction in the Taurus region, \citet{2002ApJ...580L..57P}
found that $\zeta_{p}/\zeta_{3}$ varies for $p=1$ to $20$ in a similar
manner as  in a model of supersonic
turbulence proposed by \citet{2002PhRvL..89c1102B}. In these models
the dissipation 
of supersonic turbulence occurs in sheets giving $\zeta_{p}/\zeta_{3}
= p/9 + 1-3^{-p/3}$ assuming She-Leveque
scaling \citep{PhysRevLett.72.336}. \citet{2003ApJ...583..308P} 
got a similar result observing $^{13}$CO emission lines along Perseus
and Taurus.

There is a considerable amount of effort in quantifying the
scale-invariant structures present in the ISM using fractal and 
multi-fractal analysis. Perimeter area dimension 
are measured for  the contour maps of 100 micron dust
intensity \citep{1988ApJ...333..353B}, CO
emission \citep{1991ApJ...378..186F} etc. The estimated value for the
perimeter area dimension was 
$1.2 - 1.5$ which is 
consistent with that of a slice from the 3D
structures observed in the terrestrial clouds and laboratory
turbulence \citep{1991AnRFM..23..539S}.

\citet{2001ApJ...551..712C} have performed multi-fractal analysis of the
column density maps derived from IRAS (The Infrared
Astronomical Satellite) 60 $\mu$ and 100 $\mu$ 
images. They observed the region to region diversity in
the multi-fractal spectra was significantly different from that
expected from incompressible turbulence. 

These observations not only established that the ISM has clumpy
scale-invariant structures generated by turbulence over a large
length-scale range, but also provided hints for the
characteristics of 
the ISM turbulence. Though initially some similarity was found with
the Kolmogorov energy spectrum, it is now believed that ISM turbulence is
quite different. It is a compressible and supersonic fluid
turbulence  driven at different scales by different mechanisms and
induces scale-invariant multi-fractal structures  in the density as well as
the velocity field.

\subsection{Energy sources}
There is no wide-scale effective model for ISM turbulence which can
explain all the observed features. The physical processes responsible
for turbulence also differ enormously across the length-scales. 

At smaller scales radio scintillation observations of intermittency can be 
well explained by sonic reflection of shock
waves \citep{1984ApJ...283..825I, 1988ApJ...332..984F}, cosmic ray
streaming and other
instabilities \citep{1969ApJ...156L..91W, 1980MNRAS.190..353H}, energy
cascades from larger scales etc. \citet{2004ApJ...603..180L}
suggested that moderate turbulence can be maintained in the ionized
clouds by variations in the background ionizing UV
radiation. \citet{1996ApJ...467..280N} recognized that most Type II
SN contribute to cluster winds and super-bubbles at a length-scale range
of $100-500$ pc. 

The gravitational binding energy of a galaxy disk heats the stellar
population during swing-amplified shear
instabilities \citep{1998MNRAS.294..513F}  and also heats the gas
triggering turbulence \citep{2002A&A...386..359H}. These gas clouds
contiguously collapse to feed 
more turbulence at the smaller
scales \citep{1999PhRvD..59l5021S}. This kind 
of gravitational source for turbulence is consistent with the observed
power spectra of flocculent spiral
arms \citep{2003ApJ...590..271E}. Since gravity acts at all the
length-scales, the self gravity can feed turbulence at a wide range 
of length-scales.

Since ISM extends well beyond
the stellar disk, galactic rotation driven turbulence can be effective
at the outer regions of the galaxy's disk. Magneto-rotational instability,
proposed by \citet{1999ApJ...511..660S, 2003ApJ...599.1157K}, explains how
galactic rotation pumps energy to the gas motion and hence feeds
turbulence. \citet{2004ApJ...601..905P} have considered a revised model
where they assumed that the  reduced dissipation enhances energy input 
to turbulence by magneto-rotational instability. Rotational energy also
generates spiral shocks where the fast-moving inter-spiral medium hits
the slower dense gas in the density wave
arms \citep{1969ApJ...158..123R} making the ISM turbulent at the outer
edges of  the disk. 

These physical mechanisms though could explain the observed
fluctuations or shapes of the spectral lines at different 
length-scales, their interplay in generating wide scale coherent
structures is not very well understood.

\subsection{Simulations}

Simulation play an effective role in understanding various physical
models, particularly nonlinear processes such as
turbulence. Observationally, only a static picture of ISM can be probed
since the dynamic time scales involved are large. Using numerical
hydrodynamic and magneto-hydrodynamic simulations it is possible to
understand the effect of different mechanisms that drives ISM
turbulence. Here we briefly discuss the different studies performed in
this  direction.

\citet{1980ApJ...239..173B} performed the first ever simulation of ISM
turbulence. However, the lack of computational facilities limited them
to a very small length-scale range. Higher resolution studies of
non-gravitating and non-magnetogenic compressible turbulence was first
performed by \citet{1999PhyA..263..263P, 2002PhRvE..66b6301P}. Their
models were dominated by filaments, sheets and clusters in the
compressible part of the flow and can explain the observed morphology
of diffuse and small molecular clouds.

The first nonmagnetic simulations which included gravity were
performed by \citet{1990MNRAS.243..293L} on a $128^{3}$ grid,
whereas \citet{1993PhFl....5.1092Z} did the first non-gravitating
compressible simulation to 
include magnetic field. These earlier works were expanded by considering
different aspects of ISM like magnetic and gravitational fields,
realistic heating and cooling mechanisms, stellar energy injections and galactic
rotation \citep{1995ApJ...441..702V,1996ApJ...473..881V}. These
simulations were meant to reproduce the 
ISM dynamics at the length-scale range of $\sim 1$ pc.
In recent times higher resolution simulations \citep{2003PhFl...15L..21K}
observed a constant energy flux over a large length-scale range with
significant departure from Kolmogorov energy spectrum.  

Despite the efforts simulation of ISM turbulence remained
challenging due to  the huge range of length-scales involved and high
Reynolds number. It is important to include a wide range of length
scales along with different models of energy inputs in the simulations to
understand the interplay and cascading of the energy driving the ISM
turbulence. Considering viscosity as the damping mechanism,
Reynolds number ($Re$) for the ISM is estimated as high as $10^{5}$ to
$10^{7}$. The number of degrees of freedom for three
dimensional incompressible turbulence scales like $Re ^{9/4}$. Hence
simulating turbulence with $Re \sim 10^{5}$ requires a dynamic range
of $10^{5}$ which is far beyond the reach of present day computers.  

\section{Radio-interferometric observation as a probe of the ISM  turbulence} 

In the previous section we have discussed the observation of ISM using
different atomic and molecular lines which suggest
turbulence. However, these elements, like dust, Ca$^{+}$, Na, CO
etc., constitute a very small part of the interstellar matter both by
mass and volume. These   also are not very evenly
distributed over the galaxy's disk. Almost $70\, \%$ of 
 the gas in ISM is  dominated by the atomic (H~{\sc i}) and ionized
 (H~{\sc ii}) hydrogen which is spread all over the galaxy's disk
 making it a better tracer  of the ISM dynamics. 

\subsection*{21-cm radiation}
The hyperfine structure transition radiation (21-cm radiation) of the
neutral atomic hydrogen in the ground state (H~{\sc i}) is an excellent probe
of the column density fluctuations in the
ISM. This spectral  line has a  frequency of 1420 MHz, which falls
in the radio band of the electromagnetic spectrum. The
associated transition probability being  very small, this line  is
considered as forbidden  for laboratory experiments. However, at
astrophysical scales it is possible to observe this radiation using very
sensitive radio telescopes. The
ISM has relatively small optical depth at these frequencies and the
observed intensity 
fluctuations in the sky $I(\vt, \nu)$ directly probes the column
density fluctuations  $N_{HI}(\vt, \nu)$ in ISM, i.e,
\begin{equation}
I(\vt, \nu) = \frac{3}{16 \pi} \, h \nu\, A_{21}\,
N_{HI}(\vt) \phi(\nu)
\end{equation} 
Here, $\vt$ is a vector  in the sky, $\nu$ is the observing
frequency and $A_{21}$ is the corresponding Einstein
coefficient. $\phi(\nu)$ gives the shape of the \HI spectral line. 

\subsection*{Radio-interferometric observations}
Most of our understanding regarding the ISM is based on the
observations in our Galaxy, the  Milky Way. This is because the single
dish radio telescopes 
lack the necessary resolution to resolve the structures in the external
galaxies. Radio-interferometers like the VLA (Very Large
Array\footnote{Very Large Array, NRAO, New Mexico}), GMRT (Giant
Meterwave Radio Telescope\footnote{Giant
Meterwave Radio Telescope, NCRA, TIFR, India}), WSRT (Westerbork
Synthesis Radio Telescope\footnote{Westerbork
Synthesis Radio Telescope, Netherlands Foundation for Radio
Astronomy}) on the other hand 
provides higher resolution. These are capable of probing ISM structures 
and dynamics at smaller length-scales in our Galaxy and also provides
means of observing turbulence at larger length-scales in external
galaxies. In this section we briefly discuss the different efforts
made in
probing turbulence using radio-interferometric observations.

Unlike conventional optical, infrared or ultraviolet telescopes, radio
interferometers do not measure the sky intensity, rather they measures
the visibilities which are the Fourier transform of the sky brightness
fluctuations. These visibilities can then be used to construct 
column density maps or the image of the observed field of
view \citep{1989ASPC....6.....P}. Using these column density maps 
the auto-correlation functions can be estimated. However, these
technique has limited utility due to the 
 the presence of correlated noise at the image-plane generated because
 of limited baseline coverages (see {\bf
 Appendix~\ref{app:cornois}}). 

\citet{1983A&A...122..282C} measured
the power spectrum 
of \HI emission from our galaxy squaring the directly observed
visibilities (from radio-interferometric
observations using WSRT) and subtracting the noise bias estimated
from the line-free channels. They found that the power spectrum is 
roughly a power law with a slope of $\sim -3$ up to the largest of
$10$ pc they could probe. Using similar techniques, 
\citet{1993MNRAS.262..327G} estimated the power spectrum of the \HI
emission of our galaxy from observations with  DRAO (Dominion
Radio Astrophysical 
Observatory\footnote{Dominion Radio Astrophysical
Observatory : National Research Council, Canada}) synthesis radio
telescopes. He found that the  
power spectrum is 
a power law with slope $-2.2-3.0$ up to a length-scale of $15$
pc. 

\citet{2000ApJ...543..227D} used a high dynamic range
estimator to measure the power spectrum of the optical depth map
created by observing \HI in the  Perseus spiral arm of our Galaxy in
absorption 
towards Cassiopeia A.   They found that the power spectrum can be well fitted
by a power law with slope $-2.75$ for the length-scale ranges $0.10 -
3.0$ pc. Based on their observation they concluded that the  ISM
exhibits non-Kolmogorovian turbulence at these length-scales.

Fourier transform power spectra of the Large  and Small Magellanic
Clouds (LMC, SMC) are also found to obey power
laws \citep{2001ApJ...548..749E, 2010ApJ...708.1204B} for a
length-scale range of $20-200$ 
pc. Other techniques adopted by
different authors to probe the ISM turbulence include velocity
channel analysis or velocity 
coordinate spectrum \citep{1995A&A...293..507L}, fractal dimensional
analysis \citep{1998A&A...336..697S} and  spectral correlation
function \citep{1999ApJ...524..887R} etc. These methods were limited
either by the correlated noise induced by incomplete baseline coverages of
radio-interferometric observations or by the noise bias in the power
spectrum. They lack the necessary sensitivity to probe the 
faint emissions from the nearby dwarf or external galaxies.

\section{Motivations and  objectives}

ISM of the Milky way is studied at length-scales ranging from $10^7$ m
to $\sim 10$ pc and is found to have scale invariant power spectrum
with slope $-2.0$ to $-3.0$ indicating turbulence is operational. For
the nearby galaxies (LMC, SMC), a similar power spectrum is observed
till a length-scale of $\sim 200$ pc. The proposed driving mechanisms
of these observed turbulence in our galaxy includes galactic rotation
and self gravity which operates at even larger length-scales compared
to the galaxies disk. It is necessary to probe the
ISM of the external faint dwarf and nearby large spiral galaxies 
 to investigate the existence of turbulence at large length-scales. 

Conventionally, star formation in galaxies was known to be
governed by the interplay of the gravity and magnetic pressure
modulated by the ambipolar diffusion. However, recent observations and
numerical modelings suggest that supersonic turbulent flow
rather than the static magnetic field has larger role to play in
this process \citep{2003ApJ...592..975L}. Star formation is now believed
 to be a two phase  process \citep{2003ASPC..287...81V}; first
the generation 
of the transient clumpy molecular cloud structures by turbulent
fragmentation  and then the collapse of the individual clumps to form
the proto-stars. The fragmentation governed by turbulence in the ISM
leads to cloud structures which are reflected in the fluctuations of
the \HI 21-cm intensity. Hence, these fragmentation can be traced
estimating the power spectrum of the \HI 21-cm intensity fluctuations. 
Dwarf galaxies often have irregular morphologies and peculiar
large scale kinematics. This also makes them interesting to
study the ISM turbulence beside investigating the galaxy
formation scenario. Observations of the nearby faint dwarf galaxies
with sufficient spatial resolution offers us
a good opportunity to  understand the early stages of the star  as  
well as the galaxy formation.

Observationally it is difficult to estimate the power
spectrum of the external galaxies because of the instrumental noise.
A visibility  
based estimator was used by \citet{2002MNRAS.334..569H} to estimate
the Cosmic Microwave Background Radiation (CMBR) angular power spectrum.
\citet{2005MNRAS.356.1519B}
have introduced a  visibility based power spectrum estimator which can
be used to probe the intensity fluctuations of \HI 21-cm line using 
radio-interferometric observations. A modified form of this estimator
is used by \citet{2006MNRAS.372L..33B} to estimate the power spectrum of
the nearby dwarf galaxy DDO~210. The power spectrum was found to be
power law with slope $\sim -2.8$ for a length-scale range till $500$ pc
suggesting the presence of turbulence in the ISM of the dwarf galaxies.

Using a similar estimator from radio-interferometric observations of
H~{\sc i}, it is possible to probe the ISM in external dwarfs
and spirals and investigate the existence and nature of turbulence at
kpc length-scales. The work presented in this thesis aims to use these
observations to estimate the power spectrum of  \HI column density
fluctuations in a large sample of galaxies. Earlier studies indicate
that these fluctuations, which typically have a scale-invariant power
spectrum, are possibly the outcome of turbulence in the ISM. The
results of our study are expected to shed light on  the nature of ISM
turbulence. 

The specific objectives of this thesis are highlighted below:

\begin{itemize}
\item Visibilities are the quantities directly measured in
radio-interferometric observations, and there are several advantages
if these be used to directly estimate the  power spectrum. This is
particularly important for the \HI emission from faint external
galaxies where the signal is weak and buried in noise. It is thus
necessary to have a well quantified  technique for visibility based power
spectrum estimation.  

\item Simulations play an important role in assessing  the efficacy and
limitations of any technique.  It is very important to apply the 
power spectrum estimation technique to simulated galaxies in order to
asses the dynamical range and errors in the estimate. 

\item Having established a technique, it is then appropriate to apply
this to estimate the power spectrum of the ISM of a large sample of
dwarf and spiral galaxies. 

\item Finally, it is desirable to have some understanding of the
measured power spectra in terms of the underlying ISM  turbulence and
the processes driving it. 

\end{itemize}
\section{Outline of the thesis}

An outline of the rest of the thesis is as follows. 

Throughout our different investigations presented in this thesis we
use archival radio-interferometric data observed using the GMRT or the
VLA. The directly observed quantity in
these observations is visibility, which is the Fourier transform of
the sky intensity fluctuations. In {\bf Chapter~\ref{chap:est}} we
discuss a visibility based power spectrum estimator. The application of
such an estimator to the \HI 21-cm observation of the ISM 
and the error estimation are also discussed. 

We present the power spectrum
analysis of the nearly face-on spiral galaxies NGC~628 and NGC~1058
and a harassed galaxy NGC~4254 of the Virgo cluster in {\bf
Chapter~\ref{chap:3spi}}. 
In {\bf Chapter~\ref{chap:sim}} we perform numerical simulations to
study the 
efficacy and limitations of the power spectrum estimator and access 
the dynamic range of length-scales for which it can be used. We
present the result of our power spectrum analysis of a sample of 7 
nearby external dwarf galaxies, and  a sample of 18 external spiral
galaxies from the 
THINGS\footnote{{\bf THINGS} : {\bf T}he {\bf HI} {\bf N}earby {\bf
G}alaxy {\bf S}urvey, \citet{2008AJ....136.2563W}} survey in {\bf
Chapter~\ref{chap:dwarf}} and {\bf \ref{chap:THINGS}}
respectively. Finally, in {\bf Chapter~\ref{chap:summary}}, we summarize the 
main results of the thesis and discuss about the future scopes.

%\newpage

%\clearpage{\pagestyle{empty}}\cleardoublepage} %%%%%%%%%%%%%%%%%%%%\
 %\newpage 
 \setcounter{section}{0}
 \setcounter{subsection}{0}
 \setcounter{subsubsection}{2}
 \setcounter{equation}{0}
 %\pagenumbering{arabic}

%-------------------------------------------
\chapter[Visibility Based Power Spectrum Estimator]  
{\bf \textbf {Visibility based power spectrum
estimator\footnote{A part of the work presented in this chapter is originally
published in the paper titled ``A study of interstellar
medium of dwarf galaxies using \HI  power spectrum analysis"
by \citet{2009MNRAS.398..887D}.}}}  
\label{chap:est}
\section{Introduction}
Turbulence generates scale-invariant structures in the ISM density
field. These structures can be quantified through the power spectrum
of \HI  21-cm radiation. The power spectrum is expected to be a power
law for scale-invariant fluctuations. In this chapter we discuss a
visibility based estimator which can be used to probe the \HI  intensity
fluctuation power spectrum of external faint dwarf or spiral
galaxies using radio-interferometric observations. We also present the
error estimation and limitations in this 
estimator. 

\section{Visibility-visibility correlation}
\label{sec:visicorr}
The specific intensity of  \HI emission from a  galaxy at the phase
centre of the observation can  be modeled as 
\begin{equation}
I(\vt, \nu) \ =\ W_{\nu}(\vt)\left[ \bar{I}_{\nu} \ +\ \delta
  I(\vt, \nu) \right ]\,,
\label{eq:a1}
\end{equation}
were $\vt$ is the angle  on the sky  measured in radians 
from the center of the galaxy. We assume that the
galaxy subtends a  small angle so that $\vt$ may be treated as a two
dimensional (2D) planar vector on the sky. The \HI  specific intensity
is modeled as  the sum of a  smooth component and a 
fluctuating component.  Typically, $I(\vt, \nu)$ is maximum
at the center and declines  with increasing $\mid \vt \mid$.
We model this through a  window function   $W_\nu(\vt)$ 
which is defined so that $W_\nu(0)=1$ at the center and has values $1
\ge W_\nu(\vt) \ge 0$ elsewhere. This multiplied by $\bar{I}_\nu$
gives the  smooth component of the specific intensity. 
For a face-on  galaxy, the window function $W_\nu(\vt)$  corresponds
to the galaxy's  radial profile.   

In our analysis we  consider two different models for the window
function of a galaxy of angular radius  $\theta_0$. In the ``top-hat''
model it is assumed that the specific 
intensity has a constant value within a  circular disk of
radius $\theta_0$, and it abruptly falls to $0$ outside. 
The window function, in this case,  is a Heaviside step function
\begin{equation}
W_\nu(\vt)=\Theta(\theta_0-\theta)
\end{equation}
 In the ``exponential'' model the window function has the form 
\begin{equation}
W_{\nu}(\vt) =\exp\left(-\frac{\sqrt{12} \theta}{ \theta_0}\right)\, ,
\label{eq:expw}
\end{equation}
where it falls exponentially away from the center. Note that for both
the models we have assumed the radial profile to be isotropic compared
to the centre of the galaxy.

It is also possible to use $W_{\nu}(\vt)$ to  define a normalized window
function $W^N_{\nu}(\vt)$ such that $\int d \vt
\, W^N_{\nu}(\vt)=1$. The second moment of $\theta$ defined as $\int
d \vt \, \mid \vt^{2} \mid \, W^N_{\nu}(\vt)$ provides a good estimate of
the angular extent of the window function. Here we have used the
condition that 
this should have the same value $\theta^2_0/2$ for all models 
of the window function, to 
determine the width of the exponential window function in terms of
$\theta_0$.  
Throughout we have assumed $\theta_0 \ll 1\ {\rm radian}$.

\begin{figure}
\begin{center}
\epsfig{file=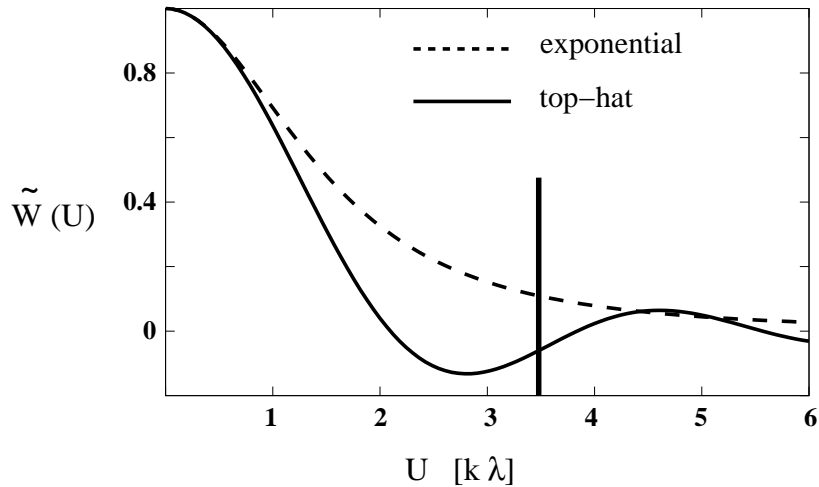,width=4.2in, angle=0}
\end{center}
\caption{$\Wt(\U)$ for top-hat and exponential window functions for
  $\theta_{0} = 1'$. The vertical line marks   $\theta_{0}^{-1}$. Note
  that the FWHM (Full Width at Half Maxima) of the two window functions
  are nearly the same and both 
  window functions have a very small value for  $U >  \theta_{0}^{-1}.$} 
\label{fig:wind}
\end{figure}
We express the fluctuating component of the specific intensity as 
$W_\nu(\vt) \, \delta I_\nu(\vt)$. Here $\delta   I_\nu(\vt)$ is a
stochastic fluctuation which is assumed to be statistically homogeneous
and isotropic. 
The \HI  emission traces these fluctuations modulated by
the window function which quantifies the large-scale \HI  distribution. 
In this thesis we use radio-interferometric observations
to  quantify the statistical properties of the fluctuations $ \delta
I_\nu(\vt)$. These fluctuations  are believed to be the outcome of
turbulence  in the ISM. 

The visibility $\V(\U, \nu)$ is the quantity that is directly observed
in the radio-interferometric 
observations\footnote{We assume the $2D$ approximation of the
visibility here. Interested readers are referred to the {\bf
Appendix~\ref{app:wterm}}.}. It is the Fourier transform of the
product of the antenna 
primary beam pattern $A(\vt, \nu)$ and the specific intensity
distribution of the galaxy. 
\begin{equation}
\V(\U, \nu) \ =\ \int d \vt \ e^{-i 2 \pi \U
{\mbox{\boldmath$\cdot$}} \vt} A (\vt, \nu) \, 
I(\vt, \nu).
\end{equation}
Here $\U$ refers to a baseline,  the antenna separation measured in
units of the observing  wavelength $\lambda$. It is common practice to
express $\U$ in units of kilo wavelength  (k$\lambda$). Throughout
this paper we follow the usual radio-interferometric convention and
express $\U$ in units of kilo wavelengths (k$\lambda$).

The angular extent of the galaxies that we consider here is
much smaller than the primary beam, and it's effect 
may be ignored. We then have 
\begin{equation}
\V(\U, \nu)\ =\ \Wt(\U, \nu) \bar{I}(\nu) + \Wt(\U, \nu)  \otimes
\tilde{\delta I}(\U, \nu)  + \N(\U, \nu)
\label{eq:vis2}
\end{equation}
where  the tilde $\tilde{\,}$ denotes the Fourier transform of the
corresponding quantity and $\otimes$ denotes a convolution. 
In addition to the signal,  each visibility
also contains a system noise contribution $\N(\U, \nu)$ which we have
introduced in  {\bf Eqn.~(\ref{eq:vis2})}. The noise in each
visibility is a Gaussian random 
variable and the noise in the visibilities at two different baselines
$\U$ and $\U^{'}$ is uncorrelated. We will discuss the property of the
noise term $\N(\U, \nu)$ in detail in a following section.

The Fourier transform of the normalized window functions  are 
\begin{equation}
\Wt(\U, \nu)  =  2 ~\frac{J_{1}(2 \pi \theta_0 U)}
{2 \pi \theta_0 U}
\end{equation}
and 
\begin{equation}
\Wt(\U, \nu)  = \frac{1}{\left[ 1 + \pi^{2} \theta_{0}^{2}U^{2}/3\right]^{3/2}} 
\label{eq:gausw}
\end{equation}
for the top-hat and exponential models respectively. Here $J_1(x)$ is the
Bessel function of order $1$. These functions,   shown in
{\bf Figure~\ref{fig:wind}}, have the property that they peak around $U=0$
and  fall off rapidly  for $U \gg  \theta_0^{-1}$. This is a generic
property of the window function, not restricted to just these two
models. At baselines $U \gg  \theta_0^{-1}$ we may safely neglect the
first term in {\bf Eqn.~(\ref{eq:vis2})} whereby 

\begin{equation}
\V(\U, \nu)= \Wt(\U)  \otimes
\tilde{\delta I}(\U, \nu)  + \N(\U, \nu)
\label{eq:vis3}
\end{equation}
 
We use the power spectrum of \HI  intensity fluctuations
$P_{HI}(U)$ 
defined as
\begin{equation}
\langle \tilde{\delta I}(\U, \nu) \tilde{\delta I}^{*}(\U', \nu)
\rangle = \delta_{D}^{2}(\U-\U') \, P_{HI}(U, \nu)
\end{equation}
to quantify the statistical properties of the intensity
fluctuations. Here, $\delta_{D}^{2}(\U-\U')$ is a two dimensional
Dirac delta function. The angular brackets denote an ensemble
average over different realizations of the stochastic fluctuation. In
reality, it is not possible to evaluate this ensemble average because
a  galaxy presents us with only a single realization. In practice we 
evaluate an angular average over different directions of $\U$. This
is expected to provide an estimate of the ensemble average for a
statistically isotropic fluctuation. Note that, the power spectrum
thus defined can in general be a function of the observation
frequency. However, for the observation of the ISM turbulence, the
power spectrum is expected to remain constant within the frequency
range of observation. We will explicitly check this assumption in a
later chapter and will find deviation to it for the ISM in a
particular galaxy. However, for the time being for most of the
galaxies, it is safe to assume that the power spectrum is independent
of $\nu$. We will drop the $\nu$ dependence henceforth. 

The square of the visibilities can, in principle, be used to estimate
$P_{HI}(U)$ 
\begin{equation}
\langle\ \V(\U, \nu)\V^{*}(\U, \nu)\ \rangle \ = \left
|\Wt(\U, \nu)  \right|^{2} \otimes   \ P_{HI}(\U) + \langle \mid
\N(\U, \nu) \mid^2 \rangle \,
\label{eq:noise}
\end{equation}
and this   has been used  in several earlier studies 
\citep{1983A&A...122..282C, 1993MNRAS.262..327G, 1995A&A...293..507L}.
This  technique has  limited utility  
to observations where the \HI signal in each visibility exceeds the
noise.  This is because the noise variance in the last term
$\langle \mid\N(\U, \nu) \mid^2 \rangle$  introduces a
positive bias in estimating the 
power spectrum. The  noise bias can be of orders of magnitude larger than
the power spectrum for the  faint external galaxies considered here.  
In principle, it may be separately estimated using  
line-free channels and subtracted. In practice this is extremely
difficult owing to uncertainties in  the bandpass response which
restrict the noise statistics to be estimated with the required
accuracy.  Attempts  to subtract out the noise-bias have shown  that
it is not possible to do this at the level of accuracy required to
detect the \HI power spectrum \citep{2006THS...BA}. 

\begin{figure}
\begin{center}
\epsfig{file=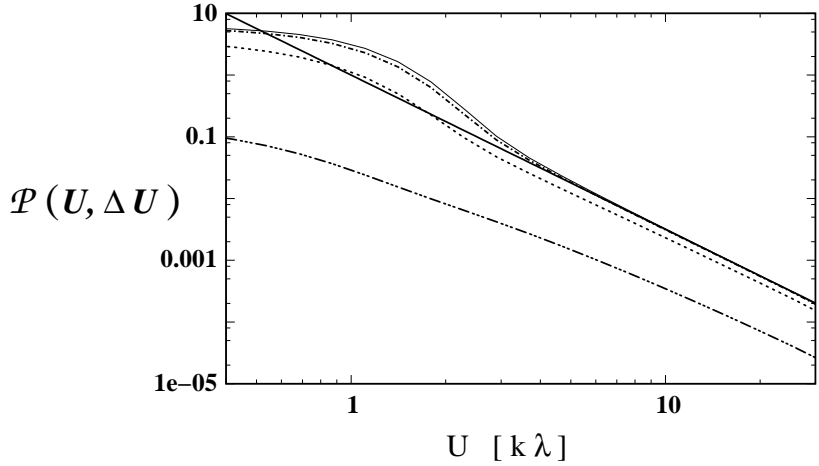,angle=0, width=4.2in}
\end{center}
\caption{This shows the convolution   ({\bf Eqn.~(\ref{eq:corr})})
of a power law $P_{HI}(U)=U^{-2.5}$  with the product of two exponential
window functions ({\bf Eqn.~(\ref{eq:gausw})})  with $\theta_0=1'$ ($(\pi
\theta_0)^{-1} \approx 1\, {\rm k}\lambda$). The bold solid curve shows the
original power law and  the thin solid curve shows $\P(\U,\Delta
\U)$ for $\Delta \U=0$. The dashed curves are for $\mid \Delta
\U\mid =0.2, 1.0 $  and $3.0 \, {\rm k} \lambda$ respectively from top
to bottom. The direction of $\Delta \U$ is parallel to that of
$\U$. Note that $\P(\U,0.2) \approx \P(\U,0)$ and 
the correlation between two different visibilities is 
small for $\mid \Delta \U \mid > (\pi \theta_0)^{-1}$.}
\label{fig:conv2}
\end{figure}

The  problem of noise bias can be avoided by correlating  visibilities
at two different baselines for which  the noise is expected to be
uncorrelated. We define the power spectrum estimator as
\begin{eqnarray}
\P(\U, \Delta \U) &=&
\langle\ \V(\U, \nu)\V^{*}(\U+\Delta \U, \nu)\ \rangle 
\nonumber  \\
&=& \int d \U' \, \Wt(\U-\U', \nu) \, \Wt^*(\U+\Delta \U-\U', \nu)
\      P_{HI}(\U').  
\label{eq:corr}
\end{eqnarray}
Since $\Wt (\U, \nu)$ falls  off rapidly for $U \gg
\theta_{0}^{-1}$ ({\bf Figure ~\ref{fig:wind}}), the 
window  functions $\Wt(\U-\U', \nu)$ and $W^{*}(\U+\Delta
\U-\U', \nu)$  in {\bf Eqn.~(\ref{eq:corr})}  have a substantial overlap only  
if $|\Delta \U| < (\pi \theta_{0})^{-1}$.  Visibilities at two  different
baselines  will be correlated only if  $|\Delta \U| <
(\pi \theta_{0})^{-1}$, and not beyond ({\bf Figure~\ref{fig:conv2}}). In  our
analysis we restrict the 
difference in baselines to  $|\Delta \U| \ll  (\pi \theta_{0})^{-1}$
so that  $\Wt(\U + \Delta \U-\U^{'}, \nu) \approx \Wt(\U-\U^{'})$ and
the estimator $\P(\U, \Delta \U)$ no  longer depends
on $\Delta \U$ ({\bf Figure~\ref{fig:conv2}}). We then
use   the visibility correlation estimator 
\begin{eqnarray}
\P(\U) &=&
\langle\ \V(\U, \nu)\V^{*}(\U+\Delta \U, \nu)\ \rangle 
\nonumber  \\
&=& \int d \U' \, \mid \Wt(\U-\U', \nu) \mid^2 \,
P_{HI}(\U')  \,.
\label{eq:corrn}
\end{eqnarray}
The measured visibility correlation $\P(\U)$
will, in general, be  complex. The real part is the power spectrum of
\HI intensity fluctuations convolved with the square of the window
function. A further simplification is possible at large baselines 
 $U \gg \theta_{0}^{-1}$, provided $\mid \Wt(U, \nu)\mid^2 $  decays
much faster than the variations in ${\rm P_{HI}}(U)$. We then have  
\begin{equation}
\P(\U)= C \,  P_{HI} (\U) 
\label{eq:corra}
\end{equation}
where $C=\int  \mid \tilde W(U, \nu) \mid^2 ~d  \U$ is a constant. 

\begin{figure}
\begin{center}
\epsfig{file=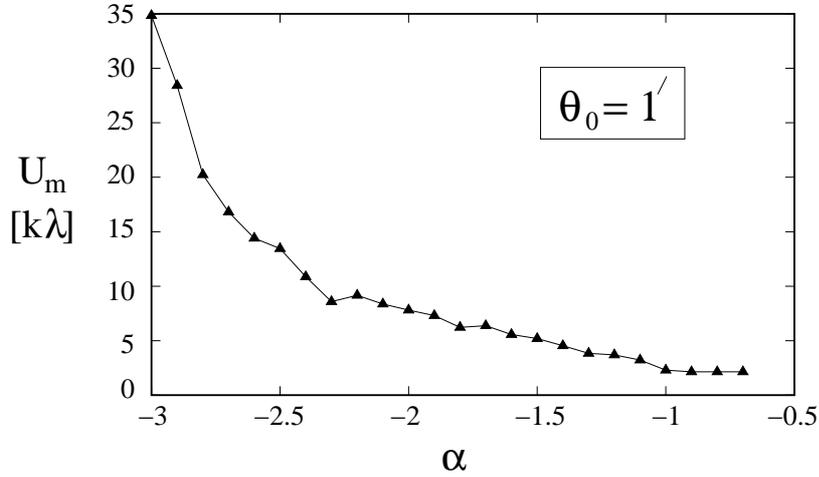, angle=0, width=4.2in}
\end{center}
\caption{This shows how $U_m$ changes with $\alpha$ for an exponential
window function with $\theta_0=1'$. Note that the discontinuities seen
in the plot appear to be genuine features and not numerical artifacts,
though the cause of these features is not clear  at present. We find
that a $4^{th}$ order polynomial $U_{m}(\alpha) = (a + b*\alpha +
c*\alpha^2 + d*\alpha^3 + e*\alpha^4)/\theta_0$ provides a good fit to
the above curve with the parameter values as $a = 35.6,\, b = 104.5,\,
c =112.5,\, d = 48.9,\, {\rm and}\,  e = 7.7$. We use this as a template for
estimating $U_{m}$ in the real observations.
}  
\label{fig:Um}
\end{figure}

\begin{figure}
\begin{center}
\epsfig{file=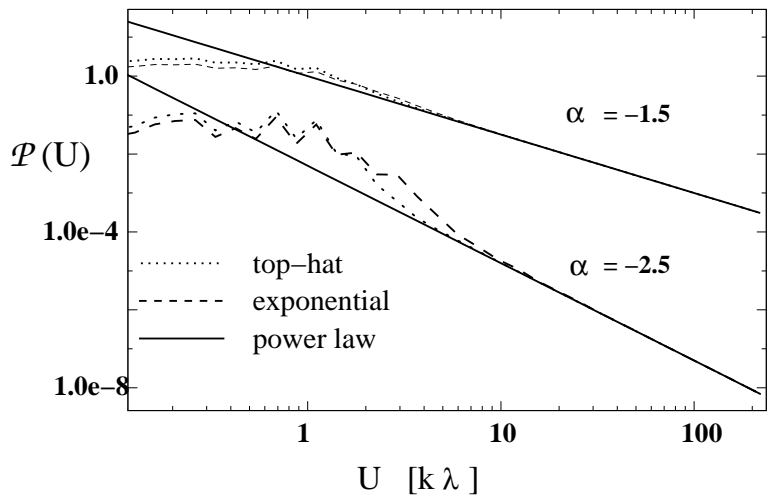, angle=0, width=4.2in}
\end{center}
\caption{Effect of window function modifies the power spectrum
  differently for different power law exponents. } 
\label{fig:conv}
\end{figure}

We use the real part of the estimator $\P(U)$ to
estimate the power spectrum $P_{HI}(U)$. Our interpretation is restricted to the $U$ range $U \gg 
(\pi \theta_0)^{-1}$ where the convolution  in
{\bf Eqn.~(\ref{eq:corrn})} does not affect the shape of the power spectrum
and {\bf Eqn.~(\ref{eq:corra})} is a valid approximation. Since
for this case the estimator is
independent of $\Delta \U$ and hence we write it as $\P(\U)$. In order to
estimate  the $U$ range where this approximation is valid we have
numerically evaluated {\bf Eqn.~(\ref{eq:corr})}
assuming $P_{HI}(U)$ to be a power law $P_{HI}(U)=A U^{\alpha}$.
{\bf Figure~\ref{fig:conv}} shows the results for both the top-hat and the
exponential models with $\theta_0=1'$ ($\theta_0^{-1}=3.4 \, {\rm
  k}\lambda$) and $\alpha=-1.5$ and $-2.5$ which roughly spans the
range of slopes usually observed for different  galaxies. 
Using $U_m$ to denote the value (in ${\rm k} \lambda$) where 
the deviation from the original power law is $10\%$, we find that 
 for the top-hat and exponential models respectively $U_m$ has values
 $(3.1,5.3)$ for $\alpha=-1.5$ and  $(4.8, 13.0)$ for $\alpha=-2.5$.
Note that  $U_{m}$  depends on two parameters, namely $\alpha$ and
$\theta_{0}$.  
 {\bf Figure~\ref{fig:Um}} shows how $U_{m}$ changes with $\alpha$ for the
 exponential model with $\theta_0=1'$.  For other  values of
 $\theta_{0}$  we  scale the value   of $U_m$ in  {\bf Figure~\ref{fig:Um}}
 using  $U_m \propto \theta_0^{-1}$. 
  For a given power law index, the estimator $\P(U)$
  gives a direct estimate of the power  spectrum   for $U \ge U_m$.

\section{Method of analysis}
\label{sec:moa}
We use radio-interferometric spectroscopic multichannel data observed
using GMRT or VLA for power spectrum analysis.
In this section we discuss the method of estimating the power spectrum
and its  errors in practice.

\subsection{Visibility correlation}
The starting point of each analysis is the radio-interferometric
spectroscopic multichannel raw $uv$ fits data. We  reduced the raw data
in the usual way using  standard tasks in 
classic AIPS\footnote{NRAO Astrophysical Image Processing System,  
a commonly used software for radio data processing.}. For each
observations, bad visibility points   
were edited out, after which the data was calibrated. The calibrated
data is then inspected for \HI emission and the line-channels are
identified.   A continuum image is then made averaging  all
the line-free channels. In the next step, this model  continuum  is 
subtracted  from the visibility data in  the $uv$ plane using the
AIPS task UVSUB. The resulting continuum subtracted data was used for
the subsequent analysis. Such a  typical spectroscopic data has $n$
emission channels. To determine if the 
\HI power spectrum changes with the width of the frequency channel, 
  we  combine $N$ successive channels to obtain a data set with $n/N$
  channels. The power spectrum is evaluated in each of this combined
  channels and then averaged over them for a particular $U$ value.
We use a range of $N$ values for the power spectrum analysis. 

We measure  $\P(U)$ by  multiplying  every visibility
$V(\U, \nu)$ with all other visibilities  $V(\U + \Delta
\U, \nu)$  within a disk $\mid \Delta \U \mid \le U_{D}$ and
then average over different $\U$ directions.  We will discuss the
considerations  for choosing the value of $U_{D}$ shortly. Note that
correlation of a visibility with itself is excluded. To increase the
signal to noise ratio we further average the correlations in
logarithmic bins of $U$ and over all  frequency channels with \HI
emission.   To choose a particular value of $U_{D}$, we perform this
analysis varying $\mid \Delta \U \mid$ from $1/\pi \theta_{0}$ to
lesser values. Note that, 
for larger values of $U_{D}$, we find 
that the imaginary part of the estimator $\P^{I}$ becomes comparable
to $\P$, while for smaller values of $\mid \Delta \U \mid$, the
amplitude of 
$\P^{I}$ decreases. For a particular observation we identify a value
of $\mid \Delta \U \mid$, for which the ratio $\P / \P^{I} > 10$ and
the amplitude 
of $\P$ is maximum. We use this value as $U_{D}$ for the subsequent analysis. 

The measured visibility correlation estimator  $\P(U)$ [{\bf
Eqn.~(\ref{eq:corrn})}] is 
the convolution of the actual \HI power spectrum with a window
function. We do not attempt to estimate this window function  to deconvolve the
power spectrum. Our approach is to estimate a baseline  $U_m$, such
that for $U \ge U_m$  the effect of the convolution can be ignored.
$\P(U)$ is  proportional to the \HI power
spectrum [{\bf Eqn.~(\ref{eq:corra})}] at baselines $U \ge U_m$. 
We estimate the \HI power spectrum using  only this range ($U \ge U_m$).

Assuming an exponential window function [{\bf Eqn.~(\ref{eq:expw})}],
the value of $U_m$ depends on 
the parameters $\theta_0$ and $\alpha$.  Note that we do not attempt
to actually fit an exponential  to the observed overall \HI
distribution and thereby determine $\theta_0$. 
The galaxies are typically oriented randomly compared to the line of
sight direction assuming different inclination angles. This offers the
window functions with elliptical shapes in the sky plane,
not circular disks as assumed in {\bf Section~\ref{sec:visicorr}}. For each of
the individual 
galaxies we generate column density maps. We 
determine the extent of the major axis and the minor axis
of the galaxy window function from these maps at a column
density value of $10^{21}$ atoms cm$^{-2}$. We use the smaller
of the two angular extents as $\theta_{0}$ in our analysis. As
$U_m\propto \theta_0^{-1}$, choosing the smaller value  gives a 
conservative estimate of $U_m$. We also estimate $\P(U)$ combining
the visibilities of the line-free 
channels. The upper limit $U_u$ where we 
have a reliable estimate of the \HI power spectrum is  determined
by the requirement that the  real part of $\P(U)$ estimated from the
line-channels should be more than
its  imaginary part and also that estimated  from
the line-free channels.  

\subsection{Error estimation}
The variance $\sigma_{\P}$ of the visibility correlation estimator $\P(U)$ is 
given by (see {\bf Appendix~\ref{app:visig}},
also \citet{2008MNRAS.385.2166A}) 
\begin{equation}
\sigma_{\P}^{2} = \frac{\P(U)^{2}}{N_{g}} +
\frac{N_{2}^{2}}{2 N_{b}},
\label{eq:sigsa}
\end{equation}
where $N_{g}$ is the number of independent estimates of $\P(U)$ in a
given $U$-bin, $N_{b}$ is the total number of visibility pairs in
the bin and $N_{2}$ is the noise variance. The first term in {\bf
Eqn.~(\ref{eq:sigsa})} is 
a contribution from the sample variance where the second term
incorporates the noise contribution. We evaluate $N_{2}$
estimating the variance in the visibility values.  Since in a
particular interferometric 
observation, the $uv$ space can be sparsely sampled, we grid the $uv$
plane.  We chose the grid size
such that each grid have same area as a circle of radius
$U_{D}$.  We determine the fraction of grids  filled with
visibility values in each annular bin ($f_{G}$). We use $\frac{2 \pi U \delta
U}{\pi U_{D}^{2}}\, f_{G}$ as an estimate for $N_{g}$. Here, $\delta
U$ is the width of a bin at $U$ from origin in the $uv$ plane. With these  
parameters the variance in the visibility correlation
estimator is given by 
\begin{equation}
\sigma_{\P} = \sqrt{\,  \frac{\P(U) U_{D}^{2}}{2 U \delta U} f_{G}
+ \frac{N_{2}^{2}}{N_{b}}}  
\end{equation}
Note that the first term under the square-root sign in the above
equation dominates for smaller 
values of $U$, since there both $U$ and $\delta U$ are small.  The
second term, however, dominates at large baselines 
where the $uv$ plane is very sparsely sampled. 

\subsection{Fitting a power law to the power spectrum}
We check if the estimated power spectrum follow a power law over a
length-scale range and evaluate the slope of this power law.
The procedure that we adopt to determine the best fit power law to the
\HI power spectrum is as follows.  We first visually identify a
baseline $U_m$ beyond which ($U \ge U_m$) the visibility correlation
estimator appears to be a power law in  $U$.  The largest baseline
$U_{max}$ till which the estimator can be used is limited by the 
requirement that the real part of $\P(U)$ estimated from the
line-channels is greater than its imaginary part as well as the real
part of the $\P(U)$ estimated from the line-free channels. 
We use the range
$U_m \le U \le U_{max}$ to  fit a power law  $\P(U) = A~U^{\alpha}$
through a $\chi^2$    minimization. The 
best fit $\alpha$ obtained by this  
fitting procedure  is used to get a revised  estimate for $U_m$ 
[see {\bf Figure~\ref{fig:Um}}] and   the 
power law fitting is repeated with this new value of $U_{m}$.  We
iterate this procedure a few times till it converges.  
To test whether the impact of the
window function  is actually small,  we convolve the best fit
power law  with $\mid  \tilde{W}(U, \nu)\mid^2$ assuming an Exponential
window.  The convolved power spectra are visually 
inspected to asses the deviations from the power law. The goodness of
fit to the data was also estimated by calculating $\chi^2$ for the
convolved power spectrum.  We accept the final fit only after
ensuring that the effect of the convolution can actually be ignored. 
We note the value of the best fit power law index $\alpha$ and the
range of $U$ values ($U_{min} - U_{max}$) for the fit. We also
determine the values 
$R_{max} = D/U_{min}$ and $R_{min} = D/U_{max}$, where $D$ is the 
distance to the galaxy. These  provide the range of
length-scales at the galaxy's disk over which the  power-law fit
holds. We can then 
conclusively state that over those length-scales in the galaxy the
power spectrum assumes power law and the ISM is turbulent.

\section{Discussions and Conclusions \label{chap2con}}
We discuss the major points of this chapter here and conclude.
\begin{itemize}
\item In this chapter we have presented a method of estimating the
power spectrum 
of \HI intensity fluctuation from radio-interferometric observations
using the visibility correlation. The effect of noise bias is avoided
correlating the visibilities at nearby baselines. This method holds the
key in measuring turbulence in the external galaxies. 
\item We have discussed an operating formula for estimating the
 error bars in the visibility correlation estimator we
presented. This estimates of the error incorporates the effect of
sample variance as well as the statistical errors and signifies the
quality of the power spectrum measurement.
\item For a turbulent ISM, the power spectrum of the \HI intensity
fluctuation is expected to be a power law. We have discussed the
procedure to obtain the best fit power law slope $\alpha$ from the
power spectrum estimates. However, since we do not exclusively evaluate the
shape of the window function here, our method can not be used to
measure the intensity of these fluctuations. 

\end{itemize}

Armed with this technique, in the following chapters we will use this
estimator for external dwarf and spiral galaxies and investigate the
relation of turbulence with  some of their dynamical and morphological
parameters. In a 
different investigation we have also used the same estimator to
evaluate the power spectrum of the supernovae remnant and opacity
fluctuation power spectrum of our galaxy. However, these works are not
relevant in the objective of the present thesis. Interested readers
may look at \cite{2009MNRAS.393L..26R}
and  \cite{2010MNRAS.404L..45R} for further information.  
%\newpage

%\clearpage{\pagestyle{empty}\cleardoublepage} %%%%%%%%%%%%%%%%%%%%
%\newpage
 \setcounter{section}{0}
 \setcounter{subsection}{0}
 \setcounter{subsubsection}{1}
 \setcounter{equation}{0}
 %\pagenumbering{arabic}

%-------------------------------------------
\chapter[Power Spectrum of Three Spiral Galaxies]{\bf \textbf
 {Power Spectrum Analysis of Three Spiral Galaxies\footnote{The works
 presented in this chapter are derived from the following three
 original publications:\\
(1) ``\HI power spectrum of the spiral galaxy NGC~628''
 by \citet{2008MNRAS.384L..34D},\\ 
(2) ``The scale-height of NGC~1058 measured from its \HI power
 spectrum.'' by \citet{2009MNRAS.397L..60D},\\
(3) ``Turbulence in the harassed galaxy NGC~4254''
 by \citet{2010MNRAS.405L.102D}. 
}}}
\label{chap:3spi}

\section{Introduction}
Power spectrum analysis is extensively used for studying the
ISM in our
galaxy \citep{1983A&A...122..282C, 1993MNRAS.262..327G,
2000ApJ...543..227D,  2001AJ....121.2706F, 2005A&A...436L..53B}. Being
observations from inside the Galaxy, these are restricted to 
length scales of $\sim 10$ AU to $10$ pc, where a power law power
spectrum with a slope of $\sim -2.8$ is
seen. This is believed to be the outcome of ISM
turbulence. The power
spectra of LMC and SMC are also
studied \citep{1999MNRAS.302..417S, 2001ApJ...548..749E,
2001ApJ...551L..53S}, where the ISM is probed at relatively larger
length-scales ($\sim 200$ pc). The power spectrum  of the nearby dwarf
galaxies are also found to be power laws of similar slope as observed for
our Galaxy. 

Recently \citet{2006MNRAS.372L..33B} have used the power spectrum
estimator defined in {\bf Chapter~\ref{chap:est}} to probe the ISM of the
external dwarf galaxy DDO~210. The power spectrum is found to be a
power law indicating the presence of turbulence. Further, they found that
the slope of the power spectrum is $\sim -2.7$, which is similar to
what was observed for our Galaxy, LMC or SMC previously. Since DDO~210
is a relatively nearby faint dwarf galaxy, they have managed to probe
a length-scale of $80$ to $500$ pc, which is also similar to what was
probed for our Galaxy.

\begin{table}[t]
\centering
\begin{tabular}{lrrrrc}
\hline
Galaxy &  Major  & Minor & $D$    & $i_{HI}$      & Reff \\  
       &  ($'$)  & ($'$) & (M pc) & ($^{\circ}$)  &       \\
& & & & &  \\
\hline \hline
& & & & &  \\
NGC~628  & $15.0$ & $11.0$ & $ 8.0$ & $13$ & 1 \\
NGC~1058 & $13.0$ & $11.0$ & $10.0$ & $11$ & 2 \\
NGC~4254 & $8.0$  & $6.5$  & $16.7$ & $42$ & 3 \\
\hline
\end{tabular}
\caption{Some relevant parameters of the 3 
  galaxies for which we do the power spectrum analysis in this
  chapter. The refferences are as follows: 
1- \cite{1992A&A...253..335K}, % NGC~628
2- \cite{2007AJ....134.1952P}, % NGC~1058
3- \cite{1993ApJ...418..113P}, % NGC~4254
}
\label{tab:t1}
\end{table}

Observing nearby spiral galaxies using radio interferometers like GMRT
or VLA, it is possible to probe the ISM at larger length-scales. Such
observations can directly estimate the operating length-scale range of
the ISM turbulence. In this chapter we present the power spectrum
analysis of 3 external spiral galaxies namely NGC~628, NGC~1058 and
NGC~4254 ({\bf Table~\ref{tab:t1}}). First two of these galaxies are
nearly face on spirals whereas
the galaxy NGC~4254 resides in a cluster environment and is believed to
be a harassed galaxy. The objective of this chapter will be to
investigate the ISM intensity fluctuation power spectrum at larger
length-scales than it is hitherto probed.  Moreover, the power
spectrum of  NGC~4254 can reveal the 
nature of ISM for the galaxies residing and interacting in a cluster
environment.

\section{Evidence of turbulence at large length-scales :
NGC~628} 
\label{sec:628}

NGC~628 (M~74) is a nearly face-on  SA(s)c spiral galaxy with an
inclination angle in the range  $6^{\circ}$ to $ 13^{\circ}$
\citep{1992A&A...253..335K}.  
It has a very large \HI disk extending out to more than 3 times the
Holmberg diameter.  \citet{2006ApJ...644..879E} have found a
scale-free size and luminosity  distribution of  star forming regions in
this galaxy, indicating turbulence to be functional here.  
The distance to this galaxy is uncertain with previous estimates
ranging from $6.5 \, {\rm Mpc}$ to $ 10 \, {\rm Mpc}$. 
\citealt{1980ApJ...238..510B} and \citealt{1992A&A...253..335K} have
used a Hubble flow distance of $\sim10$ Mpc. 
On the other hand, \citet{1996A&AS..119..499S} estimated a distance of
$7.8\pm 0.9$ Mpc from  the brightest blue star in the galaxy. This distance
estimate matches with an independent photometric distance estimate by
\citet{1996AJ....111.2280S}.  In a recent
study \citet{2004A&A...427..453V} inferred the  distance 
to be $6.7 \pm  4.5 $ Mpc by applying  the expanding  photosphere
method to the hyper novae SN~2002ap. Here we adopt the 
photometric distance of $8$ Mpc  for NGC~628. At this distance
$1^{\prime\prime}$ corresponds to $38.8$ pc. 
   
\begin{figure}
\begin{center}
\epsfig{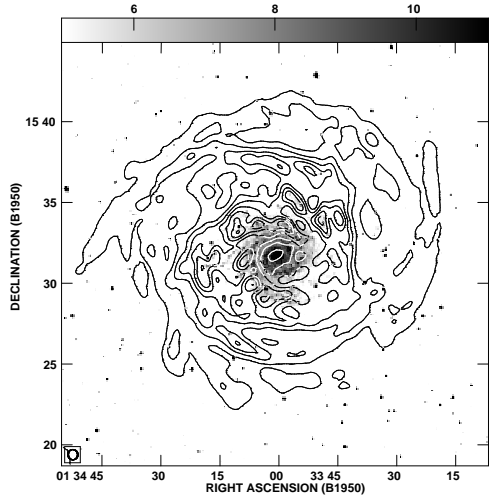}
\end{center}
\caption{The 38$^{\prime\prime} \times 36^{\prime \prime}$ resolution
integrated \HI column density map of NGC~628 (contours) overlayed on the
optical DSS image (grey scale). The contour levels are 0.24, 1.96,
3.68, 5.41, 7.13, 8.85, 10.57 and 12.29 $\times 10^{20}$ cm$^{-2}$. 
}
\label{fig:NGC628M0}
\end{figure}

\subsection{A brief description of the data used}
%\label{ref:data}

We have used  archival \HI data  of  NGC~628  from the Very Large
Array (VLA). The   observations had been carried out   on 
August 1$^{st}$ and November 14$^{th}$, 1993 respectively  in the  C and D
configurations of the VLA,  as a part of the AAO163 observing
program.  We performed the data analysis following the method described
in {\bf Section~\ref{sec:moa}}. The
multi-configuration data were downloaded from the VLA archive and 
reduced in the usual way using  standard tasks in classic
AIPS\footnote{NRAO Astrophysical Image Processing System,  
a commonly used software for radio data processing.}. For each VLA
configuration, bad visibility points  
were edited out, after which the data were calibrated. The calibrated data
for both configurations was combined using the AIPS task DBCON. 
The \HI emission from NGC~628 spans $64$ central  channels (i.e,
$n=64$) of the $256$ 
channel spectral cube (with channel width 1.29 km s$^{-1}$).  
A continuum image was made using the average of all the line-free
channels. The continuum from the galaxy was subtracted from the data in 
the $uv$ plane using the AIPS task UVSUB.
The resulting continuum subtracted data was used for the subsequent
analysis. {\bf Figure \ref{fig:NGC628M0}} shows a  total \HI column density
(Moment 0) map of NGC~628 from an image made from this data. 
The \HI disc of the galaxy is nearly face-on. 
The angular extent of the \HI distribution in {\bf
Figure~\ref{fig:NGC628M0}} is 
roughly $11' \times 15'$. Using deep VLA~D array mosaic observations,
\citet{1992A&A...253..335K} had detected a $\sim 38' \times 31'$ faint
diffuse \HI  
envelope around NGC~628. This extended envelope is not detected in
the current data set; the emission that we do detect is instead
restricted to the main \HI disc of NGC~628. The results we discuss
below are hence also relevant only to the gas in the main \HI disc
around NGC~628.

\begin{figure}
\begin{center}
\mbox{\epsfig{file=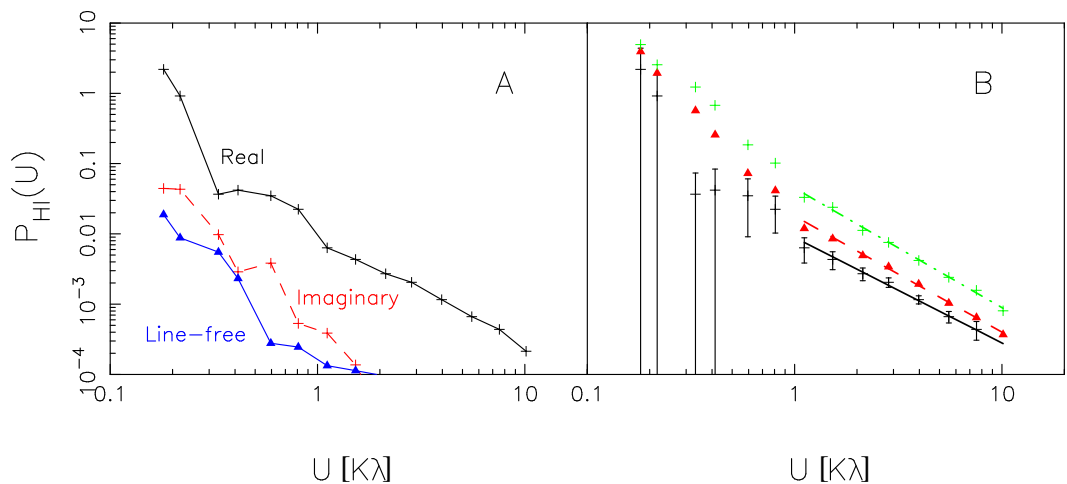,height=2.5in,angle=0}}
\caption{  A)  Real   and imaginary  parts of the
 observed value of the \HI  
 power spectrum estimator $\P(U)$ for $N=64$
{\it ie.} all frequency  channels with \HI   emission were collapsed
  into a single channel.  The real part is also shown using $64$
 line-free channels collapsed   into a single channel.
  B) Best fit power law to  $\P(U)$. The channel width is
  varied  ($N=1,32,64$; top to bottom),  $1~\sigma$ error-bars are
 shown only   for $N=64$. }
\end{center}
\label{fig:NGC628f2}
\end{figure}

\subsection{Results and discussion}
%\label{ref:discuss}

{\bf Figure \ref{fig:NGC628f2}A} shows the real and imaginary parts of 
the  observed value  of the estimator
$\P(U)$  for the $64$ channels which have \HI emission.  As expected
from the theoretical 
considerations mentioned earlier, the imaginary part is well
suppressed compared to  the real part. To test 
for  a possible contribution from residual continuum, we also show the
real part of  $\P(U)$  using $64$ line-free channels. This is
found to be much smaller than the signal   for the channels with \HI
emission. We find that a power law $A~U^{\alpha}$ 
with slope  $\alpha=-1.6 \pm 0.2$ provides a good fit to the
the observed $\P(U)$ 
 over the $U$  range $U_{min} = 1.0 \, {\rm K} \,
\lambda$ to  $U_{max} = 10.0 \, {\rm K} \, \lambda$ ({\bf
Figure~\ref{fig:NGC628f2}B})  
which corresponds to spatial scales  of $R_{min} = 800 {\rm pc}$ to 
$R_{max} = 8 \, {\rm kpc}$.  

Both \HI density fluctuations as well as spatial fluctuations in the
velocity of the \HI gas contribute to fluctuations in the \HI specific
intensity.
Considering a turbulent ISM, \citet{2000ApJ...537..720L} have shown
that  it is possible to disentangle these two contributions  by
studying the behavior of the \HI power spectrum as the thickness of the 
frequency  channel is varied. If the observed \HI power spectrum is 
due to the gas velocities, the slope of the  power spectrum 
is predicted to change as the frequency  channel thickness is
increased. 
 We have tested  this  by determining  the \HI power spectrum for
different values of  the channel width in the range
$1.29 \,  {\rm   km  \, s^{-1}}$  to 
$82.6 \,  {\rm   km \, s^{-1}}$  ({\bf Figure~\ref{fig:NGC628f2}B})  and
did not 
find any change in the slope of the \HI power spectrum ({\bf
Table~\ref{tab:NGC628R}}). As the  thickest channel that we have used
is considerably wider than the  typical \HI  velocity dispersion of
$7-10 \, {\rm  km s^{-1}}$ seen in spiral
galaxies \citep{1984A&A...132...20S}, 
we conclude that    the observed \HI power spectrum of 
NGC~628 is purely due to density fluctuations. 
 Our finding is similar to that  of  \citet{2006MNRAS.372L..33B} who
 noticed  no change of the slope with channel width
for the dwarf  galaxy DDO~210.  Further, \citet{2001ApJ...548..749E}
also reported a similar behavior for  LMC.  

\begin{table}
\centering
\begin{tabular}{lrcccccc}
\hline
N & $\Delta v$  & $U_{min}$ & $U_{max}$ &  $R_{min}$
& $R_{max}$ & $\alpha$ \\
& (km s$^{-1}$) & (k $\lambda$) & (k $\lambda$) & (k pc) & (k pc) & \\
& & & & & &\\
\hline \hline
& & & & & &\\
 $1$  & $ 1.3$  & $1.0$ & $10.0$ & $0.8$ & $8.0$ & $-1.7 \pm 0.2$ \\  
 $32$ & $42.3$  & $1.0$ & $10.0$ & $0.8$ & $8.0$ & $-1.6 \pm 0.2$ \\ 
 $64$ & $82.5$  & $1.0$ & $10.0$ & $0.8$ & $8.0$ & $-1.6 \pm 0.2$ \\ 
\hline
\end{tabular}
\caption{NGC~628 power spectrum are well fitted by a power law of
  $P_{HI}=AU^{\alpha}$. This table summarises the result for different
  channel widths. Note that the slope of the power spectrum does not
  changes with the width of the velocity channel.}  
\label{tab:NGC628R}
\end{table}

Earlier studies of the Milky-Way,  and also of the dwarf galaxies LMC,
SMC and
DDO~210 \citep{1983A&A...122..282C, 1993MNRAS.262..327G,
1999MNRAS.302..417S, 2000ApJ...543..227D, 2001ApJ...548..749E,
2006MNRAS.372L..33B} have all found 
a power law   \HI power spectrum with slope $\sim -3$. On the contrary,
we find a  slope $-1.6 \pm 0.2$ for NGC~628. This is a little more than
one in excess of  the earlier values. 
However, when comparing these values it should be noted that the
earlier works have all measured the \HI power spectrum at much
smaller length-scales in the range  $10\ \mathrm{to}\ 500$ pc 
[MW ($1$ pc - $20$ pc), SMC ($4$ pc - $30$ pc), LMC ($60$ pc -
$200$ pc), DDO~210 ($80$ pc -
  $500$ pc)] whereas the current measurement probes much larger
  length-scales from $800$ pc  to $8.0$ kpc.  
The typical \HI scale-heights within the
Milky-Way \citep{1984BAAS...16..981L,1990A&A...230...21W} 
and external galaxies (e.g. \citealt{2002A&A...390L..35N}) are 
well within $1.5$ kpc. This implies that on
the  largest length-scales which we have probed, the turbulence is
definitely confined to the plane of the galaxy's disk and is therefore
two dimensional (2D).  \citet{2001ApJ...548..749E} have found that the HI
power spectrum 
of LMC flattens  at large length-scales, which was interpreted as a
transition from three dimensional to   two dimensional turbulence. 
 Based on these we conclude
that the slope is  different in our observation because it  
probes 2D turbulence, whereas the earlier observations 
were on length-scales smaller than the scale-height where we can expect
three dimensional (3D) turbulence. 
 To the best of our knowledge our results  are the first
observational determination of the HI power spectrum 
of  an external spiral galaxy at such  large length-scales 
 which are comparable to the radius of the galaxy's disk.  

\citet{1999AJ....117..868W} have performed a fractal analysis using
the perimeter-area dimension  of  intensity contours in  HI images   
of several galaxies in the  M~81 group. Of particular interest is the
galaxy  
M~81, a spiral galaxy for which the perimeter-area dimension was found
to be $\sim 1.5$ at a length-scale $\sim 10$ kpc.  This
observation is consistent with a power law power spectrum of slope
$-1.6 \pm 0.2$ provided  the assumption  that the local dimension has
the same value as the perimeter-area dimension is valid.  
 
It is difficult to probe the \HI scale-height of external face-on galaxies. 
\citet{2001ApJ...555L..33P} presented a method to probe the scale-height
from a change in  the slope of the Spectral Correlation
Function (SCF) [\citet{1999ApJ...524..887R}], and 
applied it to \HI data for the LMC to estimate the scale-height to be
$\sim 180$ pc. \citet{2001ApJ...548..749E} suggested that one could use
a change in 
the slope of the power spectrum of the density fluctuations to measure
the scale-height of face on gas disks. Applying this method to \HI
data for the LMC they measured scale-height of $100$ pc.  To
the best of  
our knowledge, prior to this work, there hava been no observational 
constraint on the \HI scale-height of NGC~628. Since the scale-height 
is definitely less than $8$ kpc and the power spectrum is
found to have the same slope from $800$ pc to $8$ kpc, we
conclude that the scale-height must be less than $800$ pc. 
\cite{2004MNRAS.352..768K} present \HI images of a large sample of edge
on intermediate 
to late type spirals; from their data the ratio of the \HI disk height to 
the radius of the \HI disk (at a column density of $1~M_\odot$pc${-2}$) is
$\sim 0.06 \pm 0.015$. From {\bf Fig.~\ref{fig:NGC628M0}} the disk of NGC~628
has a diameter of $\sim 28$ kpc at this column density. From the average
thickness to radius ratio for edge on galaxies, one would expect
NGC~628 to have a scale-height of $\sim 840$ pc, which is consistent
with our observation. 

\section{Estimating scale-height of a face on galaxy :
NGC~1058}
\label{sec:1058} 

NGC~1058 is an almost face-on,  type Sc, spiral galaxy with an
inclination angle in the range  $4^{\circ}$ to $ 11^{\circ}$
\citep{2007AJ....134.1952P}. The distance to this galaxy is
uncertain with previous estimates ranging from $10.0$ Mpc
\citep{1981ApJS...46..177B} to $ 14.5$ Mpc 
(\citep{1974ApJ...194..559S}). Throughout this 
paper we adopt a  distance of 10 Mpc  for NGC~1058. At this distance
1$^{\prime\prime}$ corresponds to 48.5 pc. 

In this section we  present a measurement of the power spectrum of \HI
intensity fluctuations of NGC~1058. The  observed power spectrum 
shows a break with a steeper slope at length-scales smaller than the
break. We use this to estimate the scale-height  of NGC~1058
  
\begin{figure}
\begin{center}
\mbox{\epsfig{file=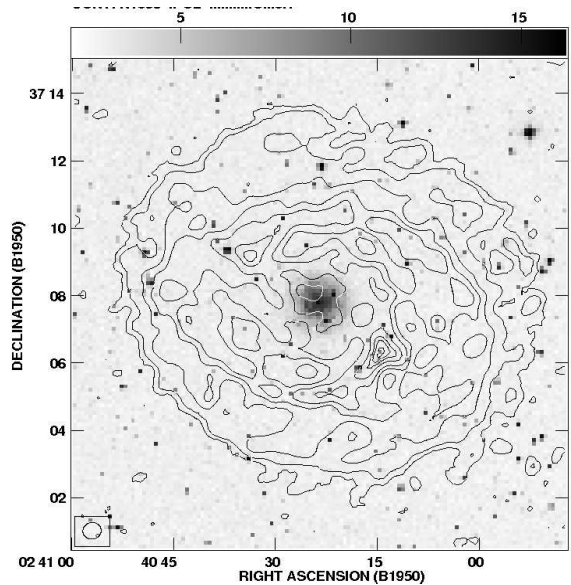,width=3.0in,angle=0}}
\end{center}
\caption{The $13.6^{\prime\prime} \times 13.6^{\prime \prime}$ resolution
  integrated \HI column  
density map of NGC~1058 (contours).  The contour levels 
are 3., 10., 30., 50., 70., 100., 120., 130., and 140. $\times
10^{20}$ cm$^{-2}$. 
}
\label{fig:NGC1058M0}
\end{figure}

\subsection{A brief description of the data used}
%\label{ref:data}

We have used  archival \HI data  of  NGC~1058  from the Very Large
Array (VLA). The   observations had been carried out   on 14$^{th}$
June 1993 in the  C configurations of the VLA.
The data was downloaded from the VLA archive and
reduced in the usual way using  standard tasks in classic
AIPS\footnote{NRAO Astrophysical Image Processing System,  
a commonly used software for radio data processing.}. We adopted 
the data analysis procedure discussed in {\bf Section~\ref{sec:moa}}
for reducing the raw data and for the power 
spectrum analysis. 
The \HI emission from NGC~1058 spans $24$ central channels (i.e, channel
$53$ to channel $76$) of the $127$ channel spectral cube. We have used only 
the central $16$ channels (i.e, channel $57$ to channel $72$) with
relatively higher \HI emission. The frequency width of each channel
corresponds to  $2.58 \, {\rm km   s}^{-1}$.   
 {\bf Figure~\ref{fig:NGC1058M0}} shows a  total \HI column density
(Moment 0) map of NGC~1058 from an image made using the continuum
subtracted data.  
The angular extent of the \HI distribution in {\bf
Figure~\ref{fig:NGC1058M0}}  
is measured to be $11' \times 13'$  at a  column density of $10^{19}
\, {\rm atoms \, cm}^{-2}$, which is $\sim 4$ times it's
Holmberg diameter \citep{2007A&A...463..481P}.

\begin{figure*}
\begin{center}
\mbox{\epsfig{file=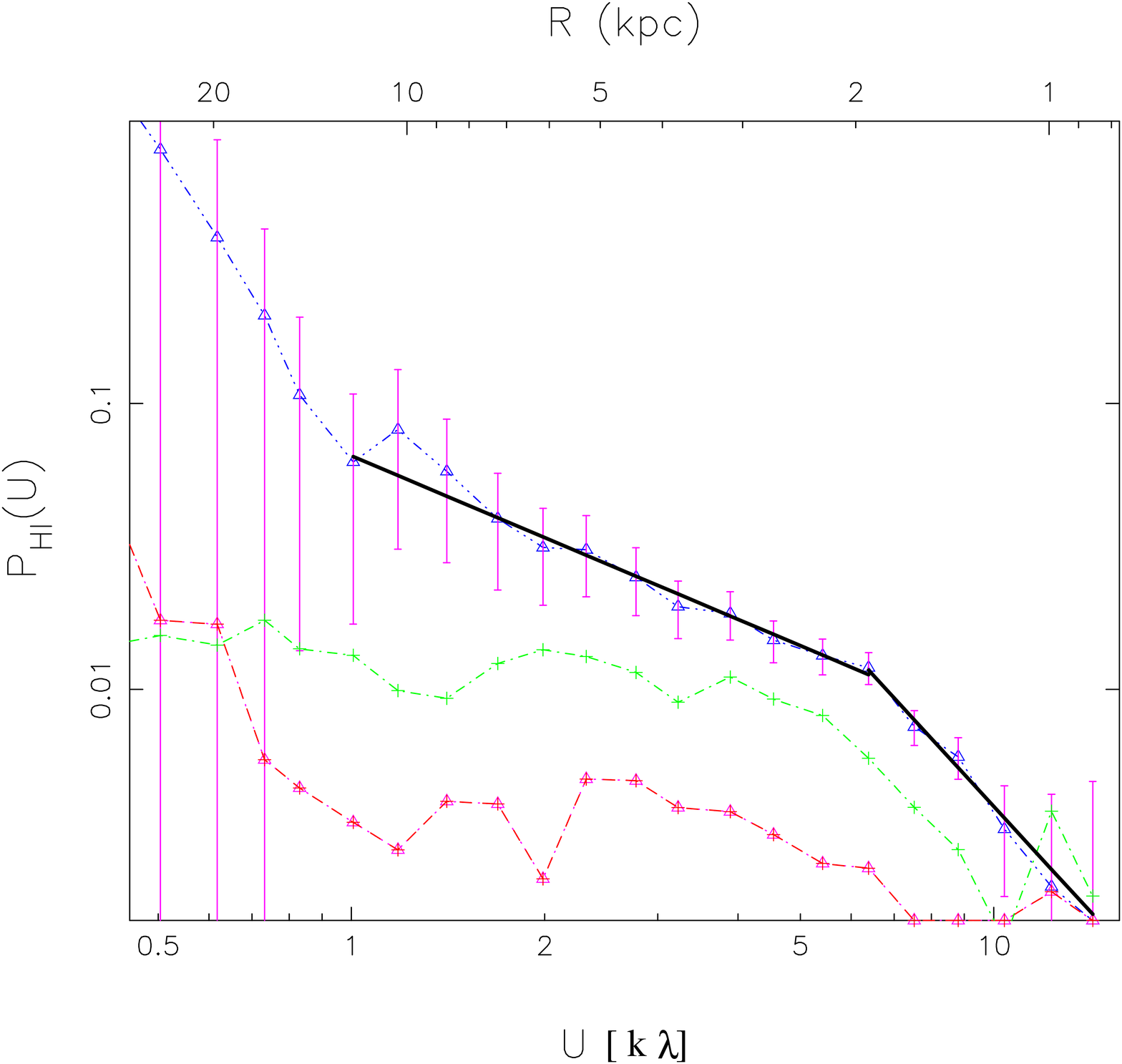, height=2.8in, angle=0}}  
\caption{ Real (blue)  and imaginary (green) parts of the
 observed value of the \HI  
 power spectrum estimator $\P(U)$ for $N=16$.
  $1~\sigma$ error-bars are  shown only   for the real part. The real
 part of the $\P(U)$ from line-free channels (red)
  is also shown. }
\end{center}
\label{fig:1058f2}
\end{figure*}

\subsection{Results and discussion}
%\label{ref:discuss}
{\bf Figure~\ref{fig:1058f2}} shows the real and imaginary parts of 
the  observed value  of the estimator
$\P(U)$  for the $64$ channels which have \HI emission. As
expected, the imaginary part and the real part of  $\P(U)$  estimated
using line-free channels is well suppressed compared to  the real part
from the line channels indicating that for these channels the observed
$\P(U)$ may be directly interpreted as the \HI power spectrum at   
$U$ values that are considerably larger than
$(\pi \theta_0)^{-1}=0.1 \, {\rm k} \lambda$.   

The power  spectrum is well fitted by
two different power laws $P(U)=AU^{\alpha}$, one  with 
$\alpha\ =-\ 1.0\pm 0.2$ for $U = 1.0
\, {\rm k}  \lambda$ to  $6.5 \, {\rm k} 
\lambda$   (large length-scales) and another with  
$\alpha\ =-\ 2.5\pm 0.6$ for  $U = 6.5
\, {\rm k}   \lambda$ to  $16.0 \, {\rm k}  \lambda$  (small  length-scales). The results are tabulated in {\bf Table~\ref{tab:NGC1058R}}.
 We also find that the  \HI power spectrum of NGC~1058 does not exhibit a
statistically significant change with increase of the channel
thickness ({\bf Table~\ref{tab:NGC1058R}}),  
indicating that the observed power spectrum is due to \HI  density
fluctuations. 

\begin{table}
\centering
\begin{tabular}{rrrr|crc}
\hline
$U_{min}$ & $U_{max}$ &  $R_{min}$ & $R_{max}$ & N & $\Delta v$  &  $\alpha$ \\
(k $\lambda$) & (k $\lambda$) & (k pc) & (k pc) &  & (km s$^{-1}$) &  \\
& & & & & &\\
\hline \hline
& & & & & &\\
       &        &       &        & 1  &  $2.68$ & $-2.5\pm0.6$ \\  
 $6.5$ & $16.0$ & $0.6$ & $1.5$  & 8  & $21.44$ & $-2.2\pm0.6$ \\
       &        &       &        & 16 & $42.88$ & $-2.2\pm0.5$ \\  
\hline
& & & & & &\\
       &        &       &        & 1  &  $2.68$ & $-1.0\pm0.2$ \\  
 $1.5$ & $6.5$  & $1.5$ & $10.0$ & 8  & $21.44$ & $-0.9\pm0.2$ \\
       &        &       &        & 16 & $42.88$ & $-0.8\pm0.2$ \\  
\hline
\end{tabular}
\caption{NGC~1058 power spectrum are well fitted by  two power laws of
  $P_{HI}=AU^{\alpha}$ in two different length scales. This table
  summerises the result for different channel width.}  
\label{tab:NGC1058R}
\end{table}

The slope  $\alpha = -1.0 \pm 0.2$ gives a good  fit to the  power
spectrum measured at length-scales $1.5 - 10.0$ kpc.  
The length-scale  $10$ kpc is definitely larger than the 
typical \HI scale-heights within the
Milky-Way \citep{1984BAAS...16..981L, 1990A&A...230...21W} and in
external spiral galaxies (e.g. \citealt{2002A&A...390L..35N}).
Considering the \HI disk of NGC~1058, it is  reasonable to conclude
that   the slope  $\alpha = -1.0 \pm 0.2$ is that of 2D turbulence
in the plane of the galaxy's disk. On the other hand we have a
slope of $\alpha\ =-\ 2.5\pm 0.6$ at the small length scales $600$ pc
to $1.5$ kpc. We expect the scale-height
of the galaxy's \HI disk to be larger than  $600$
pc \citep{2004MNRAS.352..768K}.  
The slope  $\alpha\ =-\ 2.5\pm 0.6$ may thus be interpreted 
as being that of 3D turbulence on length-scales smaller than the
thickness of the disk. We interpret the wavelength of $1.5$ kpc where
we observe the transition from 3D to 2D turbulence in 
terms of 
the scale-height  of the galaxy's \HI disk.

To  our knowledge this is
the  first observational determination of the scale-height of a nearly
face on spiral galaxy through its \HI power spectrum. Further, the
length scales at which this break in the power spectrum is seen is
measured to an accuracy of the width of the bins in baseline
($U$) for calculating the power spectrum. Based on that we can also
put $1~\sigma$ errors to our estimate, whereby we have the
scale-height $1.5 \pm 
0.3 $ kpc.  Since the disk of NGC~1058 has a diameter of $35$
 kpc ({\bf Figure~\ref{fig:NGC1058M0}}) at the column density of
$1~M_\odot$ pc$^{-2}$), following \citet{2004MNRAS.352..768K} we expect
it's scale-height to be $1.05 \pm 0.26$ kpc, consistent with 
our result. Note that, \HI scale-height is known to increase with the
distance from the centre and our value is an average over the entire disk. 

It is interesting to note that a break in the  power spectrum, like
the one seen here for NGC~1058, has also been observed in the 
shell-type supernovae remnant Cassiopeia~A \citep{2009MNRAS.393L..26R}.
The slope of the power spectrum changes from $-2.22 \pm 0.03$ at  
$1.6 - 10 \, {\rm  k}\lambda$ (large length-scales)  to 
$-3.23 \pm 0.09$ at  $11 - 30 \, {\rm  k} \lambda$ (short
length-scales).  This change in the slope is interpreted as a
transition from 2D to 3D magnetohydrodynamic turbulence. The
transition occurs   at  a length-scale  of $50$ pc which
corresponds  to  the thickness of the shell. 

%\pagebreak
\section{Power spectrum of the harassed galaxy NGC~4254}
\label{sec:4254}

Galaxy harassment (frequent high speed galaxy encounters,
\citealt{1996ApJ...457..455M})
 is believed to be an important process in driving the
morphological transformation of spiral galaxies to ellipticals inside
clusters. Typically, the first encounters convert 
a normal spiral galaxy to a disturbed spiral with dramatic
features drawn out from the dynamically cold gas.  The spiral galaxy
NGC~4254, located in the nearby Virgo cluster, is found to have a tail 
\citep{2005ApJ...622L..21M} with neutral hydrogen (\HI) mass
$2.2 \times 10^8 \ M_{\odot}$ within a 
distance of $120$ kpc from the galaxy. This gaseous tail,
without any  stellar counterpart, is believed to be
produced by an act of galaxy
harassment \citep{2007ApJ...665L..19H}. Each act of harassment  has
the potential to induce a burst of star formation and 
to  change the internal properties of the galaxy, including the 
properties of the ISM.  In this section, we
study the effect 
of the harassment on the large scale structure of the ISM. We use the
power spectrum of \HI intensity fluctuations to quantify these
structures. 

NGC~4254 is a  lopsided spiral galaxy (morphological type SA(s)c),
with  an  inclination of $\sim
42^{\circ}$ \citep{1993ApJ...418..113P}. The galaxy is   
located  at a distance of $1$ Mpc from the core of Virgo cluster and
is  believed to be falling into the cluster with a relative velocity 
of $1300 \ {\rm km\ s}^{-1}$ \citep{2005A&A...439..921V}. The distance
to this  galaxy is estimated to be $16.7$
Mpc \citep{2007ApJ...655..144M}; at this distance $1^{\prime\prime}$
corresponds to $81$ pc.  
\begin{figure}
\begin{center}
\mbox{\epsfig{file=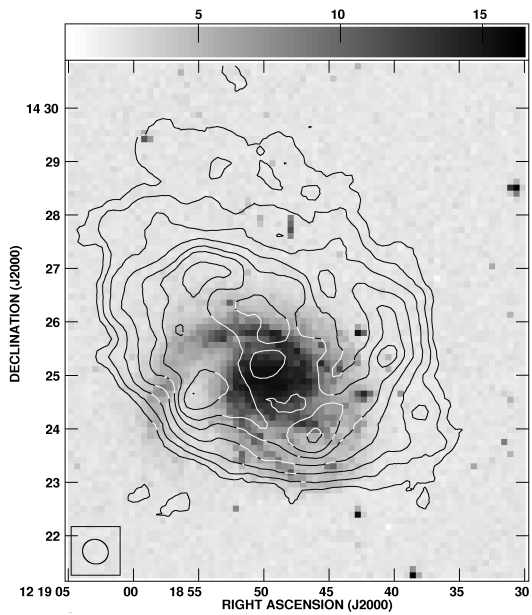,width=2.8in,angle=0}}
\end{center}
\caption{
The $6.5^{\prime\prime} \times 8.0^{\prime \prime}$ resolution
integrated \HI column  density map of NGC~4254 (contours) is overlayed
with the optical image of the galaxy (density).  The contour levels 
are  3., 5., 10., 20., 25., 35., 40. and 45. $\times
10^{20}$ atoms cm$^{-2}$. }
\label{fig:NGC4254M0}
\end{figure}

\subsection{A brief description of the data used}
%\label{sec:data}

We have used  archival \HI data  of  NGC~4254  from the Very Large
Array (VLA). The   observations had been carried out   on
$4^{th}-5^{th}$ of March, 1992 using the  C configuration of the VLA
\citep{1993ApJ...418..113P}.  
The data was downloaded from the VLA archive and
reduced in the usual way using  standard tasks in classic
AIPS\footnote{NRAO Astrophysical Image Processing System,  
a commonly used software for radio data processing.}. As usual, we
adopted the data analysis procedure discussed in {\bf
Section~\ref{sec:moa}} for reducing the raw
data and power spectrum analysis. 
The \HI emission from NGC~4254 spans over $30$ central  channels i.e,
($18$ to $47$) from $1401.7$ MHz ($2253$ km s$^{-1}$) to $1416.4$ MHz
($2562$ km s$^{-1}$) of the $63$ channel spectral cube.  A frequency
width of $48.8$ k Hz for    each channel in the data cube corresponds
to  the velocity resolution of $10.32~ \, {\rm   km~s}^{-1}$.
{\bf Figure~\ref{fig:NGC4254M0}} shows a  integrated \HI 
column density (Moment 0) map of NGC~4254 made from the continuum
subtracted data.  
The angular extent of the \HI distribution in {\bf Figure~\ref{fig:NGC4254M0}} 
is measured to be $6.5' \times 8.0'$  at a  column density of $10^{19}
\, {\rm atoms \, cm}^{-2}$, which is comparable to it's
optical diameter \citep{1991S&T....82Q.621D}.

\begin{figure}
\begin{center}
\mbox{\epsfig{file=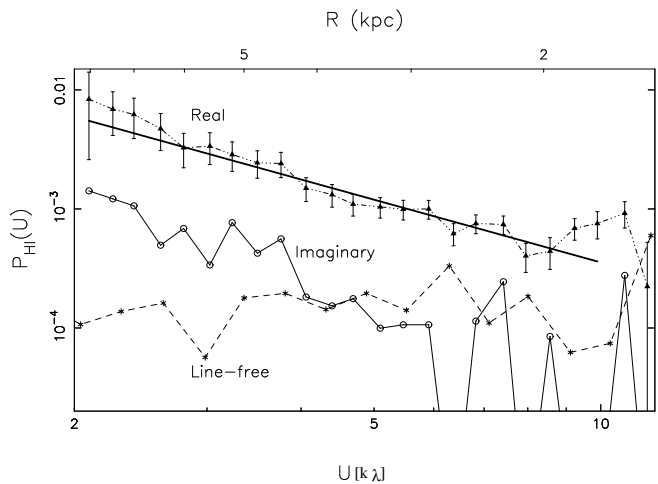,height=2.5in,angle=-0}}
\caption{Real and imaginary parts of the
 observed value of the \HI   power spectrum estimator for channels
 27-42 with $N=16$.
  $1~\sigma$ error-bars are  shown only   for the real part. The real
 part of the estimator from line-free channels is also shown. }
\end{center}
\label{fig:NGC4254f2}
\end{figure}

\subsection{Results and discussion}
%\label{sec:discuss}
{\bf Figure \ref{fig:NGC4254f2}} shows the real and imaginary parts of 
the  observed value  of the estimator 
$\P(U)$  evaluated using $16$ channels from $27$ to $42$ with
relatively \HI emission. The
imaginary part is well suppressed compared to  the real part.  We also 
estimate $\P(U)$ using  the line-free  channels 
to test  for  any  contribution from the residual continuum.
This is found to be much smaller than the signal
({\bf Figure~\ref{fig:NGC4254f2}}) indicating that the continuum has been
adequately  subtracted out. 

The power law $A~U^{\alpha}$, with  $\alpha\ =-\ 1.7\pm
0.2$  is found to 
give a good fit to the  \HI power  spectrum for the $U$ range 
$2.0 \, {\rm k}  \lambda$ to $10.0 \, {\rm k}  
\lambda$ ({\bf Table~\ref{tab:t4}}). Corresponding largest
length-scale   ($8.4$ kpc)  
is  definitely larger than the  typical \HI scale-heights of the
galaxies. It  is thus 
quite reasonable to  conclude that   the slope  $\alpha = -1.7 \pm
0.2$ is  of 2D  turbulence  in the plane of the galaxy's disk. The
fact that we do not observe the transition to 3D turbulence   allows
us to place an upper limit on the galaxy's scale-height as $2.6$ kpc.  

\begin{figure}
\begin{center}
\mbox{\epsfig{file=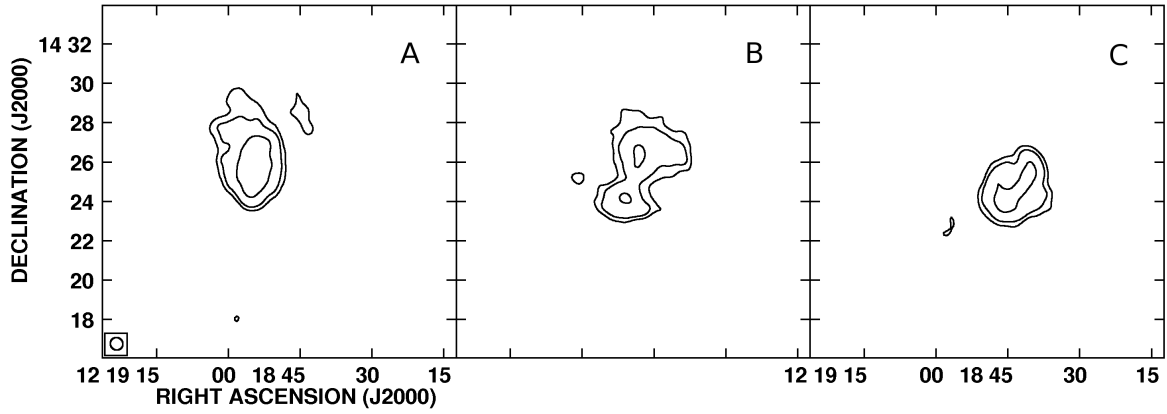,width=6.2in,angle=0}}
\caption{Integrated \HI column density maps of the galaxy NGC~4254 using
  data cubes A, B and          
   C. Note the diagonal movement of the centroid of emission from North
  east (A) to South west (C). The contour levels are  5., 8. and
  12. $\times 10^{20}$ atoms cm$^{-2}$.} 
\label{fig:abc}
\end{center}
\end{figure}
\begin{figure}
\begin{center}
\mbox{\epsfig{file=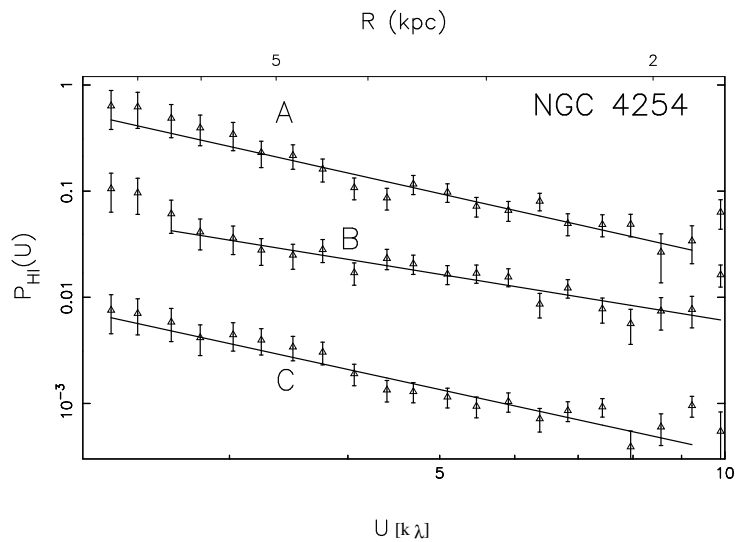,height=2.8in,angle=-0}}
\caption{\HI power spectrum for data cubes A, B and C plotted with
  arbitrary offsets to prevent them from overlapping. }
\label{fig:aps}
\end{center}
\end{figure}

\begin{figure}
\begin{center}
\mbox{\epsfig{file=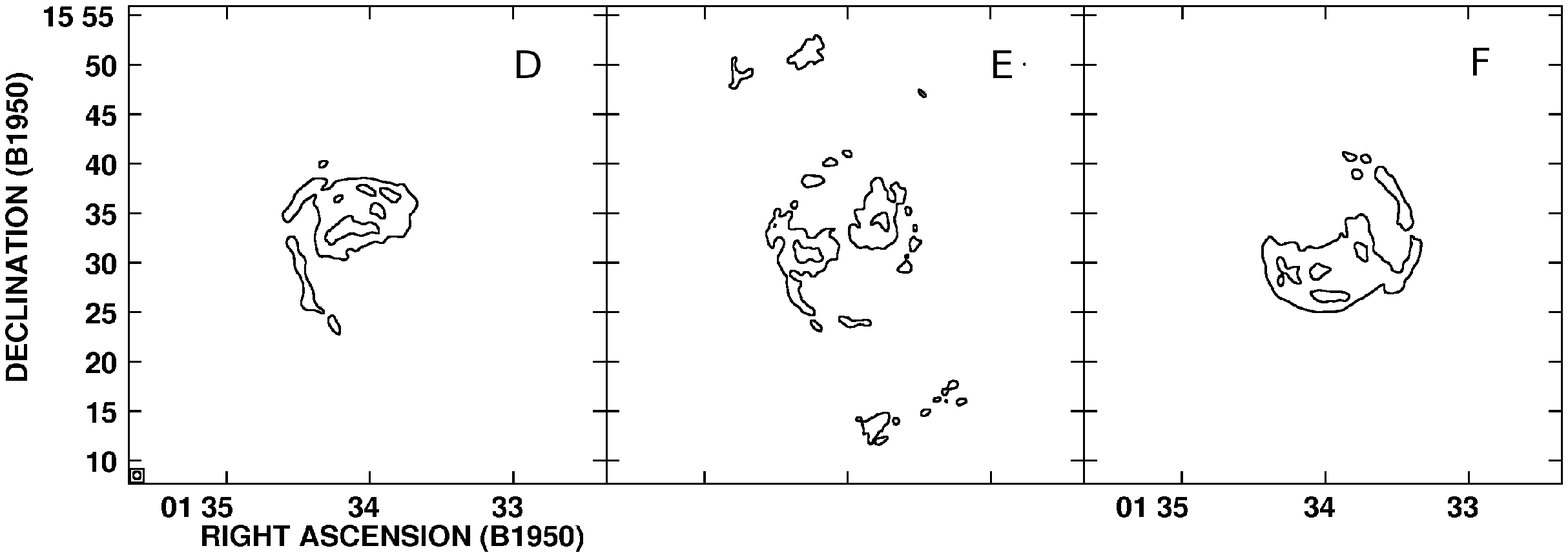,width=6.2in,angle=0}}
\caption{Integrated \HI column density maps of the galaxy NGC~628 using
  data cubes D, E and  
  F. Note the diagonal movement of the centroid of emission from North
  east (D) to South west (F). The contour levels 
are  6.,  12. and 18. $\times 10^{20}$ atoms cm$^{-2}$. } 
\label{fig:DEFM}
\end{center}
\end{figure}

\begin{figure}
\begin{center}
\mbox{\epsfig{file=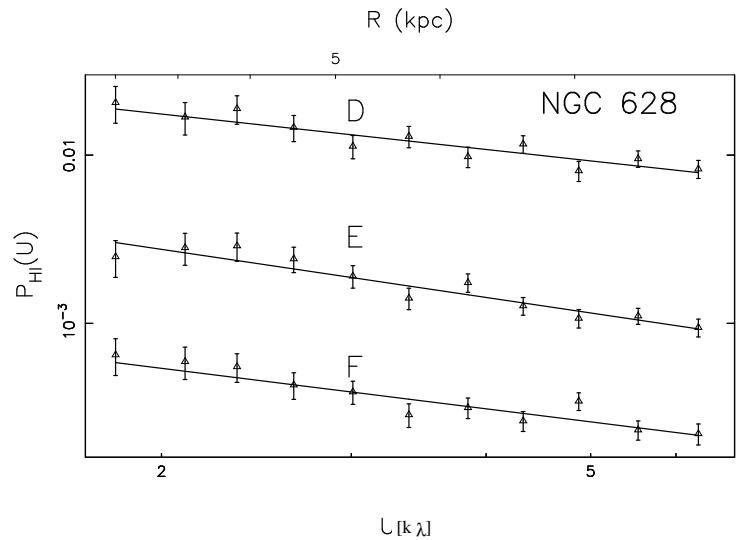,height=2.8in,angle=-0}}
\caption{Power spectra of the \HI emission for the galaxy NGC~628 is
  shown for D (channels 108-119), E (channels 120-131) and F (channels
  132-143) with  arbitrary offsets to prevent them from
  overlapping. }
\label{fig:628PS}
\end{center}
\end{figure}

\begin{table}
\centering
\begin{tabular}{cc|rrcc|}
\hline 
Data & Channels & $U_{min}$ & $U_{max}$ & $\alpha$   \\
& & (k $\lambda$) & (k $\lambda$) & \\
\hline
& & & & \\
16 central  & $27-42$ & $2.5$ &  $10.0$ & $-1.7\pm 0.2$ \\
channels &&&& \\
\hline
& & & & \\
A & $23-30$ &  $2.0$ &  $10.0$ & $-2.0\pm 0.3$ \\
B & $31-38$ &  $2.5$ &  $10.0$ & $-1.5\pm 0.2$ \\
C & $39-46$ &  $2.0$ &  $10.0$ & $-2.0\pm 0.3$ \\
 \hline 
\end{tabular}
\caption{Results of the power spectrum analysis of NGC~4254 for
  different data cubes spanning over different velocity ranges.}  
\label{tab:t4}
\end{table}

Galaxy harassment is expected  to have  different effects on the
inner and outer parts of the galaxy. While gas is stripped 
from the outer parts, the inner part looses angular momentum and
gradually collapses to the center through repeated galaxy encounters. 
We can selectively study different parts of NGC~4254, whose rotation
axis is tilted at  $42^{\circ}$  to the line of sight, by considering
different velocity channels. Our analysis till now has used only the
central $16$ channels, we now use the central $24$ channels ($23-46$)
for the subsequent analysis.  We  construct  
$3$  different data cubes namely A, B and C containing channels $23-30$,
$31-38$ and $39-46$ respectively. We can now separately probe the 
North east,  Central and South west parts of the galaxy
({\bf Figure~\ref{fig:abc}}) using these three data cubes. For each data
cube,  we  evaluate the \HI power spectrum using individual channels
and then average over the channels in the respective cubes. We estimate 
$U_m$ separately for A, B and C and the best fit
power law is obtained for $U \ge U_m$ only. We find
that for each  data cube the \HI power spectrum is well fitted by a
power law ({\bf Figure~\ref{fig:aps}}), the details being shown in  
{\bf Table~\ref{tab:t4}}. We find that the slope $\alpha$ is 
$-2.0\pm 0.3$ for A and C which probe the outer parts of the disk
while it is  $-1.5\pm 0.2$ in B which probes the central region. 
To verify if this change in slope is generic to all galaxies, 
we also consider the \HI power spectrum of the  spiral galaxies NGC~628
and NGC~1058  which are not undergoing galaxy
harassment.  {\bf Figure~\ref{fig:DEFM}} shows  the regions  of the
spiral galaxy    NGC~628  corresponding to  each of the channel range
108-119 (D),  120-131 (E) and  132-143 (F).  We find that the 
power spectra of these three data cubes ({\bf Figure~\ref{fig:628PS}})
all  have the same slope $\sim -1.6$,  which is also similar to the
slope of the central part of NGC~4254. The results are similar for 
NGC~1058 and hence we do not explicitly show these here. 
 Based on this we conclude that the difference in slope 
between the inner and outer parts of NGC~4254 is not a generic feature
of the spiral galaxies rather is a consequence of the 
galaxy harassment.  This suggests that not only does galaxy harassment
affect the global morphology of the galaxy, it also affects the fine
scale structure in the ISM as reflected by  \HI power spectrum.

We note that NGC~4254 has an inclination of $42^{\circ}$, on the other
hand NGC~628 and NGC~1058 are more face-on galaxies (inclination
$\sim 10^{\circ}$). In our analysis we selectively study different
parts of a galaxy by considering different velocity channel
ranges. However, since face-on galaxies do not have much
range in radial velocity from the rotation curve, the spatial extent
of \HI in NGC~628 and NGC~1058 will not be very different for the three
velocity channel ranges, which is unlike the case for NGC~4254.
Hence, it is likely that we failed to find a change of slope in NGC~628
and NGC~1058 because of this effect of inclination.
However, as seen in {\bf Figure~\ref{fig:DEFM}} for NGC~628, all three data
cubes D, E and F have a significant contribution of \HI from the outer
region of NGC~628, leading us to believe that the power spectrum
has a slope of $\sim 1.6$ in the outer parts of NGC~628. This is similar
to the value of slope seen in the central parts of NGC~4254 and
significantly different from the slope in the
outer parts of NGC~4254. Hence it indicates an impact of harassment in
NGC~4254. Further analysis of spiral galaxies with large inclination
angles would possibly be able to resolve this issue.

We currently do not have  an understanding of how galaxy harassment
caused a steepening of the \HI power spectrum in the outer parts of the
galaxy.  Theoretical modeling and the analysis of other Virgo cluster 
spiral galaxies are  needed for further progress in this direction.

\section{Summary and conclusions}
We have presented the result of power spectrum analysis of the
external spiral galaxies NGC~628, NGC~1058 and NGC~4254 in this
chapter. Here we summarize the main points of this
analysis. 
\begin{itemize}
\item  For all the three galaxis we have studied here,  the power
spectrum is well fitted by a power law
indicating the presence of turbulence in the ISM. The slope of the power
law is found to be $\sim -1.7$ for NGC~628 and NGC~4254 and $\sim
-1.1$ for NGC~1058 at large length scales. We have interpreted this as
a result of the 2D turbulence in the scale of the galaxy's disk. 

\item Our present observation also suggests that the spiral galaxies
exhibit  scale-invariant density fluctuations that extend to
length-scales  of  $\sim 10$ kpc  which is
comparable to the diameter of the \HI disk. While a large variety of
possible energy sources like proto-stellar winds, supernovae, shocks, etc. 
have been proposed to drive turbulence  seen at the smaller
scales \citep{2004ARA&A..42..211E},  it is still to 
be seen whether these are effective  on length-scales as  large
as $10$ kpc. 

\item We have successfully estimated the scale-height of the  galaxy
NGC~1058. This, to the best of our knowledge, is the first
determination of the scale-height  of any external face-on spiral galaxy.
However, for the  two other galaxies we could not detect any change in
slope of the power spectrum and hence could only put limits to their
scale-height. 

\item For all three galaxies in our sample we could not detect any
change in the slope of the power spectrum with increasing thickness of
the velocity channels. This definitely shows that for these galaxies
velocity does not noticeably modify the power spectrum. The result
presented in \citet{2000ApJ...537..720L} shows a possible modification
of the power spectrum of the ISM intensity fluctuation of our
Galaxy. 

\item We observe that the outer part of the galaxy NGC~4254 has a
different power law slope compared to the inner part. This difference, we
propose, is because of galaxy harassment. We do not find any
such difference in slope for the other two galaxies presented in this
chapter. However, since this is a single observation of a harassed
galaxy, it is not possible to conclusively report this as a generic feature
of galaxy harassment. A systematic study of the other harassed
galaxies in the nearby Coma or Virgo cluster may provide us more interesting
information about the ISM turbulence in such environments.
\end{itemize}

Since it is now evident that the external spiral galaxies also have
turbulent ISM, we may now proceed in estimating the power spectrum of
a spiral galaxy sample. This will help us understand the nature of turbulence,
its origin and the energy input to it. In {\bf Chapter~\ref{chap:THINGS}} we
will consider such a sample and investigate how the turbulence is
correlated with other dynamical and morphological properties of the 
spiral galaxies.

%\newpage

%\clearpage{\pagestyle{empty}\cleardoublepage} %%%%%%%%%%%%%%%%%%%%
%\newpage
\setcounter{section}{0}
\setcounter{subsection}{0}
\setcounter{subsubsection}{2}
\setcounter{equation}{0}
%\pagenumbering{arabic}

%-------------------------------------------
\chapter[Simulating Galaxy Observations]{\bf \textbf {Simulating
Galaxy Observations\footnote{The work presented in this chapter is originally
published in the paper titled ``A study of interstellar
medium of dwarf galaxies using \HI power spectrum analysis"
by \citet{2009MNRAS.398..887D}.}}}
\label{chap:sim}

We have presented a visibility based power spectrum estimator in {\bf
Chapter~\ref{chap:est}} which we have  used 
 to perform power spectrum analysis of three
external spiral galaxies and the results are presented in {\bf
Chapter~\ref{chap:3spi}}. The observed bi-modality in the value of the
best fit power law slope $\alpha$ of these spiral galaxies with the
same from Milkyway or nearby galaxies have been explained  in terms of
3D and 2D turbulence at small and large scales respectively. The power
spectrum of the galaxy NGC~1058 is 
found to be a broken power law showing a transition from 3D to 2D
turbulence. The break was seen at a wavelength of $1.5$ kpc which we
interpreted as the scale-height of the galaxy. However, the actual
length-scales associated with the scale-height could not be
definitively estimated.
  
In this chapter we perform numerical simulations to asses the impact
of the overall galaxy structure on our estimates of the power spectrum
of \HI intensity fluctuations. We are particularly interested in the
effect of galaxy's radial and transverse profile, inclination angle and
the large scale density field in 
the power spectrum estimator we have been using.

\section{Simulation} 
The starting point of our simulation is a three dimensional (3D)
$512^3$ mesh, labeled using coordinates $\vr$,  on which we
generate a statistically homogeneous  and isotropic Gaussian random
field $\hat{h}(\vr)$  with power spectrum $P(k)= C k^{\gamma}$. We
have used $\gamma=-2.5$ throughout our simulations, 
the results can be easily generalized to other $\gamma$
values. The 3D power spectrum is shown in  {\bf Figure \ref{fig:sim1}}.   
For reference we also show the two dimensional (2D) power spectrum of
$\hat{h}(\vr)$ evaluated on a $512^2$ planar section of the cubic
mesh. As expected, the 2D power spectrum also is a power law with
$P(k) \propto  k^{\gamma+1}$.  In all cases we have generated five
independent realizations of the Gaussian random field, and averaged
the power spectrum over these to reduce the statistical
uncertainties.  

In our simulations 
we  use the Gaussian random field $\hat{h}(\vr)$ as a  model  for 
the 3D  \HI  density fluctuations  that would
arise from homogeneous and isotropic 3D turbulence. The $\hat{h}(\vr)$
values on a 2D section through the cube  serves as a model for the \HI
density fluctuations in the limiting situation where we ignore 
the  thickness of the galaxy and treat it as a 2D disk. Note that the
resulting $\hat{h}(\vr)$ is a statistically homogeneous and isotropic
Gaussian random field on the 2D section. We use this to represent the
density fluctuations that would arise from homogeneous and isotropic
2D turbulence.  

We next embed a galaxy in the middle of the 3D cube. 
 The overall,  large scale  3D structure of 
the galaxy is  introduced  through a function $G(\vr)$ so that the \HI 
density at any position is $G(\vr) [h_0 + \hat{h}(\vr)]$. Here $G(\vr)
h_0$ is the  smoothly varying component of the galaxy's 
\HI density and $G(\vr)  \hat{h}(\vr)$ is its fluctuating component. 
 It is assumed that the observer's line of sight is  along the
$z$ axis. The \HI density    $G(\vr) [h_0 + \hat{h}(\vr)]$ is projected
on the $x-y$ plane. We interpret the projected values 
as  the \HI specific intensity $I(\vt, \nu)$  in the plane of the
sky [{\bf Eqn.~(\ref{eq:a1})}]. The $512^2$ mesh in the $x-y$ plane of our
simulation corresponds  to $4' \times 4'$ on the sky. 

To understand the effect of different parameters such as the shape of
the window function, the inclination angle etc. on the power spectrum
estimation we consider two models for the galaxy's \HI density. 

\subsection{Case I : Without a radial profile}
We first consider a face-on galaxy, and use $G(\vr)= \exp(-z^2/z_h^2)$.
This incorporates only  the finite thickness of the disk which is
characterized by the scale-height $z_h$ and  ignores  the galaxy's
radial profile  in the plane of the disk. Note that, for $z<z_h$ this
function closely matches the function ${\rm sech}(-z^2/z_h^2)$  
which is also used to model the scale-height
profile \citep{1998gaas.book.....B}. 
For the scale-height, we have used the values  $z_h=4$ and $32$
mesh units 
which we refer to as the ``thin' and ``thick'' disk respectively. 

\subsection*{Results}

\begin{figure}
\begin{center}
\epsfig{file=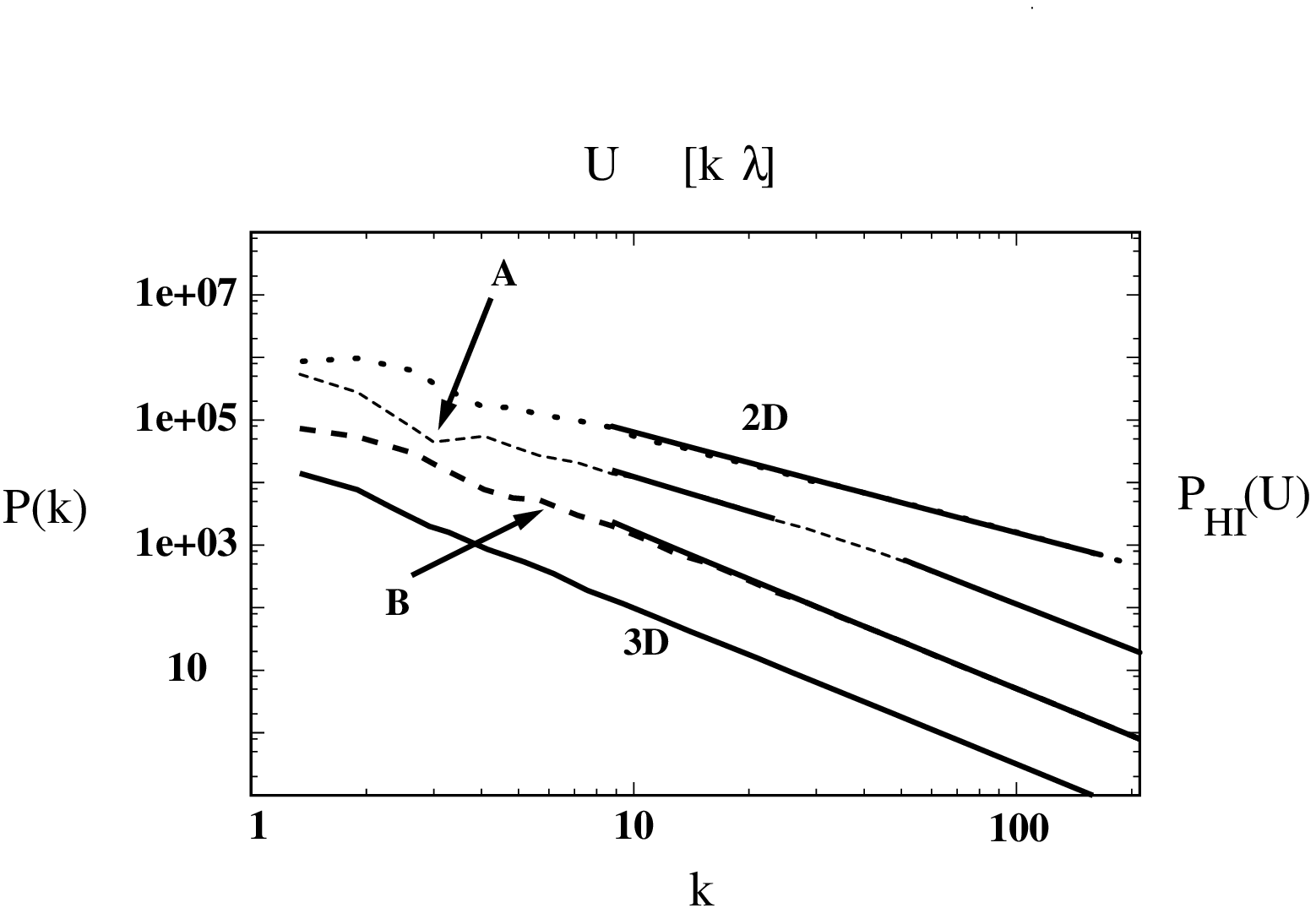, width=3.8in, angle=0}
\end{center}
\caption{The 3D and 2D power spectrum $P(k)$.  
The simulated \HI power spectrum $P_{HI|}(U)$
for the thin (A) and thick (B) disk without the radial profile are
also shown.  The $P(k)$ and $k$ values (left and bottom axes) have
been arbitrarily scaled to match the $P_{HI}(U)$
and $U$ axes (top and right). The power-law fits are shown by  solid
lines.  The different curves have been plotted with arbitrary offsets
to make them distinguishable.
}
\label{fig:sim1}
\end{figure}

{\bf Figure ~\ref{fig:sim1}}  shows the  power spectrum 
of the simulated \HI specific intensity $I(\vt, \nu)$
 for both the thin and thick disk.
  We find that the \HI
power spectrum of the thin disk has a slope $-2.5$, same as the 3D
power spectrum, at large $U$.  There is a break
at $U \sim 35 \, {\rm k}\lambda$ which corresponds to $ 1/\pi  z_h$,
and the slope is  $\sim -1.9$ at smaller $U$.  We interpret this
change in the slope in terms of a transition from 3D fluctuations on
scales smaller than the disk thickness (large $U$) 
to 2D fluctuations on scales larger than disk thickness (small $U$).
We expect the slope to approach $-1.5$, the 2D slope,   at very small
$U$.  Note that in our simulation it is not possible to evaluate the
slope at  
very small $U$ values where the sample variance is rather large. 
We find that the \HI power spectrum of the thick disk  is well fit by a
single power law $P_{HI}(U)=A U^{-2.5}$ which has the 3D slope. In
this case we expect the break corresponding to  the transition from 3D
to 2D  at $U \sim 4 {\rm k}\lambda$. This break lies in the  sample
variance dominated region which explains why we do not detect it.

\subsection{Case II : With a radial profile}
We next incorporate the galaxy's radial profile  using 
\begin{equation}
G(\vr) = \exp \left [-\frac{\sqrt{12}\theta}{\theta_{0}}\right
]\exp(-z^2/z_{h}^2),
\label{eq:galprof}
\end{equation}
where $\theta=\sqrt{x^2+y^2}$  is the radial coordinate  in the
plane of the disk. Here $\theta$ coincides with the angle in the sky
measured from the center of the galaxy.   We have used $\theta_0=1'$
which  corresponds to $128$ mesh units . Note that the radial profile
used in the simulation is  exactly the same as 
the  window function $W(\vt)$ of the exponential model introduced 
in {\bf Chapter~\ref{chap:est}}.   The relative amplitude of $h_0$ and
$\hat{h}(\vr)$ is a free parameter  which  decides the respective
contributions from the  smooth and the fluctuating \HI components.   
Using the ratio $R=\sqrt{\langle \hat{h}^2 \rangle}/h_0$
of the root mean square value of  $\hat{h}(\vr)$   to $h_0$   
to quantify this, we have carried out simulations for 
$R=0.35$ and $0.7$.  We have also carried out simulations where the
disk is tilted, and the normal to the disk make an angle of
$60^{\circ}$ to the line of sight. 

\subsection*{Results}
\begin{figure}
\begin{center}
\epsfig{file=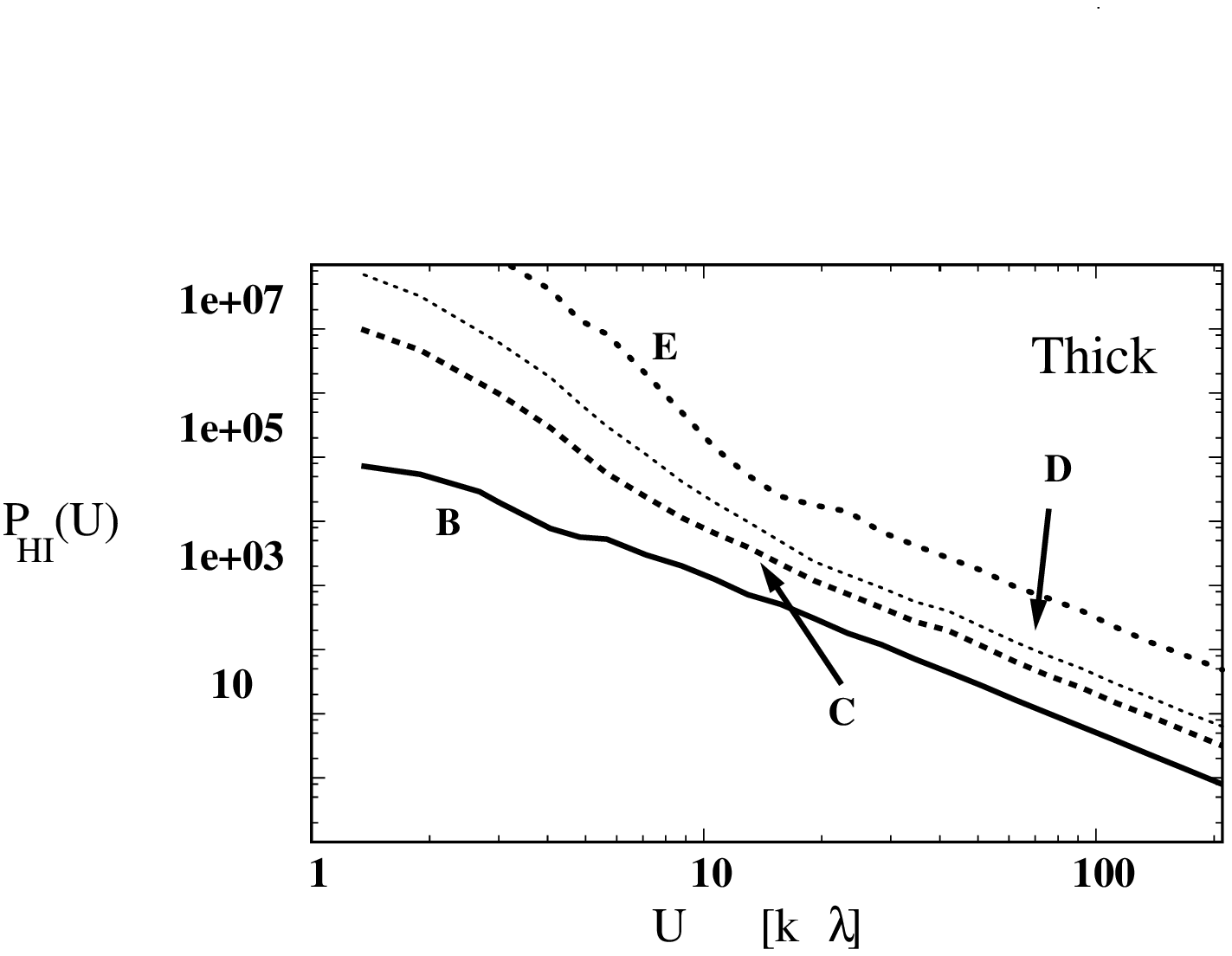,width=4.in, angle=0}
\end{center}
\caption{The simulated \HI power spectrum for the face-on thick disk
  with  (B) no radial profile and  (C), (D)  with radial profile using 
  $\theta=0.35,0.7$ respectively.  (E) is same as (C) with the disk
 tilted  at
  $60^{\circ}$.  The different curves have been plotted with arbitrary offsets
to make them distinguishable.
} 
\label{fig:sim2}
\end{figure}

\begin{figure}
\begin{center}
\epsfig{file=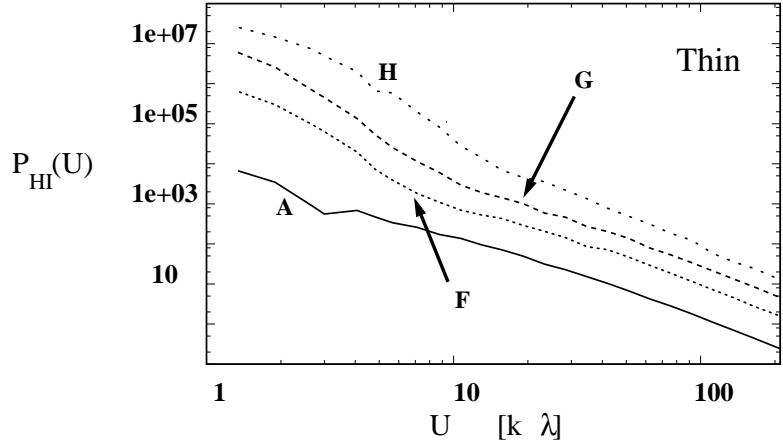,width=4.in, angle=0}
\end{center}
\caption{The simulated \HI power spectrum for the face-on thin disk
  with  (A) no radial profile and  (F), (G)  with radial profile using 
  $\theta=0.35,0.7$ respectively.  (H) is same as (F) with the disk
  tilted
  at 
  $60^{\circ}$.  The different curves have been plotted with arbitrary
  offsets 
to make them distinguishable. 
} 
\label{fig:sim3}
\end{figure}

The  simulated \HI power spectra are  shown in 
{\bf Figures~\ref{fig:sim2}} and {\bf \ref{fig:sim3}}  for the thick and thin
disks respectively.
We find that for a face-on disk with $\theta_0=1'$, for both the thick
and thin disks, 
the results of  our simulations are in good agreement with our
analytic  estimate presented in  the {\bf Section~\ref{sec:visicorr}},
which predicts  that  the effect of   
the galaxy's  radial profile is  contained within  a  limited
baseline range  $U \le U_{m}$, which comes out to be 
$U \le U_{m} \sim 3.5  \ \theta_0^{-1}$, for $\alpha = -2.5$ ({\bf
Figure~\ref{fig:Um}}). 
The \HI power spectrum is insensitive to the galaxy's  radial  
profile at $U > U_m$, and  the shape of the power spectrum is the same
whether we include the galaxy's radial profile or not. 
We  find that the value of $U_m$ is not very sensitive  
to changes in $R$, the ratio of the fluctuating component to the smooth 
component. For a disk of the same size, the value of $U_m$ increases
when the disk  is tilted.  The image of the disk  tilted  by
$60^{\circ}$ has an anisotropic window function $W(\vt)$ with angular
radius  $1'$ and  $0.5'$ along the major and minor axes respectively.  
We find that for  a tilted disk,
 the value of $U_m$ is determined by the smaller of the
 two angular diameters, $0.5'$ in this case.

\section{Scale-height estimation}
In {\bf Chapter~\ref{chap:3spi}} we have observed a break in the power
spectrum of the galaxy NGC~1058 at a baseline of $6.5$ k $\lambda$
(or at a wavelength of $1.5$ kpc) and interpreted it as arising due to
a transition from 3D to 2D turbulence. The results of our simulation
suggests that a break in the  power spectrum  at a baseline $U$
corresponds to an angular scale of $z_h=1/\pi U$ and a length-scale of
$D/\pi U$ at the plane of the sky ($D$ is the distance to the
galaxy). Following this for  an 
adopted distance of $10$ Mpc to the galaxy NGC~1058, its scale-height
is $490\pm90$ pc. We have also observed that
the galaxy NGC~628 exhibit 2D turbulence up to a baseline of $1.0$ k
$\lambda$ with a single power law of slope $\sim -1.6$. This imposes
an upper limit of $320$ pc on its scale-height. Note that the
scale-height   of 320 pc for NGC~628 differs from what is presented 
in \citet{2008MNRAS.384L..34D} because of the extra factor of $\pi$
indicated by our simulations. We do not attempt to set a bound for the
scale-height of the galaxy NGC~4254 here, since it has a considerably
large inclination angle. Finally, we note that we have not considered
the fact that the scale-height varies with the distance
from the centre of the galaxy's disk. Hence our estimate gives an
average value of the disk scale-height.

\section{Summary and conclusions}
We discuss the main findings of this chapter here.
\begin{itemize}
\item We find that if the \HI distribution is a thin disk, the power 
spectrum is well fitted using a broken power law with the change in
scope at a length-scale corresponding to the line of sight thickness
of the disk. We interpret this as a geometrical effect arising because
of a transition from 3D to 2D fluctuation. Since the spiral galaxies
have disk like geometry,   a break in the power spectrum at a $U$ value
greater than  $U_m$   is indicative of a transition from 3D to 2D at
the length-scale corresponding to the scale-height $z_h=D/\pi U$.

\item Our simulations demonstrate that for baselines $U>U_m$ the
galaxy's radial profile has no effect on the power spectrum of \HI
fluctuations. In other words, the power spectrum would be unchanged if
the same fluctuations were present in an uniform disk with  no
radial profile.
\item We find that the inclination angle of the Galaxy further
restricts the usable range of baselines for the power spectrum
estimation. Our simulation shows that the semi-minor axis of the
projected image of the galaxy has to be used to determine the $U_{m}$
value over which the  power spectrum can be estimated.

\end{itemize}

Finally, we note that we have ignored velocity fluctuations and 
the galaxy's rotation throughout our simulations. This would 
rearrange the \HI emission amongst the  different frequency channels.
Our simulation corresponds to the situation where there is just a
single frequency channel whose width encompasses the entire HI
emission, and the velocities   have no  effect in this
situation.

%\newpage

%\clearpage{\pagestyle{empty}\cleardoublepage} %%%%%%%%%%%%%%%%%%%%
 %\newpage
 \setcounter{section}{0}
 \setcounter{subsection}{0}
 \setcounter{subsubsection}{2}
 \setcounter{equation}{0}
 %\pagenumbering{arabic}

\chapter[Power Spectrum and Dwarf Galaxies] {\bf \textbf {Power
 Spectrum as a Probe of Dwarf Galaxy Properties\footnote{The work presented in this chapter is originally
published in the paper titled ``A study of interstellar
medium of dwarf galaxies using \HI power spectrum analysis"
by \citet{2009MNRAS.398..887D}.}}} 
\label{chap:dwarf}

\section{Introduction}
 This chapter presents the results of our analysis
of a sample of 7 nearby gas rich faint ($M_{B} > -13.0$ mag) dwarfs.
The corresponding \HI radio-interferometric data used here is from
GMRT. The observation parameters and other details are 
given in the {\bf Table~\ref{tab:obs}}. Note that the galaxy GR~8 is known
to have peculiar kinematics \citep{2003A&A...409..879B} whereas the spiral
dwarf galaxy NGC~3741 has an extended \HI
disk \citep{2005A&A...433L...1B}. These were included in 
our sample to see the effect of the peculiar dynamics and morphology
of the ISM in the power spectrum. 
\begin{table}[t]
\centering
\begin{tabular}{lcccccc}
\hline
Galaxy & & $\alpha$(J2000)  & $\delta$(J2000) & Date of &
Time on & References\\  
&  & (h m s) & ($^{\circ}\, '\, ''$) & observation & source &\\
& & & & & &\\
\hline \hline
& & & & & &\\
DD0~210  & & $20\, 46\, 53.00$ & $-12\, 50\, 57$ & 13-15 July 02 & 16 hrs   &  2,5 \\ 
NGC~3741 & & $11\, 36\, 06.40$ & $+45\, 17\, 07$ & 22 Jul \& 26 Aug 04
&  8 hrs  &  3,5 \\  
UGC~4459 & & $08\, 34\, 06.50$ & $+66\, 10\, 45$ & 15,23,24 Nov 02 &
14 hrs &  4,5 \\   
GR~8     & & $12\, 58\, 40.40$ & $+14\, 13\, 03$ & 16-18 Nov 02 & 16 hrs &  1,5 \\  
AND~IV   & & $00\, 42\, 32.30$ & $+40\, 34\, 19$ & 1 Jan, 07 &      & 6\\  
KK~230   & & $14\, 07\, 10.70$ & $+35\, 03\, 37$ & May 8, 6 Jun, 26
Nov 01 & 18 hrs & 4,5 \\  
KDG~52   & & $08\, 32\, 56.00$ & $+71\, 01\, 46$ & 21-23, 27 Jun 02 &
18 hrs & 4,5 \\  
\hline
\end{tabular}
\caption{Some observation parameters of the galaxies in our
  sample. References are as follows:
1- \cite{2003A&A...409..879B}, %GR8
2- \cite{2004A&A...413..525B}, %DDO210
3- \cite{2005A&A...433L...1B}, % NGC3741
4- \cite{2006MNRAS.365.1220B}, %UGC4459, KK230, KDG52
5- \cite{2008MNRAS.386.1667B}, %ALL DWARF - ANDIV
6- Chengalur et al. 2009 (in preparation)%ANDIV
}
\label{tab:obs}
\end{table}

\section{Notes on the individual galaxies}
We briefly discuss some of the relevant   properties of 
the galaxies in our sample. A few  of the galaxy parameters are also 
summarized in {\bf Table~\ref{tab:profile}} which  contains - 
Column (1): Name of the galaxy; (2) and (3): major and minor axis in arc min
calculated at a column density of $10^{19}$ atoms~cm$^{-2}$; (4):
distance in Mpc; (5):  \HI inclination angle; (6): Log[SFR] in
M$_{\odot}$yr$^{-1}$  measured from H$\alpha$ emission; (7): 
\HI velocity dispersion in km s$^{-1}$; (8): total dynamical mass in
$10^{8} \, M_{\odot}$; (9): references. 

\subsection*{DDO~210}
DDO 210 is the faintest ($M_{B}\sim -10.9$) relatively close (at a
distance $950\pm50$ kpc; \citealt{1999AJ....118..853L}) gas-rich
member of the Local 
Group. The \HI disc of the galaxy is nearly face-on. On large scales,
the \HI distribution is not axisymmetric; the integrated \HI column
density contours are elongated towards the east and south. No
H$\alpha$ emission 
was detected indicating a lack of on-going star formation in this
galaxy. 
\begin{table}[t]
\centering
\begin{tabular}{lcccccccc}
\hline
Galaxy & Major  & Minor & $D$ & $i_{HI}$ & log(SFR) &  $\sigma_{v}$
& M$_{dy}$ & Ref. \\  
&  ($'$) & ($'$) & (M pc) & ($^{\circ}$) & (M$_{\odot}$ yr$^{-1}$) & (km
s$^{-1}$) & ($10^{8}$ M$_{\odot}$) & \\
& & & & & & & & \\
\hline \hline
& & & & & & & & \\
DD0~210 & $\ 5.0$ & $\ 4.5$ & $1.0$ & $26.0$ & $>-5.4$ & $6.5(1.0)$ &
$0.34$ & 2,5 \\
NGC~3741 & $27.0$ & $12.0$ & $3.0$ & $68.0$ & $\ \ \ \, -2.47$ & $\sim8.0$ &
$40.3$ & 3,5 \\
UGC~4459 & $\ 4.5$ & $\ 3.6$ & $2.1$ & $30.0$ & $\ \ \ \,  -2.04$ & $9.0(1.6)$ &
$1.40$ & 4,5 \\
GR~8 & $\ 4.2$ & $\ 4.0$ & $2.1$ & $27.0$ & $\ \ \ \, -2.46$ & $9.0(0.8)$ &
$0.78$ & 1,5 \\
AND~IV & $15.0$ & $\ 8.5$ & $6.7$ & $55.0$ & $\ \ \ \,  -3.0$ & $\sim 8.0$ &
$32.5$ & 5,6 \\
KK~230 & $\ 3.0$ & $\ 2.0$ & $1.9$ & $50.0$ & $\ >-5.53$ & $7.5(0.5)$ &
- & 4,5 \\
KDG~52 & $\ 3.5$ & $\ 3.0$ & $\ 3.55$ & $23.0$ & $>-5.1$ & $9.0(1.0)$ &
- & 4,5 \\
\hline
\end{tabular}
\caption{Some relevant morphological and dynamical parameters of the
  galaxies in our sample. The references are as follows:
1- \cite{2003A&A...409..879B}, %GR8
2- \cite{2004A&A...413..525B}, %DDO210
3- \cite{2005A&A...433L...1B}, % NGC3741
4- \cite{2006MNRAS.365.1220B}, %UGC4459, KK230, KDG52
5- \cite{2008MNRAS.386.1667B}, %ALL DWARF - ANDIV
6- Chengalur et al. 2009 (in preparation)%ANDIV
}
\label{tab:profile}
\end{table}

\subsection*{NGC~3741}
NGC~3741 is a nearby dwarf irregular galaxy ($M_{B}\sim -13.13$) with
a gas disk that extends to $\sim 8.8$ times the Holmberg radius
(\citealt{2005A&A...433L...1B, 2008MNRAS.383..809B}). The galaxy is
fairly edge-on with 
kinematics  inclination varying from $\sim 58^{\circ}-70^{\circ}$.
NGC~3741  appears to have a \HI bar  and is dark matter dominated
with a dark to luminous mass ratio of  
$\sim 149$. Further, this galaxy is undergoing significant star
formation in the center. An interplay between the neutral ISM and star
formation in this 
galaxy is studied in detail by \citet{2008MNRAS.383..809B}. 
 The rotation curve flattens beyond 
$300^{\prime \prime}$ to a value $\sim 50\ {\rm  km \,
s}^{-1}$ \citep{2007MNRAS.375..199G}. They find  that the ISM in this
galaxy shows radial  motion of   $\sim 5 -  13 \, {\rm km \,  s}^{-1}$.  
 \citet{2004AJ....127.2031K} have estimated the  distance 
to this galaxy as $3.0\pm0.3$ Mpc using the tip of
the red giant branch (TRGB) method. 

\subsection*{UGC~4459}
The faint dwarf ($M_{B}\sim-13.37$) galaxy UGC~4459 is a member of the 
M~81 group of galaxies. It is fairly isolated  
from its nearest neighbor UGC~4483 at a projected distance of
$3.6^{\circ}~(\sim 223$ kpc) and a velocity difference $135\ $km 
s$^{-1}$.  
UGC~4459 is a relatively metal poor galaxy, with $12\ +\ \log($O/H$)\sim
7.62$ \citep{2000A&ARv..10....1K}. The optical appearance 
of  this galaxy is dominated by bright blue clumps, which emits
copious amount of H$\alpha$, indicating high star formation.  
The velocity field of UGC~4459 shows a large scale gradient across the
galaxy with an average of $\sim 4.5$ km s$^{-1}$ kpc$^{-1}$, 
though this gradient is not consistent with that expected from a
systematic rotating disk. 
This galaxy has  a TRGB distance of $3.56$ Mpc \citep{2004AJ....127.2031K}.

\subsection*{GR~8}
This  is a faint ($M_{B}\sim -12.1$) dwarf irregular galaxy with very
unusual \HI kinematics \citep{2003A&A...409..879B}. 
The \HI distribution in the galaxy is very clumpy, and  shows 
substantial diffuse, extended gas.   
The high density \HI clumps in the galaxy are associated with 
optical knots.  In optical, it has  
a patchy appearance with the emission dominated by  bright blue
knots  which are sites of active star
formation \citep{1967ApJ...148..719H}.  
Both radial and circular   motions are  present in this
galaxy.  It also possesses a  faint extended emission in H$\alpha$. 
 The distance to this galaxy is estimated to be  $2.10$ Mpc
 \citep{2004AJ....127.2031K}.

\subsection*{AND~IV}
This  is a dwarf irregular galaxy with a moderate surface brightness 
($\bar{\mu}_{V} \sim 24$) and a very blue color ($V-I \leq 0.6$)
\citep{2000AJ....120..821F}. 
It  is  at a projected distance of $40^{'}$  from the
center of M~31 \citep{2006MNRAS.369..120M} and  has  a very low
ongoing star formation rate (SFR) of $\sim 0.001 M_{\odot} $yr$^{-1}$.  Its   
 \HI disk  extends to  $\sim 6$ times the Holmberg 
diameter and also shows large scale, purely gaseous spiral arms
[Chengalur et al. 2008 (in preparation)]. The distance to this galaxy is
estimated to be $6.7\pm 1.5$ Mpc \citep{2000AJ....120..821F}.  

\subsection*{KK~230}
KK~230, the faintest($M_{B}\sim -9.55$) dwarf irregular galaxy in our
sample, lies at the periphery  of the Canes Venatici~I cloud of  
galaxies \citep{2004AJ....127.2031K}. The velocity field  shows a
gradient in the direction roughly perpendicular to the \HI and 
optical major axis with a magnitude of $\sim 6\ $km s$^{-1}$
kpc$^{-1}$ \citep{2006MNRAS.365.1220B}. There is no measurable ongoing
star formation   
in this galaxy as inferred from the absence of any detectable 
H$\alpha$ emission. \citet{2004AJ....127.2031K} has found a tidal
index of   $-1.0$ indicating the galaxy to be  fairly isolated. They
have estimated the TRGB distance  to be  $1.9$ Mpc.

\subsection*{KDG~52}
KDG~52 (also called M~81DwA), another member of the M81 group, is a
faint dwarf  galaxy ($M_{B}\sim-11.49$)  with  a clumpy  \HI 
distribution in  a broken ring surrounding the optical
emission \citep{2006MNRAS.365.1220B}.  
The \HI hole is not exactly centered around the optical emission.
This galaxy does not have any detectable ongoing star formation. 
The distance to this galaxy is estimated to be $3.55$
Mpc \citep{2004AJ....127.2031K} derived from the TRGB  method.

\section{Results and discussions}
We adopt the data analysis procedure discussed in {\bf
Section~\ref{sec:moa}}.  
All the data are reduced in the usual way using standard tasks in
classic AIPS \footnote{NRAO Astrophysical Image Processing System, a
commonly used software for radio data processing.}. For each galaxy,
after calibration, the frequency channels with \HI emission are
identified and a  continuum image is made by combining the line-free
channels. The continuum is hence subtracted from the data in the $uv$
plane using the AIPS task UVSUB. The number of channels with \HI 
emission ($n$) is different for each galaxy. To determine 
if the \HI power spectrum changes with the width of the frequency
channel, $N$ successive channels are combined to obtain n/N
channels. Power spectrum analysis is performed for 
different values of $N$ in the range $1\le N\le n$ . The analysis is
initially carried out for $N =  1$ and unless mentioned otherwise the
results refer to this value. 
                             
\begin{figure*}
\begin{center}
\epsfig{file=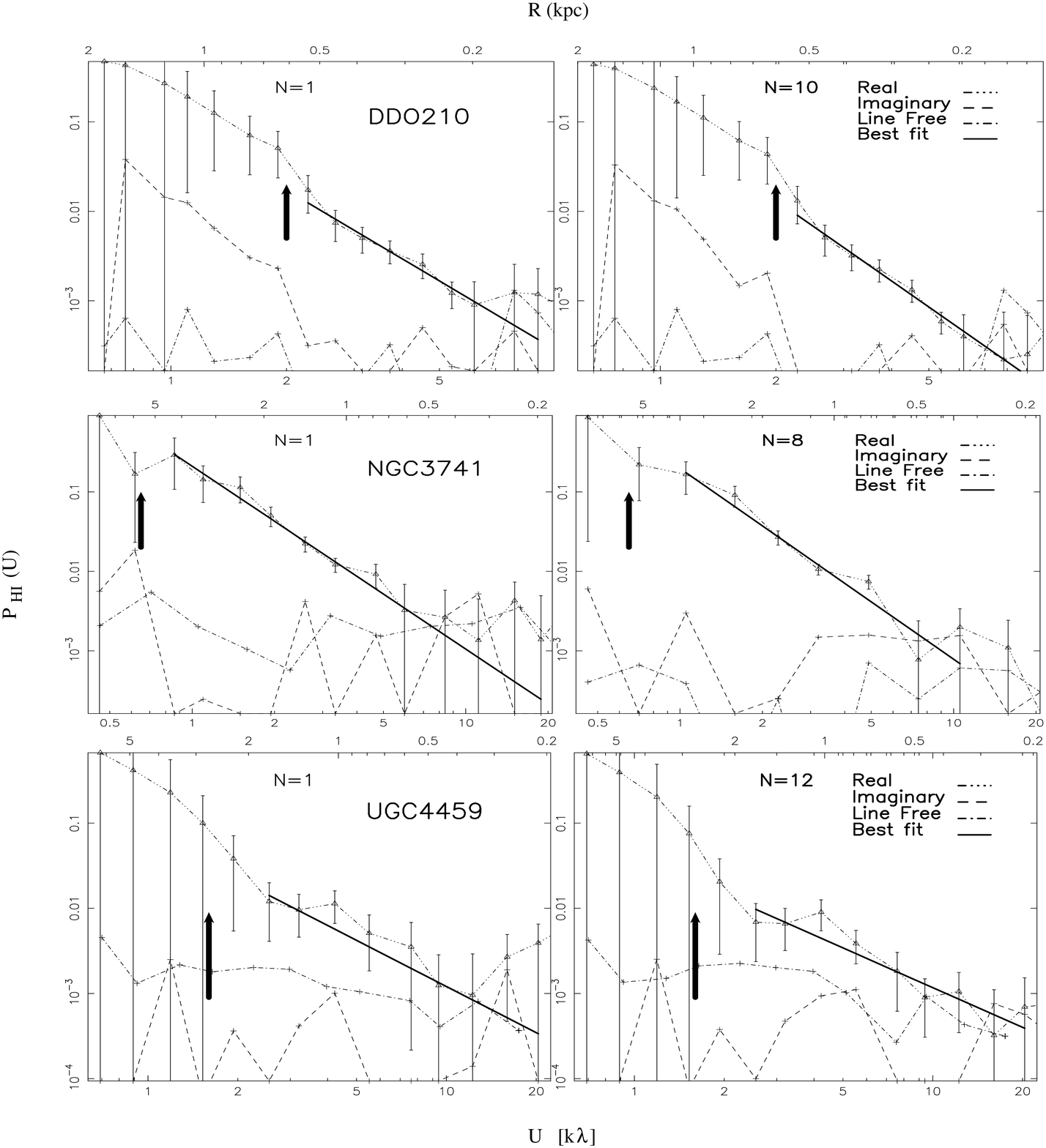,width=4.8in, angle=0}
\end{center}
\caption{Power spectrum of the galaxies DDO~210, NGC~3741 and
  UGC~4459. The real and imaginary parts   of $\P(U)$
estimated after averaging $N$ channels with \HI emission, and the real
  part from $N$ line-free channels are shown together for $N=1$  (left
  panel) and $N=n$ (right panel). The error-bars are for the real part
  from channels with \HI emission.    The best fit power law is shown
  in bold.  In each case $U_{m}$ is marked with  a  bold-faced arrow
  and   the fit is restricted to $U > U_m$ where the
  effect of the convolution with the window function can be
  ignored.}
\label{fig:dpsfig1}
\end{figure*}

\begin{figure*}
\begin{center}
\epsfig{file=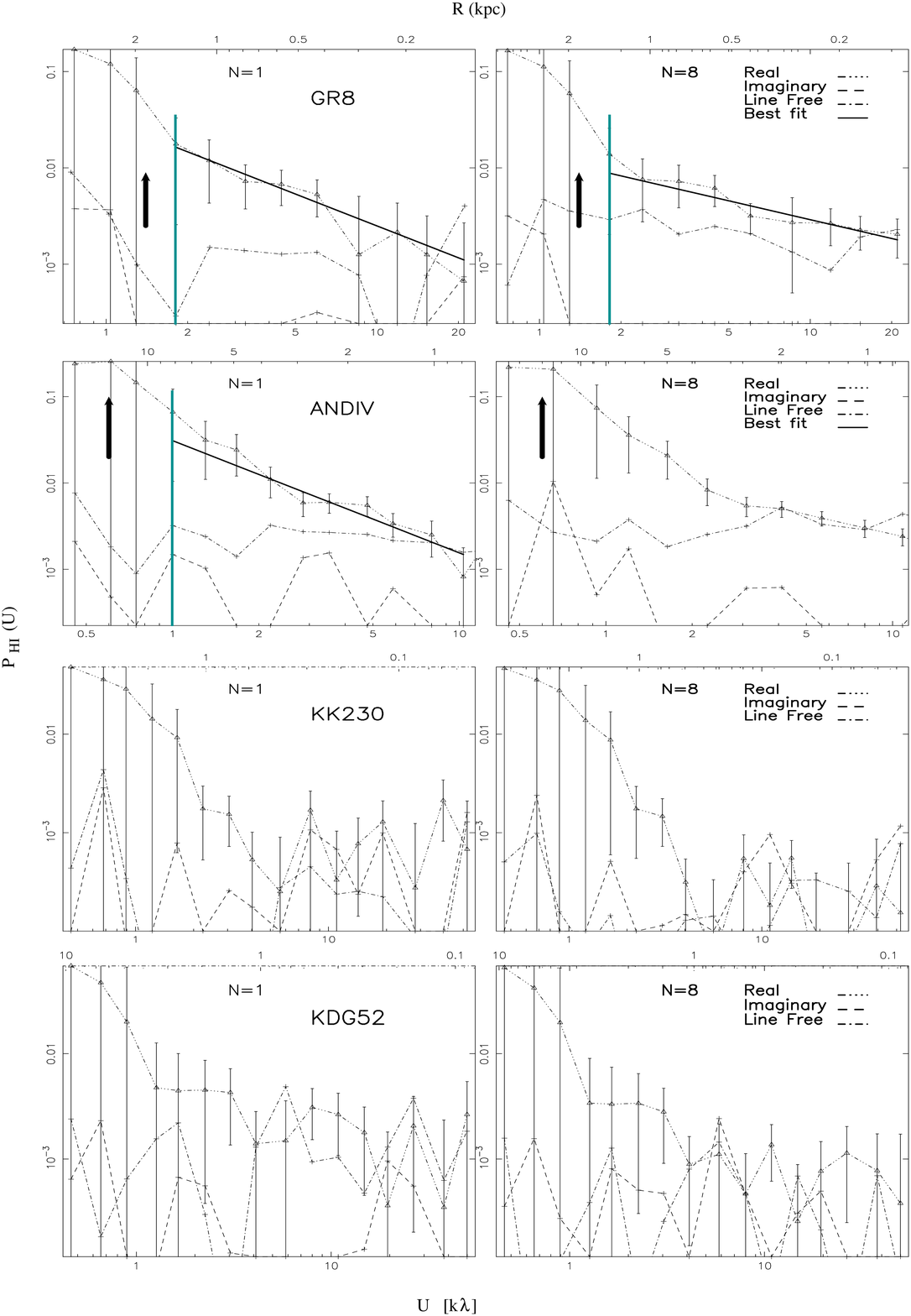,width=4.8in, angle=0}
\end{center}
\caption{Power spectrum of the galaxies  GR~8, AND~IV, KK~230 and
  M81DWA.    The real and imaginary parts   of $\P(U)$ 
estimated after averaging $N$ channels with \HI emission, and the real
  part from $N$ line-free channels are shown together for $N=1$  (left
  panel) and $N=n$ (right panel). The error-bars are for the real part
  from channels with \HI emission.    The best fit power law is shown
  in bold only where such a fit is possible. For those,    $U_{m}$ is
  marked with  a  bold-faced arrow   and   the fit is restricted to $U
  > U_m$ where the   effect of the convolution with the window
  function can be   ignored. }
\label{fig:dpsfig2}
\end{figure*}

{\bf Figure~\ref{fig:dpsfig1}}  and  {\bf \ref{fig:dpsfig2}} show the
results of our 
analysis. The results are also summarized in {\bf Table~\ref{tab:result}}.
  We have detected the \HI power spectrum of  
the galaxies  DDO~210, NGC~3741, UGC~4459, GR~8
and AND~IV. For these galaxies we find a range of baselines $U$,
where the 
real part of $\P(U)$ estimated from the channels with \HI emission
is larger than the imaginary part estimated  from the same channels
as well as  the real  part estimated from the line-free channels. 
This is not true  for KK~230 and KDG~52 (and also for AND~IV with
$N=n$),  for these galaxies   all three curves
({\bf Figure~\ref{fig:dpsfig2}}) lie within the $1~\sigma$ errors bars
with very small offset and the interpretation is not 
straight-forward.   For  the subsequent analysis  we use only the real
part of  $\P(U)$ estimated from the 
channels with \HI emission and interpret it as power spectrum.

  We find that a power law  $\P(U)=A\  U^{\alpha}$   
provides  a good fit at a baseline range $U_{min}$ to $U_{max}$ with a
 reasonable  $\chi^{2}/\nu$  for  
DDO~210, NGC~3741, UGC~4459, GR~8 and AND~IV.  The presence
 of a scale-invariant, power law power spectrum indicates that
 turbulence  is operational in the ISM of these galaxies. 
The range of  length-scales for the power law  fit differs from galaxy
 to galaxy and in total  it covers  $100$ pc to $6.2$ kpc. The details are
 summarized in {\bf Table~\ref{tab:result}}. 

\begin{table}
\centering
\begin{tabular}{lcccccccc}
\hline
Galaxy & N & $\Delta v$  & $U_{min}$ & $U_{max}$ &  $R_{min}$
& $R_{max}$ & $\alpha$  &$\zeta_{2}$\\  
&  & (km s$^{-1}$) & (k $\lambda$) & (k $\lambda$) & (k pc) & (k pc) & &
\\
& & & & & & & &\\
\hline \hline
& & & & & & & &\\
DD0~210  & $1$ & $1.65$ & $3.7$ & $13.0$ & $0.06$ & $0.27$ & $-2.3 \pm
0.6$ & $0.0\pm1.2$\\ 
NGC~3741 & $1$ & $1.65$ & $1.6$ & $16.0$ & $0.19$ & $1.90$ & $-2.2 \pm
0.4$ & $0.0\pm0.8$\\
UGC~4459 & $1$ & $1.65$ & $3.1$ & $22.0$ & $0.18$ & $1.16$ & $-1.8 \pm
0.6$ & $0.0\pm1.2$ \\
GR~8     & $1$ & $1.65$ & $1.6$ & $23.0$ & $0.10$ & $1.30$ & $-1.1 \pm
0.4$ & $0.0\pm0.8$\\
AND~IV   & $1$ & $1.65$ & $0.6$ & $\ 6.7$& $0.56$ & $6.20$ & $-1.3 \pm
0.3$ & $0.0\pm0.6$\\ 
\hline
& & & & & & & &\\
DD0~210  & $10$ & $16.5$ & $3.7$ & $13.0$ & $0.06$ & $0.27$ & $-2.1
\pm 0.6$ &  $0.0\pm1.2$\\ 
NGC~3741 & $\ 8$& $13.2$ & $1.6$ & $16.0$ & $0.19$ & $1.90$ & $-2.5
\pm 0.4$ & $0.0\pm0.8$\\ 
UGC~4459 & $12$ & $19.8$ & $3.1$ & $22.0$ & $0.18$ & $1.16$ & $-1.7
\pm 0.4$ & $0.0\pm1.2$  \\
GR~8     & $\ 8$& $13.2$ & $1.6$ & $23.0$ & $0.10$ & $1.30$ & $-0.7
\pm 0.3$ & $0.0\pm0.8$ \\
AND~IV   & $16$ & $26.4$ & $-$   & $-$    & $-$    & $-$    & $-$ &
$0.0\pm0.6$ \\ 
\hline
\end{tabular}
\caption{The results for  the 5 galaxies in our sample.  Columns 1-9
  give (1) Name of the galaxy, (2) $N$:  the number of channels
  averaged over, (3) $\Delta V$: the corresponding  velocity width,
  (4) and (5) $U_{min}$ and $U_{max}$: the baseline range for which
  the power spectrum is determined, (6) and (7)  $R_{min}$ and $R_{max}$:
corresponding range of length scales in the galaxies disk,  (8) 
$\alpha$: the best fit slope for the HI power spectrum, (7)
$\zeta_{2}$: possible limits for the spectral slope of the velocity
structure function.}  
\label{tab:result}
\end{table}

Both \HI density fluctuations as well as spatial fluctuations in the
velocity of the \HI gas contribute to fluctuations in the \HI specific
intensity.
%Considering a turbulent ISM, \citet{2000ApJ...537..720L} have shown that 
%it is possible to disentangle these two contributions  by studying
%the behavior of the \HI power spectrum as the thickness of the
%frequency  channel is varied. If the observed \HI power spectrum is 
%due to the gas velocities, the slope of the  power spectrum 
%is predicted to decrease with increasing thickness of the 
%frequency  channel.   
To test if the power spectrum has any contribution from the gas
velocity fluctuations,  we have   repeated the power
spectrum analysis 
increasing  the channel thickness $N$ from $N=1$ to $N=n$.   In
addition to $N=1$,   {\bf Figures~\ref{fig:dpsfig1}}  and
{\bf \ref{fig:dpsfig2}}, 
and  {\bf Table~\ref{tab:result}}  also show the  results for $N=n$
  where the channel thickness spans the entire frequency range that
  has   significant  \HI emission.   We do  not find a
  significant change in the slope of the power spectrum   for any of
  the galaxies.    
Since for all the galaxies   the thickest channel  is considerably
larger  than the velocity dispersion, we conclude that    
 the \HI power  spectrum is  purely due to  density
fluctuations and not gas velocities. The fact that the slope does not
change with channel thickness can be used to  
constrain the value of the  slope of the
velocity structure function $\zeta_{2}$   (see \citealt{2000ApJ...537..720L}).  
The $\zeta_2$ values   are tabulated in  {\bf Table~\ref{tab:result}}.  

The galaxies in our sample have slope $\alpha$ ranging from $-2.6$ 
to $-1.1$.  The two galaxies DDO~210 and NGC~3741 have slope $\sim
-2.5$, while the slope is  $\sim -1.5$ for  UGC~4459 and
AND~IV,   and $-1.1$ for   GR~8.   We have proposed a possible
explanation  for this dichotomy in the  values of $\alpha$ in
{\bf Section~\ref{sec:moa}}.  This was based on the fact that DDO~210,
where the power spectrum was measured across length-scales $100-500$
pc, had a slope  of $-2.6$  while NGC~628, a nearly face-on galaxy 
where  the power spectrum was measured across length-scales $0.8-8$
kpc had a slope of $-1.6$. We have interpreted the former as three 
dimensional (3D) turbulence operational at small scales whereas the
latter was interpreted as two dimensional (2D) turbulence in the plane
of the galactic disk. In {\bf Chapter~\ref{chap:sim}} we have performed
numerical simulations which tend to support this interpretation.
Continuing with this  implies that we have also measured  
3D turbulence in NGC~3741,  and 2D turbulence in   UGC~4459,  GR~8 and
AND~IV. Our dwarf galaxy  sample contains   3 nearly face-on  
galaxies with $i_{\rm HI} < 30^{\circ}$, namely  
 DDO~210, UGC~4459 and GR~8. DDO~210 has 3D turbulence across the 
baselines $2.0-10.0 \ {\rm k \lambda}$. The absence of a break in the
power spectrum  places a lower limit of $160$   pc 
on the  scale-height.  The galaxies 
UGC~4459 and GR~8  exhibit 2D turbulence for the entire $U$ range 
({\bf Table~\ref{tab:result}}) in the  measured \HI power spectrum. This 
imposes  the upper limit of $51$ pc
and $30$ pc respectively on  the
scale-height.  

\begin{figure*}
\begin{center}
\epsfig{file=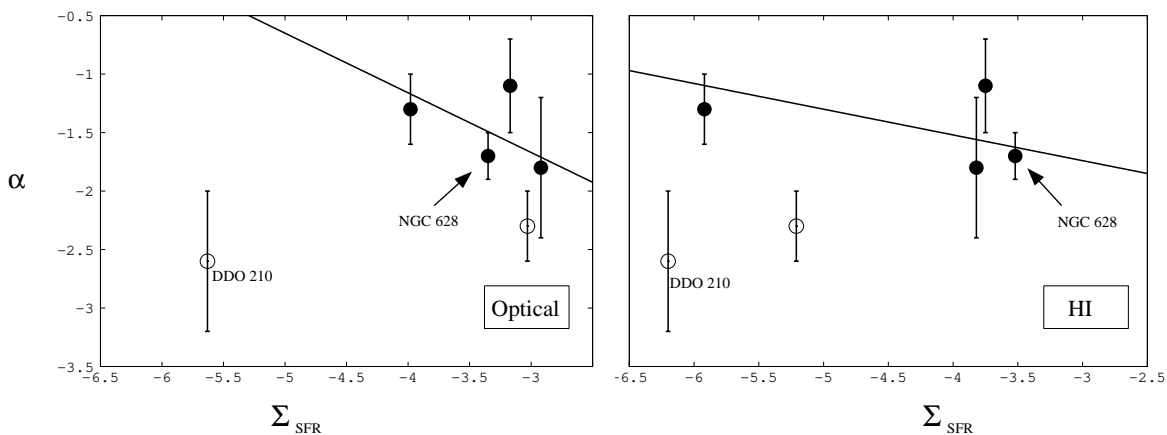, angle=0, width=6.0in}
\end{center}
\caption{The slope $\alpha$ of the \HI power spectrum plotted against
  $\Sigma_{{\rm SFR}}$, the   SFR per unit area. The area has been 
determined from optical images and \HI images in the left and right
panels respectively. The  galaxies with  3D and 2D
turbulence are shown using empty and filled circles respectively. 
Note that the SFR for DDO~210, marked in the figure, is only 
 an upper limit. We also include the data for  NGC~628, a
 spiral galaxy. It is 
 marked with an arrow.  The straight lines show  the linear
 correlation that we find (discussed in the text)   between the
 slope of the  \HI power spectrum and  the SFR per unit area.   
} 
\label{fig:SFR}
\end{figure*}

The energy input from star formation is believed to be a major driving
force for the turbulence in the ISM \citep{2004ARA&A..42..211E}. Also
first stage of 
the star formation process is known to be driven by the ISM turbulence.
Hence, it is very 
interesting to check whether the slope of the power spectrum of
intensity fluctuations in these galaxies has any correlation with the
SFR. Any correlation, if present, will provide
an insight into the 
relation between the star formation and ISM turbulence.
In a recent study \citet{2005AJ....129.2186W} calculates the SFR per
unit area for  9 irregular 
galaxies  and investigates the correlation with the 
V-band and  H$\alpha$ power spectra.   The length-scales they  probe
are  $10-400$ pc. In  H$\alpha$ they  find that  the power spectra
becomes steeper as the SFR per unit area  increases. However  they do
not find  any   correlation in the V-band.  In this paper we probe
 length-scales $100$ pc to $3.5$ kpc. 
We test for a   possible correlation between the slope 
of the power spectrum  and the  SFR per unit area and per unit \HI mass
for  the 5 dwarf 
galaxies in our sample ({\bf Figure~\ref{fig:SFR}}). We also include
the spiral galaxy NGC~628. 
 We report results using  the area estimated
from  both,  the \HI disk and the optical disk.  
 The data for the SFR, angular extent,  inclination in \HI and \HI mass   for
 these galaxies are from \citet{2005A&A...433L...1B}
 and  \citet{2008MNRAS.386.1667B}. The SFR is 
 determined from H$\alpha$ emission,  and the 
 angular extent in \HI is determined from  the column density maps at
 the value of  $10^{19}$ atoms~cm$^{-2}$.  For the angular extent and
 inclination  of the  optical disk   we use the   parameters from
 \citet{2006MNRAS.365.1220B}  and \citet{2008MNRAS.384.1544S}.
The  linear correlation coefficient   is found to have  values $0.35$
and  $0.34$ for the \HI and optical disks respectively, indicating the
absence of any correlation between the slope of the power spectrum and
the SFR per unit area. We have similar result for SFR per unit \HI mass
for these 5 galaxies. 

Turbulence can also possibly be related to  different parameters
of the  galaxy. We test  for correlations between the slope of the \HI
power spectrum and   the following parameters: total \HI
mass,  \HI mass to light ratio, total dynamical mass, total baryonic
mass,  gas fraction, baryon fraction. We use the estimates of these
observable  quantities from \citet{2008MNRAS.386.1667B}. In all the
cases the linear 
correlation coefficient is found to lie between $-0.5$  and $0.5$,
indicating the absence  of correlation.  

In the above analysis we  have considered galaxies with both 3D and 2D
turbulence 
taken together, which in turn can suppress the correlations.   
Hence, we further investigated the correlations by considering the 
 3 galaxies in our sample  with 2D
turbulence.   In
this case the correlation coefficient comes out to be $-0.70$ and
$-0.74$ for SFR per unit area of optical and \HI disks respectively. This
indicates a strong correlation and the result is similar to that of
\citet{2005AJ....129.2186W}, namely that the  surface density of star
formation rate 
is larger  for galaxies with steeper \HI power spectrum. These findings
indicate  a possible link between star formation and 
the nature of turbulence in the ISM. However, it would not be 
realistic  to speculate on a possible cause and effect relation
between these two. It is quite possible   that the observed correlation
is an outcome of  extraneous factors which influence both.

\section{Summary and conclusions}
In this chapter we have performed and presented the power spectrum
analysis of 7 faint dwarf galaxies. For 5 of these galaxies we can
conclusively estimate the power spectrum. Below we summarize the main
points of this chapter.

\begin{itemize}
\item ISM of the dwarf galaxies have a scale-invariant power spectrum
consistent with turbulence. Power law power spectrum of slope
$-1.1$ to $-2.6$ provides good model to these power spectrum.
\item We have observed dichotomy in the slope of the power spectrum which
we have  explained by 3D and 2D turbulence. Based on this we
could also provide limits to the scale-height of these
galaxies. 
\item We have investigated possible correlation between the power spectrum
slope and the dynamical parameters of the galaxies. Our null result
suggests that these parameters does not influence turbulence in the
dwarf galaxies.
\item We have found a correlation between the SFR per unit area (of the
optical disk and the \HI disk) with the power spectrum slope for the
galaxies with 2D turbulence in our sample. This is a
possible indication of the relation between the star formation and
turbulence in the ISM. However, the
H$\alpha$ emission is  not a good tracer of SFR in low mass galaxies
due to stochastic star formation   \citep{2007AJ....133.1883K,
2007ApJ...671L.113L}.  Also, the 
length-scales across which the  power spectrum  has been estimated for  
the  dwarf galaxies in our sample are substantially larger than the
optical disk where  star formation occurs.  These points must be
taken into consideration while interpreting our results.
\end{itemize}
Finally, we note that the total number of galaxies in  our
analysis is rather small  for a statistical conclusion. A larger
galaxy sample  is required for a better  understanding of the  generic
features,  if any,  of turbulence  in  the ISM of faint dwarf
galaxies. We considered the possibility of performing the power
spectrum analysis of a larger sample derived from the 
FIGGS \footnote{Faint Irregular Galaxies GMRT
Survey \citep{2008MNRAS.386.1667B}}. We tried to estimate the power
spectrum of the galaxies  DDO~43, DDO~125, DDO~189, UGC~7605 and
UGC~8508. We found that the FIGGS data are noise limited for these
galaxies which suggests the necessity of deeper observations.

%-------------------------------------------------------------------
%-------------------------------------------------------------------

%-------------------------------------------------------------------
%-------------------------------------------------------------------

%\newpage

%\clearpage{\pagestyle{empty}\cleardoublepage} %%%%%%%%%%%%%%%%%%%%
\chapter[Power Spectrum Analysis of the THINGS Galaxies]{Power
  Spectrum Analysis of the THINGS Galaxies} 
\label{chap:THINGS}

\section{Introduction}
In {\bf Chapter~\ref{chap:3spi}} we have performed power spectrum analysis
of three spiral galaxies and have found that the power spectrum shows
the existence of scale-invariant structures in the ISM at a wide range
of length-scales starting from $\sim 10$ kpc to lower. It will be
interesting to 
probe ISM of  a sample of spiral galaxies to investigate the nature of
turbulence therein and possible relation of turbulence with the other
dynamical properties of the galaxy. This requires systematic
observations of a hand full of nearby spiral galaxies with good spatial
resolution.  

THINGS, “The \HI Nearby Galaxy Survey”, is a  survey
using  \HI 21-cm emission of 34 nearby galaxies
obtained using the VLA
\citep{2008AJ....136.2563W}. The main  aim of this survey is to
investigate the nature of the ISM, galaxy morphology, star formation
and mass distribution etc. across 
the Hubble sequence. THINGS provides homogeneous, high
quality 21-cm data cubes, with  high spatial ($\sim 6′′$)  and
velocity ($\leq 5.2$ km s$^{{-1}}$) resolution,
for a representative sample  of  local spiral and dwarf galaxies. 
\citet{2008AJ....136.2648D} have presented high
resolution rotation curve for the galaxies in the THINGS sample. Star
formation rate and efficiency of the THIGNS galaxies are also very
extensively studied \citep{2008AJ....136.2846B, 2008AJ....136.2782L}.  
This is  an ideal data set for a comparative study of  turbulence
in the neutral ISM of  spiral galaxies. 
In this chapter we present the result of power spectrum analysis
of a  sample of $18$ spiral galaxies drawn from the
THINGS survey. 

\section{Data and analysis}
\label{sec:data}

\begin{table}
\centering
\begin{tabular}{lrrcrrrrr}
Galaxy & Major & Minor & D & $i_{HI}$ & SFR & $M_{HI}$ &  $M_{dy}$ &
$\sigma_{v}$ \\ 
  & $(')$ & $(')$ & (\small{Mpc}) & $(^{\circ})$ & (\scriptsize{M$_{\odot}$ yr$^{-1}$}) & (\scriptsize{$10^{8}$ M$_{\odot}$}) & (\scriptsize{$10^{11}$ M$_{\odot}$}) & (\scriptsize{km s$^{-1}$}) \\ 
\hline \hline \\ 
 NGC~628  & $22.0$ & $20.0$ & $ 7.3$ &   $ 15.0$ & $1.2$ & $38.8$ & $6.3$& $7.2$ \\
 NGC~925  & $16.0$ & $10.0$ & $ 9.2$ & $50.0$ & $1.1$ & $45.8$ & $
 1.7$ & $9.4$ \\
 NGC~2403 & $25.0$ & $22.0$ & $ 3.2$ & $55.0$ & $0.9$ & $25.8$ & $
 3.1$ & $9.6$ \\
 NGC~2841 & $22.0$ & $26.0$ & $14.1$ & $69.0$ & $0.2$ & $85.8$ &
 $31.2$ & $21.6$ \\
 NGC~2903 & $25.0$ & $15.0$ & $ 8.9$ & $66.0$ & $  -$ & $43.5$ & $
 4.3$ & $15.8$ \\
 NGC~3031 & $38.0$ & $24.0$ & $ 3.6$ & $59.0$ & $1.1$ & $36.4$ & $ 5.2$ & $12.9$ \\
 NGC~3184 & $17.0$ & $14.0$ & $11.1$ & $29.0$ & $1.4$ & $30.7$ &
 $6.4$& $9.1$ \\
 NGC~3198 & $22.0$ & $ 7.5$ & $13.8$ & $72.0$ & $0.9$ & $101.7$& $ 8.1$ & $11.6$ \\
 NGC~3521 & $22.0$ & $ 7.5$ & $10.7$ & $69.0$ & $8.4$ & $80.2$ & $12.2$ & $29.8$ \\
 NGC~3621 & $25.0$ & $15.0$ & $ 6.6$ & $62.0$ & $2.1$ & $70.7$ & $12.8$ & $22.6$ \\
 NGC~4736 & $18.0$ & $12.0$ & $ 4.7$ & $44.0$ & $0.4$ & $ 4.0$ & $ 1.2$ & $11.4$ \\
 NGC~5055 & $30.0$ & $25.0$ & $10.1$ & $51.0$ & $2.4$ & $91.0$ & $12.8$ & $12.0$ \\
 NGC~5194 & $16.0$ & $12.0$ & $ 8.0$ & $30.0$ & $6.1$ & $25.4$ & $2.7$ & $14.4$ \\
 NGC~5236 & $30.0$ & $24.0$ & $ 4.5$ & $31.0$ & $2.5$ & $17.0$ & $2.6$ & $10.9$ \\
 NGC~5457 & $30.0$ & $25.0$ & $ 7.4$ & $30.0$ & $2.5$ & $141.7$& $5.9$ & $15.6$ \\
 NGC~6946 & $35.0$ & $25.0$ & $ 5.9$ & $35.0$ & $4.8$ & $41.6$ & $ 7.3$ & $9.3$ \\
 NGC~7793 & $12.0$ & $ 9.0$ & $ 3.9$ & $43.0$ & $0.5$ & $ 8.9$ & $ 0.7$ & $9.9$ \\
 IC~2574  & $14.0$ & $ 8.0$ & $ 4.0$ & $51.0$ & $0.1$ & $14.8$ & $0.5$ & $9.9$ \\
\hline
\end{tabular}
\caption{Some parameters of the galaxies used for the power spectrum analysis.
 Columns 1-9 gives 1) Name of the galaxy, 2) and 3) Major and Minor
 axis at a column density of $10^{19}$ atoms cm$^{-3}$, 4) Distance to
 the galaxy, 5) average HI inclination angle,
 6) Star Formation Rate,  7) HI mass, 8)
 Dynamical mass,
 9) average velocity dispersion. evaluated from the Moment II
 map. These values are obtained from the following refferences:
\citet{2008AJ....136.2563W},  % Walter et al 2008.
\citet{2008AJ....136.2648D},  % de block et al 2008.
\citet{1973A&A....29..425B},  % NGC 5236
\citet{1979A&A....74..138H},  % NGC 5457
\citet{1985A&A...143..216H},  % NGC 3184
\citet{1992A&A...253..335K},  % NGC 628
\citet{1996ApJ...458..120S}.  % NGC 5194
}
\label{tab:sample}
\end{table}

Here we  restrict our analysis to the spiral galaxies
in the  THINGS sample with minor axis greater than $6'$.
{\bf Table~\ref{tab:sample}} gives different parameters of the galaxy
 sample analyzed here. The values for the parameters like the  distance
 to the 
galaxy, SFR and the total \HI mass is taken from 
\citet{2008AJ....136.2563W}, 
whereas values for the inclination angle are taken from
\citet{2008AJ....136.2648D}. 
\cite{2008AJ....136.2648D} also present rotation curves for the $13$ 
galaxies in our sample. We use these
rotation curves to estimate 
the dynamical mass for those galaxies. For the rest of the
galaxies the dynamical mass values are taken from
\citet{1973A&A....29..425B, 1979A&A....74..138H, 1985A&A...143..216H,
  1992A&A...253..335K, 1996ApJ...458..120S}  and then   
rescaled for the adopted distances noted in {\bf Table~\ref{tab:sample}}.
We have estimated the \HI velocity dispersion using a method described  in
{\bf Section~\ref{sec:results}}. Note that, although the galaxy NGC~4826 fits
our selection criteria, we have not used it in our analysis. It has a
very bright \HI core which makes the window function complicated  
and the power spectrum estimation is not straight forward. 
 
The observations and data reduction of the THINGS galaxies are
discussed  in 
\citet{2008AJ....136.2563W}\footnote{We are indebted to Fabian Walter 
for providing us with calibrated  \HI data from the THINGS survey.}.
For our analysis, we  start with the calibrated $uv$ data prior to
continuum subtraction.
In the standard THINGS pipeline, the continuum is subtracted by fitting a
linear polynomial to each visibility (i.e. AIPS task UVLIN), whereas 
for our analysis it is important that the strong continuum sources 
be modelled and then subtracted from the visibilities. For each galaxy,
the frequency channels with \HI emission are identified and a 
continuum image is made by combining  all the line-free channels. 
This continuum is then  subtracted from the data in the 
$uv$ plane using the AIPS  task UVSUB. The resulting continuum-subtracted
data are used  subsequently. We follow the pipeline
discussed in {\bf Section~\ref{sec:moa}} for the rest of the analysis. 

\section{Result and discussion}
\label{sec:results}

\begin{table*}
\centering
\begin{tabular}{lrrrrrrrr}
\hline
Galaxy & & $\Delta \ v$ & $U_{min}$   & $U_{max}$     & $R_{min}$ & $R_{max}$ & $\alpha$  & $\alpha$\\ 
       & & ($km\ s^{-1}$) & (k $\lambda$) &  (k $\lambda$) & (kpc)      & (kpc)
& ($n=N$) & ($n=1$)  \\ 
\hline \hline \\ 
 NGC~628 & & $41.6$ & $1.0$ & $10.0$ & $0.8$ & $7.5$ & $-1.6 \pm 0.1$  & $-1.6 \pm 0.1$\\ 
 NGC~925 & & $41.6$ & $1.0$ & $10.0$ & $0.9$ & $9.2$ & $-1.0 \pm 0.2$  & $-1.0 \pm 0.2$\\ 
 NGC~2403 & * & $83.2$ & $0.7$ & $7.0$ & $0.6$ & $4.0$ & $-1.1 \pm 0.1$  &  $-2.1 \pm 0.3$\\ 
 NGC~2841N & & $83.2$ & $1.0$ & $10.0$ & $1.4$ & $14.0$ & $-1.7 \pm
 0.2$  & $-1.7\pm 0.1$\\ 
 NGC~2841S & & $83.2$ & $1.0$ & $10.0$ & $1.4$ & $14.0$ & $-1.5 \pm
 0.2$  & $$ $-1.5\pm0.1$\\ 
 NGC~2903 & & $83.2$ & $0.8$ & $8.0$ & $1.1$ & $11.1$ & $-1.5 \pm 0.2$  & $-1.5\pm0.1$\\ 
 NGC~3031N & & $41.6$ & $2.0$ & $10.0$ & $0.4$ & $1.8$ & $-0.7 \pm 0.1$  & $-0.7\pm0.1$\\ 
 NGC~3184 & & $41.6$ & $0.7$ & $7.0$ & $1.6$ & $15.8$ & $-1.3 \pm 0.2$  & $-1.3\pm0.1$\\ 
 NGC~3198 & * & $83.2$ & $1.6$ & $10.0$ & $1.4$ & $8.6$ & $-0.4 \pm
 0.3$  & $-1.6\pm0.3$ \\ 
 NGC~3521N & & $83.2$ & $1.0$ & $17.0$ & $0.6$ & $10.7$ & $-1.8 \pm 0.1$  & $-1.8\pm0.1$\\ 
 NGC~3521S & & $83.2$ & $1.0$ & $17.0$ & $0.6$ & $10.7$ & $-1.6 \pm
 0.2$  & $-1.6\pm0.2$\\ 
 NGC~3621 & * & $83.2$ & $1.0$ & $12.0$ & $0.6$ & $6.6$ & $-0.8 \pm 0.2$  & $-1.9\pm0.2$\\ 
 NGC~4736 & * & $83.2$ & $0.6$ & $10.0$ & $0.5$ & $7.8$ & $-0.3 \pm 0.2$  & $-0.9\pm0.2$\\ 
 NGC~5055 & & $83.2$ & $1.0$ & $10.0$ & $1.0$ & $10.0$ & $-1.6 \pm 0.1$  & $-1.6\pm0.1$\\ 
 NGC~5194 & & $83.2$ & $1.0$ & $8.0$ & $1.0$ & $8.0$ & $-1.7 \pm 0.2$  & $-1.7\pm0.2$\\ 
 NGC~5236 & & $83.2$ & $0.6$ & $6.0$ & $0.8$ & $7.5$ & $-1.9 \pm 0.2$  & $-1.9\pm0.2$\\ 
 NGC~5457 & & $83.2$ & $0.6$ & $12.0$ & $0.6$ & $12.3$ & $-2.2 \pm
 0.1$  & $-2.2\pm0.1$\\ 
 NGC~6946 & & $20.8$ & $1.5$ & $10.0$ & $0.3$ & $4.0$ & $-1.6 \pm 0.1$  & $-1.6\pm0.1$\\ 
 NGC~7793 & & $41.6$ & $0.6$ & $6.0$ & $0.6$ & $6.5$ & $-1.7 \pm 0.2$  & $-1.7\pm0.1$\\ 
 IC~2574 & & $41.6$ & $1.8$ & $10.0$ & $0.4$ & $3.3$ & $-1.7 \pm 0.3$  & $-2.0\pm0.2$\\ 
\hline
\end{tabular}
\caption{Result of the power spectrum analysis with $N=n$ and
  $N=1$. Column 1 to 8 gives 1)   name of the galaxy, 2) width of the
  channel used to estimate the   power spectrum, 3) and 4) the range
  of $U$ value for which the    power spectrum is evaluated, 5) and 6)
  correspoindin length scales   and 7) the power law index $\alpha$
  with $1-\sigma$ error for $N=n$ and Column 8) gives the power law index $\alpha$
  with $1-\sigma$ error for $N=n$. We found the power law index
  $\alpha$ changes with channel thickness for the   galaxies marked by
  (*) in the table. The values for the $U_{max},\, U_{min},\, R_{max}$
  and $R_{min}$ are given for $N=n$ only}. 
\label{tab:result}
\end{table*}

\begin{figure*}
\begin{center}
\epsfig{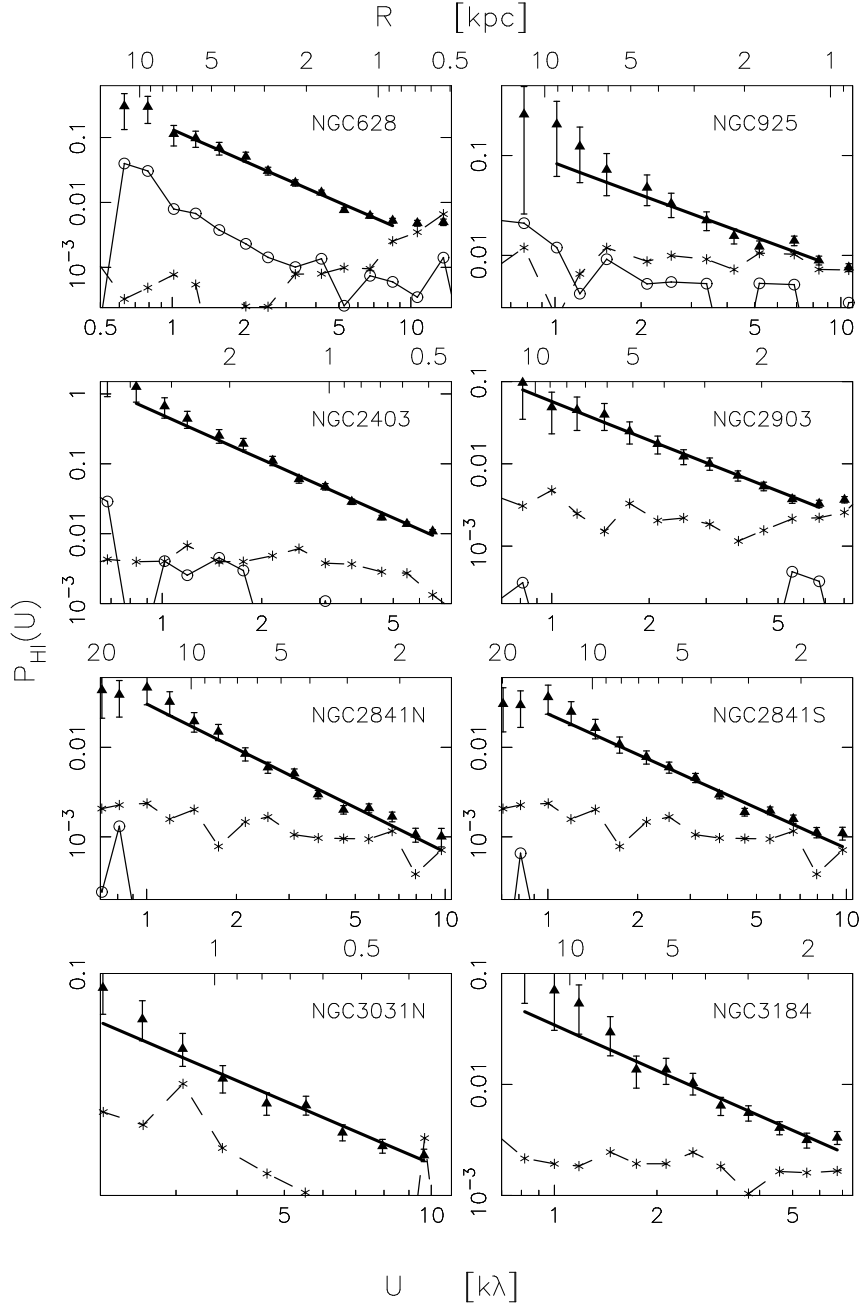}
\end{center}
\caption{The real (triangle) and imaginary (circle) part of the
  estimator evaluated using the channels with HI emission is plotted
  with the $1~\sigma$ error bars in a log log 
  scale against the baseline values for the galaxies NGC~628, NGC~925,
  NGC~2403, NGC~2903, NGC~2841, NGC~3031 and NGC~3184. The real
  part evaluated with the   line-free channels (star) is also
  shown. The best fit power law is given by the bold solid line. Note
  that in the $U$ range the power spectrum is deduced, the real part
  of the estimator has a higher value compared to the imaginary part
  or the real part of it from the line free channels.
} 
\label{fig:fig1}
\end{figure*}

\begin{figure*}
\begin{center}
\epsfig{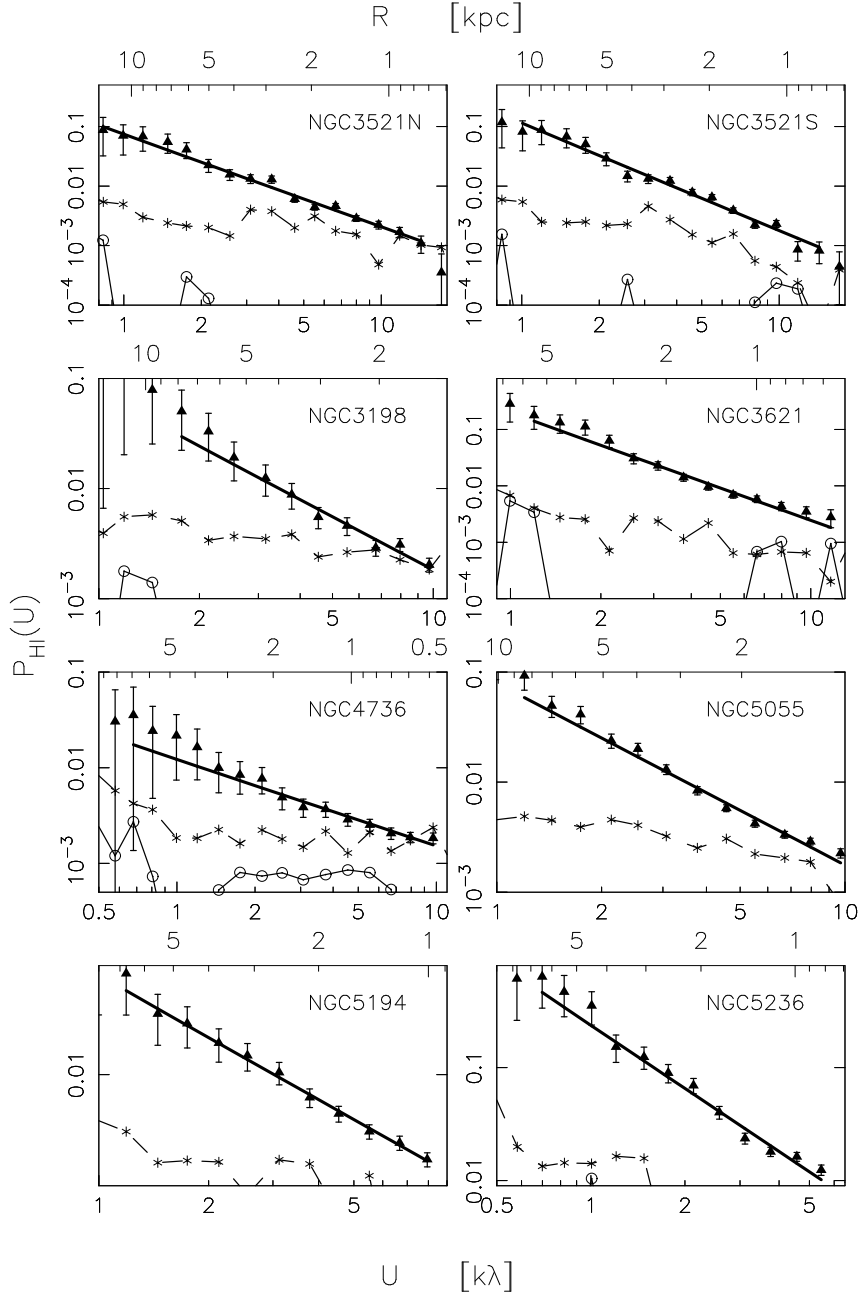}
\end{center}
\caption{The real (triangle) and imaginary (circle) part of the
  estimator evaluated using the channels with HI emission is plotted
  with the $1~\sigma$ error bars in  log-log 
  scale against the baseline values for the galaxies NGC~3521,
  NGC~3198, NGC~3621, NGC~4736, NGC~5055, NGC~5194 and NGC~5236. The real
  part evaluated with the   line-free channels (star) is also
  shown. The best fit power law is given by the bold solid line. Note
  that in the $U$ range the power spectrum is deduced, the real part
  of the estimator has a higher value compared to the imaginary part
  as well as  the real part of it from the line free channels.}
\label{fig:fig2}
\end{figure*}

\begin{figure*}
\begin{center}
\epsfig{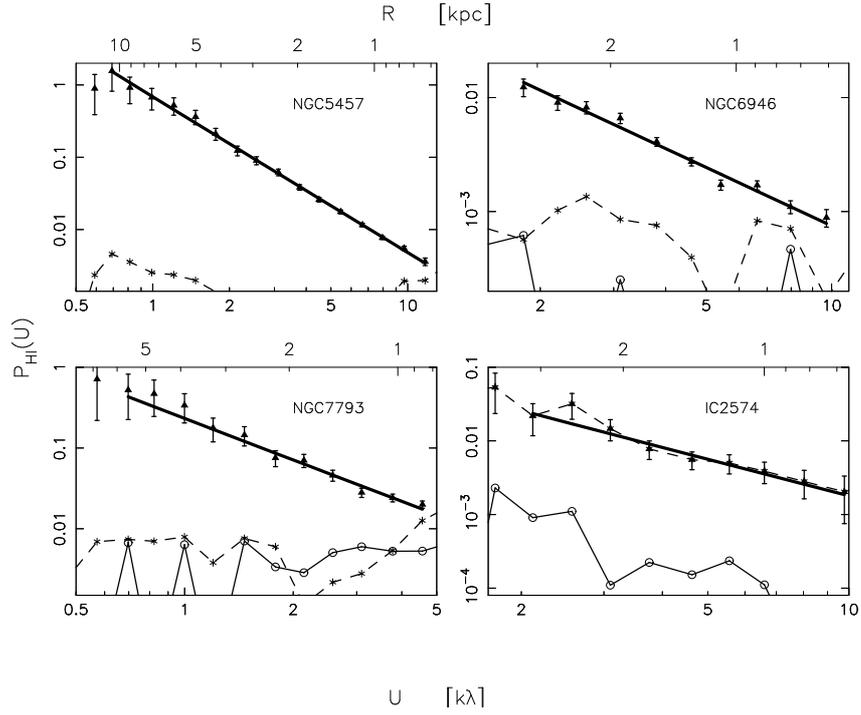}
\end{center}
\caption{The real (triangle) and imaginary (circle) part of the
  estimator evaluated using the channels with HI emission is plotted
  with the $1~\sigma$ error bars in log-log 
  scale against the baseline values for the galaxies NGC~5457,
  NGC~6946, NGC~7793 and IC~2574. The real
  part evaluated with the   line free channels (star) is also
  shown. The best fit power law is given by the bold solid line. Note
  that in the $U$ range the power spectrum is deduced, the real part
  of the estimator has a higher value compared to the imaginary part
  as well as  the real part of it from the line free channels.}
\label{fig:fig3}
\end{figure*}

For each galaxy, we
have identified a baseline range ($U_{min}$ to $U_{max}$) over which  the
real part of $\P(U)$ estimated from the channels with \HI emission is
large compared to both its imaginary part and the real part of $\P(U)$
estimated from the line-free
channels. A power law of the form $A~U^{\alpha}$ is found to provide a
good fit to the observed power 
spectra.  
{\bf Figure~\ref{fig:fig1}}, {\bf \ref{fig:fig2}} and {\bf
  \ref{fig:fig3}} shows 
the result of our power spectrum analysis. The results are also
summarized in {\bf Table~\ref{tab:result}}, where Column (2) to (7)
gives the value for the velocity channel thickness, range of baseline
for the fit, corresponding range of length-scales and the power law
index $\alpha$ respectively for $N=n$ (see {\bf
  Section~\ref{sec:moa}}). Column (8) gives the  power law index  
$\alpha$ for $N=1$. Unless otherwise stated we will use the values of
$\alpha$ with $N=n$ for further analysis. The presence of a
scale-invariant, power law power spectrum indicates that   
turbulence  is operational in the ISM of these galaxies.

For the galaxies NGC~2841,
NGC~3031 and NGC~3521, the angular extent,  
$\theta_{0}$, is comparable to the telescope field of view and hence 
two observations with two different pointing centers (North and South)
were used.  We refer these two different directions using N or S after
the galaxy's name. For  NGC~3031S, the power spectrum could be
estimated over a very limited baseline range and hence we do not
consider it here.

%\pagebreak
%\pagebreak
%\pagebreak

%\begin{figure}
%\begin{center}
%\epsfig{file=./chapter6/fig4.eps,width=2.6in, angle=-90}
%\end{center}
%\caption{The real (dot-dash) and imaginary (dash) part of the
%  estimator evaluated using the channels with HI emission is plotted
%  with the $1-\sigma$ error bars in log-log 
%  scale against the baseline values for the galaxy NGC~3031S. 
%The real  part evaluated with the   line free channels (solid) is also
%  shown. We could fit a power law (bold solid
%  line) to the power
%  spectrum over a baseline range of $0.6 - 2.5$ k $\lambda$. However,
%  since this is a very small range of $U$ value, we do not use this
%  result for our analysis. }
%\label{fig:3031S}
%\end{figure}

 The length-scales over which we could fit the power  law ({\bf
   Table~\ref{tab:result}}) covers in total a range from $300$ 
 pc to  $15.8$ kpc across the entire sample. The power spectrum of  the
 galaxy NGC~5457 is found to be a power law for  the length-scale
 range of $0.6$ kpc to $12.0$ kpc which  is the largest dynamical range
 obtained for an individual galaxy in our analysis. Note that,
 NGC~3184 exhibits a  power law power spectrum up to a length-scale 
 of $\sim16$ kpc, where the same for the galaxy NGC~2841 is $14$
 kpc. These could be understood as the existence of turbulence in the
 ISM of the spiral galaxies at a scale comparable to the radius of the
 galaxy's disk.

\begin{figure}
\begin{center}
\epsfig{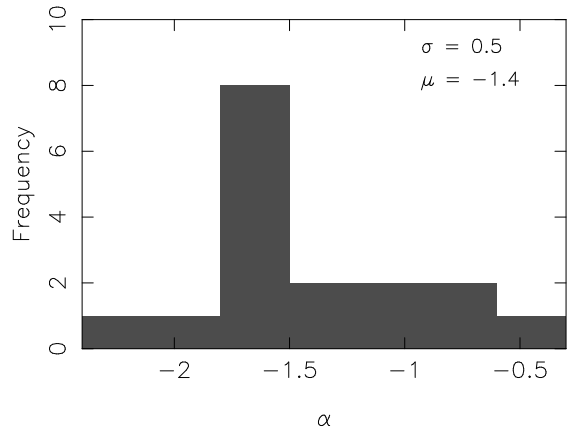}
\end{center}
\caption{Shaded regions shows the histogram of the distribution of the
power law index $\alpha$. Note that the histogram has 7 bins for 18
data points. Sample mean $\mu$ and sample standard deviation $\sigma$
is also shown. } 
\label{fig:hist}
\end{figure}

{\bf Figure~\ref{fig:hist}} shows a histogram of the distribution of
the estimated power law index $\alpha$. The histogram is sharply
peaked in the range $\alpha = -1.8$ to $-1.5$. The mean and the
standard deviations are $-1.4$ and $0.5$ respectively. We
do not find a break in the power spectrum of any of the galaxies in
our sample. For all the galaxies in our sample, the largest length-scales over
which the power law is estimated  are larger than the typical
scale-height of galaxies. Following  the results of our simulation
discussed 
in the {\bf  Chapter~\ref{chap:sim}}, we conclude that we are dealing
with 2D turbulence  for all these galaxies. 

\begin{figure}
\begin{center}
\epsfig{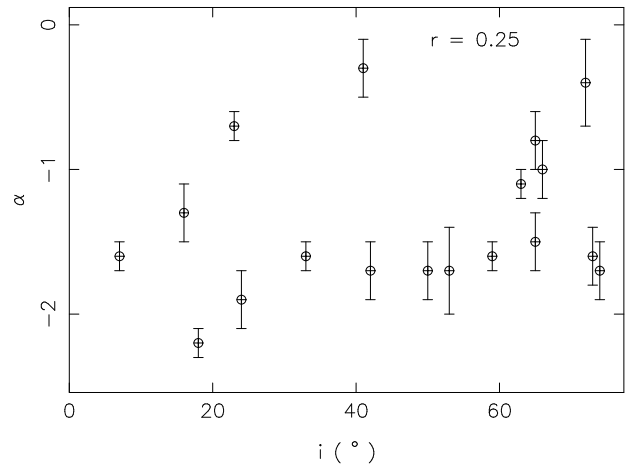}
\end{center}
\caption{Scatter plot of the average inclination angle $i$ with power law index
  $\alpha$. The $1~\sigma$ error bars of $\alpha$ are also
  shown. The value of the linear correlation coefficient $r$ is
  given at the top right corner.} 
\label{fig:inc}
\end{figure}

The galaxies in our sample have a wide range of inclination angles
`$i$', ranging from $7^{\circ}$ for NGC~628 to $74^{\circ}$ for NGC~2841.
To investigate whether the spread in the value of $\alpha$ seen here
is an effect of the difference in $i$, we evaluate the linear 
correlation between the power law  index $\alpha$ and the inclination 
angle of the galaxy ({\bf Figure~\ref{fig:inc}}). The resultant linear
correlation  
coefficient of $0.25$ is a clear indication that the dispersion in the
values of $\alpha$ is not  due to the inclination angles. 

\begin{figure*}
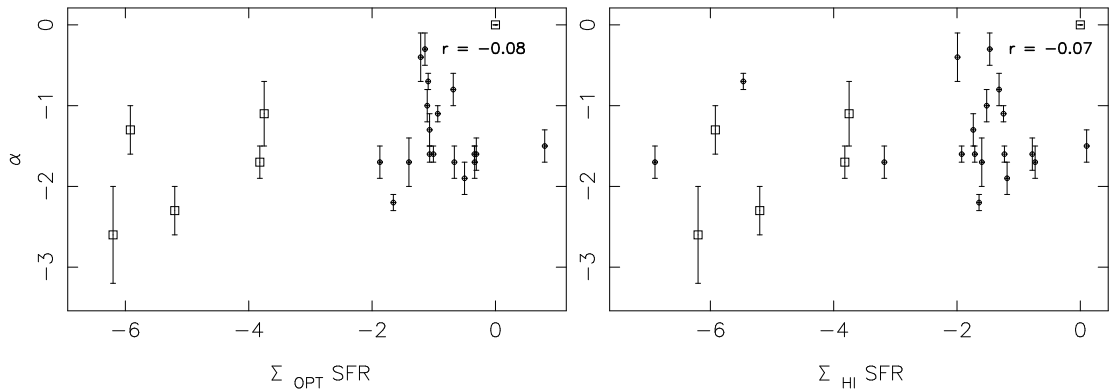

\begin{center}
\epsfig{file=./chapter6/SFRO.eps,width=2.in, angle=-90}
\epsfig{file=./chapter6/SFRR.eps,width=2.in, angle=-90}
\end{center}
\caption{Scatter plot of the surface density of the star formation
  rate  ({\bf Table~\ref{tab:sample}})
  with power law index 
  $\alpha$. The $1~\sigma$ error bars of $\alpha$ are also
  shown. The left and right pannels shows the SFR per unit area of the optical
  and HI disk of the galaxy respectively.  The value of the Linear Correlation Coefficient $r$ is
  given at the top right corner of the each panel.}
\label{fig:sfr}
\end{figure*}

At small scales, a major driving force of the  ISM turbulence is the
energy input from star formation processes. For a sample of irregular
galaxies \citet{2005AJ....129.2186W} finds that on scales of $10-400$ pc
the power law index of the H$\alpha$ power spectra becomes steeper as
the SFR per unit area increases. In  {\bf
  Chapter~\ref{chap:dwarf}} we have reported a weak correlation
between the SFR and the power law index 
of the \HI intensity fluctuations in dwarf galaxies. We have calculated the 
linear correlation coefficient of $\alpha$ with the  SFR per unit area of the 
optical disk as well as the SFR per unit area of the \HI disk. The
result, shown in {\bf Figure~\ref{fig:sfr}}, suggests that 
on large-scales at least, there is no correlation between the
SFR and the power law spectral index.

\begin{figure*}
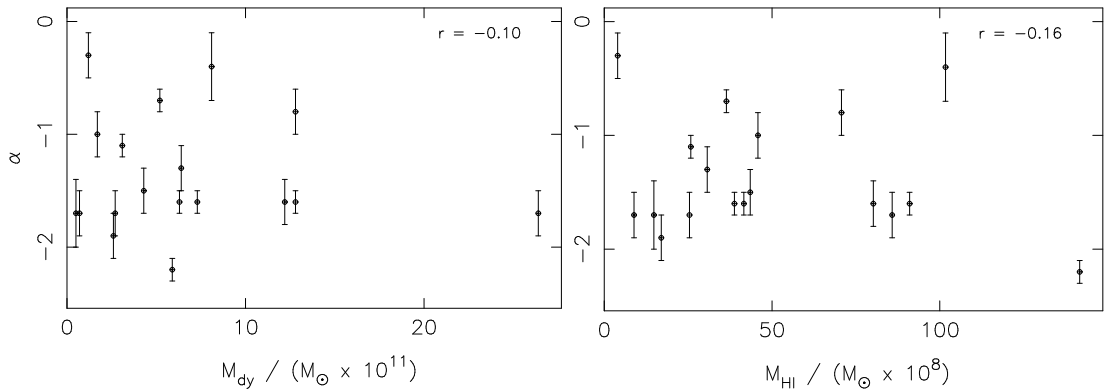

\begin{center}
\epsfig{file=./chapter6/MDY.eps, width=2.in, angle=-90}
\epsfig{file=./chapter6/MHI.eps, width=2.in, angle=-90}
\end{center}
\caption{Scatter plot of the dynamical mass $M_{dy}$ and the total HI
  mass $M_{HI}$ with power law index 
  $\alpha$. The $1~\sigma$ error bars of $\alpha$ are also
  shown. The value of the Linear correlation coefficient $r$ is
  given at the top right corner.} 
\label{fig:ms}
\end{figure*}

We next investigate the linear correlation of the power law index
$\alpha$ with the large scale dynamical parameters of the galaxies
namely the \HI mass ($M_{HI}$) and the dynamical mass ($M_{dy}$). We
use the values of the  maximum rotational velocity ($V_{max}$) and the
maximum
radius ($r_{max}$) from  the rotation curves presented in 
\citet{2008AJ....136.2648D} to evaluate the dynamical mass enclosed 
within that radius, viz. ($M_{dy} = V_{max}^2 r_{max} /G$). Note 
that \citet{2008AJ....136.2648D} have given rotation curves 
for only $13$ galaxies in our sample. For the other $5$ galaxies,
namely NGC~68, NGC~3184, NGC~5194, NGC~5236 and NGC~5457, we have taken the
$M_{dy}$ estimates from earlier references, viz.
\citep{1973A&A....29..425B, 1979A&A....74..138H, 1985A&A...143..216H,
  1992A&A...253..335K,1996ApJ...458..120S}.   The $M_{HI}$
values are taken from \citet{2008AJ....136.2563W}. {\bf Figure~\ref{fig:ms}}
shows the scatter  plot of $\alpha$ with $M_{dy}$ (left) and $M_{HI}$
(right) respectively. Our analysis shows that $\alpha$ is not
correlated with the dynamical  mass or the total \HI mass.

\begin{figure}
\begin{center}
\epsfig{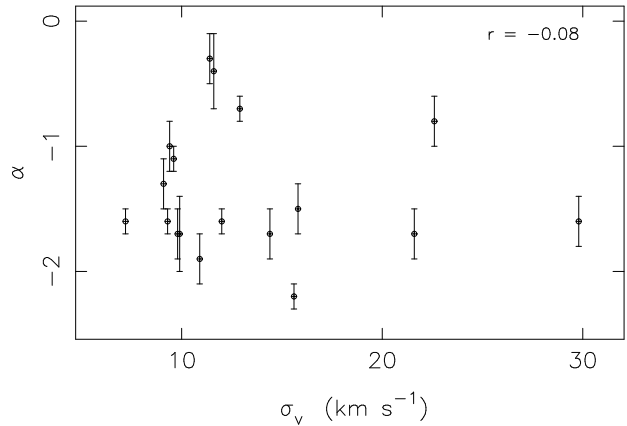}
\end{center}
\caption{Scatter plot of the average velocity dispersion $\sigma_{v}$
  with power law index 
  $\alpha$. The $1~\sigma$ error bars of $\alpha$ are also
  shown. The value of the Linear correlation coefficient $r$ is
  given at the top right corner.} 
\label{fig:disp}
\end{figure}

ISM turbulence can also be driven by the kinetic energy of the \HI gas. 
In this case, one would expect that the \HI velocity dispersion
($\sigma_{v}$) would be correlated with
$\alpha$. We use {\bf Eqn.~(7)} in \citet{2008AJ....136.2563W} to define the
average \HI velocity dispersion. In estimating $\sigma_{v}$, we perform a 8 pixel
smoothing on the THINGS archival Robust Weight velocity dispersion
(MOMENT 2) maps \footnote{http://www.mpia-hd.mpg.de/THINGS/Data.html}
and then  resample it to a resolution of $6''$. The  overall velocity
dispersion of the galaxies in the 
sample is evaluated using the values in individual pixels of the
resampled map ({\bf Table~\ref{tab:sample}}, Column (9). {\bf
  Figure~\ref{fig:disp}} 
shows the  scatter plot of the $\sigma_{v}$ values with $\alpha$. We find
no correlation between $\alpha$ and  $\sigma_{v}$.

In summary, the slope $\alpha$ is not correlated with any of the
galaxy parameters that we have considered. While $\sim 40 \%$ of the
$\alpha$ values are in the narrow range $-1.8$ to $-1.5$, the
remaining galaxies have $\alpha$ values spread over the broad range
$-2.2$ to $-0.3$. It is unclear if this arises from statistical
fluctuations in the distribution of $\alpha$, or if there is a
physical mechanism underlying this spread.

\begin{figure*}
\begin{center}
\epsfig{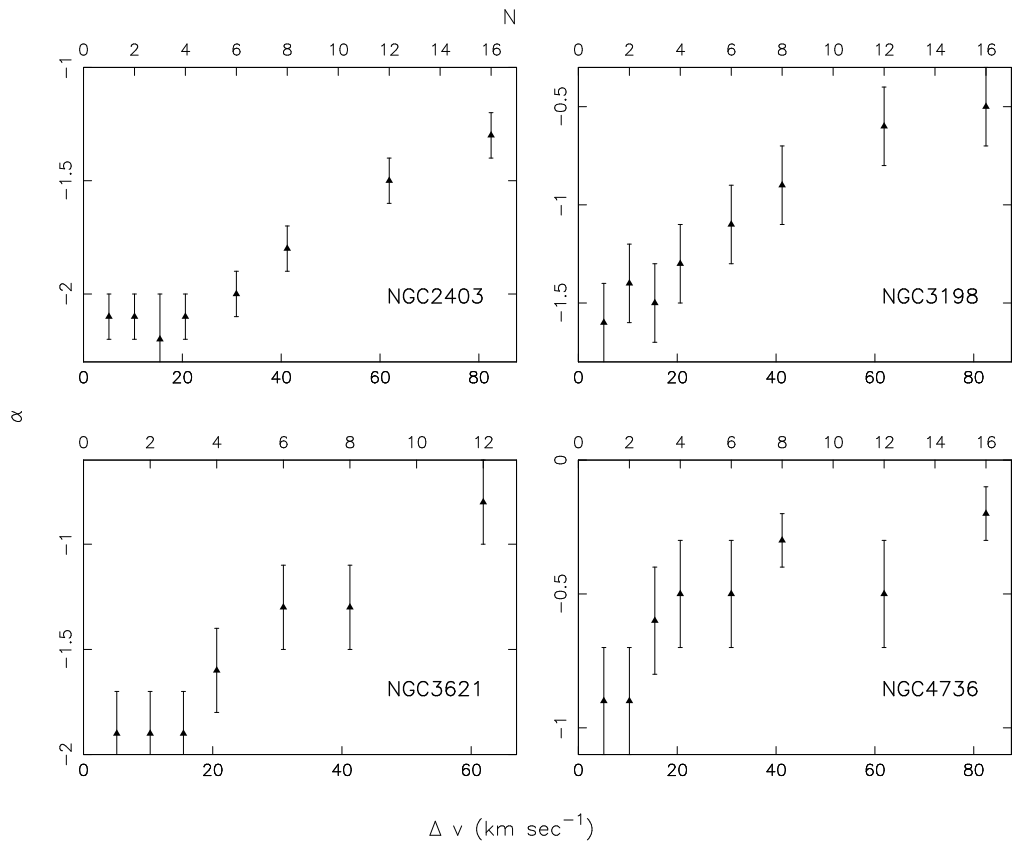}
\end{center}
\caption{Change in the slope of the power law index $\alpha$ with
  velocity channel thickness is shown for the four galaxies NGC~2403,
  NGC~3198, NGC~3621 and NGC~4736. Levels at the bottom shows the
  width of the velocity channels whereas the top levels show the number
  of channels averaged together before evaluating the power
  spectrum.} 
\label{fig:fig4}
\end{figure*}

Both \HI density fluctuations as well as spatial fluctuations in the
velocity of the \HI gas could contribute to observed fluctuations in
the  \HI specific intensity. \citet{2000ApJ...537..720L} have shown
that  it is possible to disentangle these two contributions  by studying
the behavior of the \HI power spectrum as the thickness of the
frequency  channel is varied. If the channel thickness  $\Delta v$ is
greater than the velocity dispersion for a galaxy, the velocity fluctuations 
are averaged out and the intensity fluctuation power spectrum becomes a  
direct probe of the density fluctuation alone. We investigate the 
possible effect of the velocity fluctuation in the observed intensity 
fluctuation power spectrum by estimating $P_{HI}(U)$  increasing  the channel
thickness ($\Delta v$) beyond the velocity dispersion $\sigma_{v}$ of the
galaxy. We find the slope $\alpha$ of the power spectrum of the galaxies
 NGC~2403, NGC~3198,NGC~3621 and NGC~4736 increase as the channel 
thickness is varied beyond a value $\Delta v = v_{c}$. These galaxies
are marked by an Astrix in the {\bf Table~\ref{tab:result}}. We show the
variation of $\alpha$ with the increasing channel thickness for these
galaxies in  {\bf Figure~\ref{fig:fig4}}. It is interesting to note that
for all four galaxies $v_c \sim 20$km s$^{{-1}}$ and the change
in slope is independent of the velocity dispersion $\sigma_{v}$ of the
galaxy in question. We interpret the value 
of $\alpha$ for the largest $\Delta v$ to be the power law index of the 
density fluctuation power spectrum
({\bf Figure~\ref{fig:fig4}}). The slope of the
power law is found to remain constant as the thickness of the channels
is varied for the rest of the galaxies in the sample. Since for all
the galaxies   the thickest  channel  is considerably larger  than the
velocity dispersion, we conclude that     for these galaxies, the \HI 
power  spectrum is purely due to  density fluctuations and not the velocity
fluctuations in the \HI gas. 

\section{Summary and conclusion}
\label{sec:conc}
This paper summarizes the power spectrum analysis of 18 spiral
galaxies from the THINGS sample. The estimated 
power spectra can be well fitted with a  power law, indicating the presence of
turbulence in the ISM. This analysis is the first comprehensive study
of turbulence in spiral galaxies  with a moderate sample size.
In this section we give a brief summary of the
important findings from our power spectrum analysis.

\begin{itemize}
\item We find that the galaxies in our sample shows a power
  law power spectrum over length-scales ranging from $\sim \ 300$ pc to
  $\sim \ 16$ k pc with several galaxies having $R_{max} > 10$
  kpc. Presently there is no known mechanism which can drive
  turbulence at such large length-scales. 

\item We have observed that  for most of the galaxies in our sample,
  $\alpha \sim -1.6 $ ({\bf 
    Figure~\ref{fig:hist}}). This value of $\alpha$ is similar to that
  is found for  2D power spectrum of the  galaxies in {\bf
    Chapter~\ref{chap:3spi}}.  

\item In {\bf Chapter~\ref{chap:dwarf}} we have found no significant correlation
  between $\alpha$ and the SFR, dynamical mass, total \HI mass of 
  dwarf galaxies.  Here we estimate the power spectrum
  of 18 external spiral 
  galaxies which provide us a moderate sample size. However, our 
  analysis also shows that the $\alpha$ has  no 
  correlation with the inclination angle, SFR, dynamical and total \HI
  mass of these galaxies. 

\item For most of the galaxies in our sample the
  width of the frequency channel is found to have no effect on the power law
  index. The slope of the power spectrum remains the same as the
  channel-thickness becomes greater than the average velocity
  dispersion 
  of the galaxies. This suggests that the intensity fluctuation power
  spectrum is a direct probe of the column-density fluctuations in the
  galaxy's disk. Interestingly, we notice that for four galaxies NGC~2403,
  NGC~3198,   NGC~3621 and NGC~4736
  the slope of the power spectrum decreases as the channel width
  $\Delta v$ is  increased, and beyond a certain value of $\Delta v
  \sim 20$ km s$^{{-1}}$
  it remains   constant. However,   the channel width where this
  transition occurs has no   correlation   with the velocity
  dispersion of the galaxy considered.  We found this to be different
  from \citet{2000ApJ...537..720L} and presently do not have any physical
  understanding of the reason behind this.
\end{itemize}

%\clearpage{\pagestyle{empty}\cleardoublepage} %%%%%%%%%%%%%%%%%%%%
\chapter[Summary and Future Scopes]{Summary and Future Scopes}
\label{chap:summary}

This thesis presents our effort of probing turbulence in the ISM of
the external galaxies by estimating the power spectrum using
radio-interferometric observations of \HI 21-cm radiation. Here we
summarize our main results and discuss the future scopes.

\begin{itemize}
\item We have estimated the power spectrum of 25 external galaxies
  including 5 nearby dwarf and 20 spirals. For every galaxy, the power
  spectrum is found to follow a power law over a wide range of
  length-scales which spans $60$ pc to $16$ kpc in total. This
  indicates that 
  the ISM is turbulent at length-scale comparable to the radius of the
  galaxy's disk. 

\item We have observed a dichotomy in the estimated value of the
  power spectral index and have explained it in  terms of 2D and 3D
  turbulence at  large and small scales in the galaxy's disk
  respectively. Based on a break in the power spectrum of the galaxy
  NGC~1058 we have estimated its scale-height as $450\pm60$ pc. This is
  the first determination of the scale-height of any external spiral
  galaxy. 

\item We investigated whether the turbulence is related to the
  different dynamical parameters like dynamical and \HI mass, star
  formation rate etc. of the galaxies. For the dwarf galaxies we found
  a weak correlation between surface density of star formation rate
  and the power law index. There was no such correlation observed for
  the spiral galaxies. 

\item The observed power spectrum is found to have no effect of the
  velocity fluctuations for 21 of the 25 galaxies we have
  analyzed. For the rest of the galaxies we have observed power law
  index changes with the channel thickness indicating an effect of the
  velocity fluctuation.
 
\item Our power spectrum analysis of the harassed galaxy NGC~4254 has
  shown that the galaxy harassment effects the ISM structures. The
  central part of such a galaxy exhibit different dynamics compared to
  that in the outer disk.

\item Given a power law power spectrum, its  estimator introduced in
  this thesis is sensitive to the slope but not to the amplitude of
  the power law. On the other hand, for smaller length scales in our
  galaxy,  the amplitude and slope of the power spectrum is very well
  constrained. If one can carry out observations with more sensitivity
  with better uv-coverage at the larger baseline range, it would be
  possible to probe similar length scales in the external nearby
  galaxies with the existing radio interferometers like VLA or
  GMRT. This will provide a way of calibrating the amplitude of the
  estimator we have used and then a direct comparison between the
  large and small scale powers can be done. Also note that we have
  been restricted to the nearby galaxies because of the limited
  angular resolution and hence linear resolution in the disk of the
  galaxies that we can achieve with the present telescopes. For an
  even better uv-coverage and more dynamic range in the length scale
  within which the power spectrum can be estimated, upcoming radio
  telescopes like SKA will come be very useful.    
\end{itemize}

The known energy sources of ISM turbulence include galaxy rotation,
self gravity, supernovae etc., which are believed to drive turbulence
at parsec scales. It is not clear if these are equally effective in
explaining the large scale turbulence observations reported in this
thesis. A detailed physical modeling is required to have better
understanding of ISM dynamics. It may be possible to estimate the ISM
velocity fluctuation power spectrum, which  can independently test the
existence and nature of turbulence in the ISM. We wish to perform
numerical simulation to understand the effect of velocity fluctuations
etc. in our power spectrum estimation. Finally, it will also be
interesting to study the ISM turbulence in a sample of galaxies
residing in the cluster environment.

%\clearpage{\pagestyle{empty}\cleardoublepage} %%%%%%%%%%%%%%%%%%%%
%\input{chapter8/chap8.tex}
%\input{tempbib.tex}

%\clearpage{\pagestyle{empty}\cleardoublepage} 
\newpage
\addcontentsline{toc}{chapter}{References}
\singlespacing
\bibliographystyle{ifacconf}         %{klunamed}       %{acm}
\bibliography{references}
\clearpage
\onehalfspacing
%\clearpage{\pagestyle{empty}\cleardoublepage} %%%%%%%%%%%%%%%%%%%%
\clearpage
\chaptermark{Appendix}
\addcontentsline{toc}{chapter}{Appendix } 
\appendix
\thispagestyle{empty}
%\setcounter{section}{0}
%\setcounter{subsection}{0}
%\setcounter{subsubsection}{0}
%\setcounter{equation}{0}
%\pagenumbering{arabic}
\hspace{-.8cm}{\huge{\bf Appendix}}

\section{Incomplete $uv$ coverage and correlated noise in the
  radio-interferometric observations} 
\label{app:cornois}
Noise in the visibility plane $\N(\U)$ corresponds to $I^{(N)}(\vt)$
given by
\begin{equation}
I^{(N)}(\vt) \ =\ \int d \U \ e^{i 2 \pi \U . \vt} \N(\U)
\label{eq:nois1}
\end{equation}
Fourier transform is a linear operation and in practice a discrete
Fourier transform is used. We can write {\bf Eqn.~(\ref{eq:nois1})} as 
\begin{equation}
I^{(N)}_{j} \ =\ \sum_{k}\mathrm {F}_{jk} N_{k},
\label{eq:nois2}
\end{equation}
where $\mathrm {F}_{jk}$ are the components of the transformation
matrix $\mathrm {F}_{jk} = \exp(\frac{i 2 \pi jk}{N})$. In a typical
radio-interferometric observation  the $uv$ plane is not
completely sampled. The
actual inverse-transformation from the measured visibility to the
specific intensity is given by
\begin{equation}
I^{(N)}_{j} \ =\ \mathrm {F}_{jk}\Gamma_{kl} N_{l}\, ,
\label{eq:nois3}
\end{equation}
where $\Gamma_{kl} N_{l}$ gives the measured visibility points. We
call this matrix $\Gamma$ as the projection matrix. It is to note
here, that $\Gamma$ is an Identity matrix if all the points in the
$uv$ plane are sampled, otherwise it is a diagonal matrix with
diagonal elements as $0$ or $1$. In the latter case the
inverse of this matrix does not exist. Components of the noise
covariance matrix (assumed to be diagonal) can be written as
\begin{equation}
\mathrm{N}_{ij} \ =\ \delta_{ij} \sigma^{2}_{i}
\label{eq:noiscorr}
\end{equation}
We calculate the auto-correlation function as
\begin{eqnarray}
\xi^{(N)}_{ij} &=& \langle I^{(N)}_{i}\, I^{(N)^{\dagger}}_{j}\rangle \nonumber \\
&=& \mathrm {F}_{ik}\, \Gamma_{kp}\, \sigma^{2}_{p}\, \Gamma_{pn}^{\dagger}\, 
  \mathrm {F}_{nj}^{\dagger}.
\label{eqn:noisxi}
\end{eqnarray}
 In case of complete $uv$ coverage,  we have 
\begin{eqnarray}
\xi^{(N)}_{ij} &=& \mathrm
   {F}_{ik}\  \delta_{kp}\  \sigma^{2}_{p}\  \delta_{pn}\  
   \mathrm {F}_{nj}^{\dagger} \nonumber \\
&=& \sigma^{2}_{i}\ \delta_{ij} ,
\label{eqn:noisxi1}
\end{eqnarray}
since $\mathbf{F}$ is unitary. However for real observations with
limited $uv$ coverage  $\Gamma_{kp}$  is not a identity matrix. The noise
power spectrum  is given by,
\begin{eqnarray}
P^{(N)}_{mj} &=& \mathrm {F}_{mi}^{\dagger}\ \xi_{ij} \nonumber \\
&=& \Gamma_{mp}\, \sigma^{2}_{p}\  \Gamma_{pn}^{\dagger}\ \mathrm
{F}_{nj}^{\dagger}   
\label{eqn:pownois}
\end{eqnarray}
This clearly shows that for limited baseline coverage, there exists
correlated noise in the image plane. Every attempt to estimate power
spectrum from the Fourier transform estimators hence has to take care of
this correlated noise effect.

\newpage

\def\un{(\U, \nu)}
\def\tn{(\vt, \nu)}
\section{Validity of the 2D approximation.}
\label{app:wterm}
In {\bf Chapter~\ref{chap:est}} we have assumed the baseline $\U$ as a 2D
vector. This is rather an approximation and in reality the vector $\U$
has three components. Here we discuss the validity of the 2D
assumption.\footnote{The Appendix is adopted from the originally
published in the paper titled ``The 
Effect of  $w$-term on Visibility Correlation and Power Spectrum
Estimation" by \citet{2010MNRAS.406L..30D}. Note that we have used
different symbols here compared to what is defined in rest of the
thesis for reasons given in the text.} 

\subsection{2D and 3D visibility, the effect of   \lowercase{$w$-term}
on the aperture} 

The direct observable  in the radio-interferometric observations is the
complex visibility $\V^{3D} \un$. For a pair of antennae  separated by
$\d$, with each antenna pointing along the direction of the unit
vector $\k$  (referred to as the phase center) we have 
\begin{equation}
{\V}^{3D} \un = e^{2 \pi i w} \ \int d \Omega_{\n}\,   e^{2 \pi i \,  \U \cdot
(\n -\k)}  A(\n - \k,\nu)   I(\n - \k,\nu) 
\label{eq:v3d} 
\end{equation}
where $\n$ denotes the unit vector to different directions of the sky,
baseline  $\U = \d /\lambda $,  $A(\n - \k,\nu)$  denotes the
primary beam and $I(\n - \k,\nu)$ is  the specific intensity.
Writing $\U = \U_{\perp} + w\k$, where $\U_{\perp}$ is a 2D vector, 
and defining $ \n - \k = \vt $,  we have,  for
$ \mid \vt\mid \ll   1$, \  $\vt \cdot \k \approx 0 $, implying that 
$\vt$ is a $2D$  vector. In this limit $\vt$ gives
the position of any point  on the sky with
respect to the phase centre in a 2D tangent plane. This is known as
the flat-sky approximation. The term $ w\k $
quantifies deviation from this.
\begin{figure}
\begin{center}
\epsfig{file=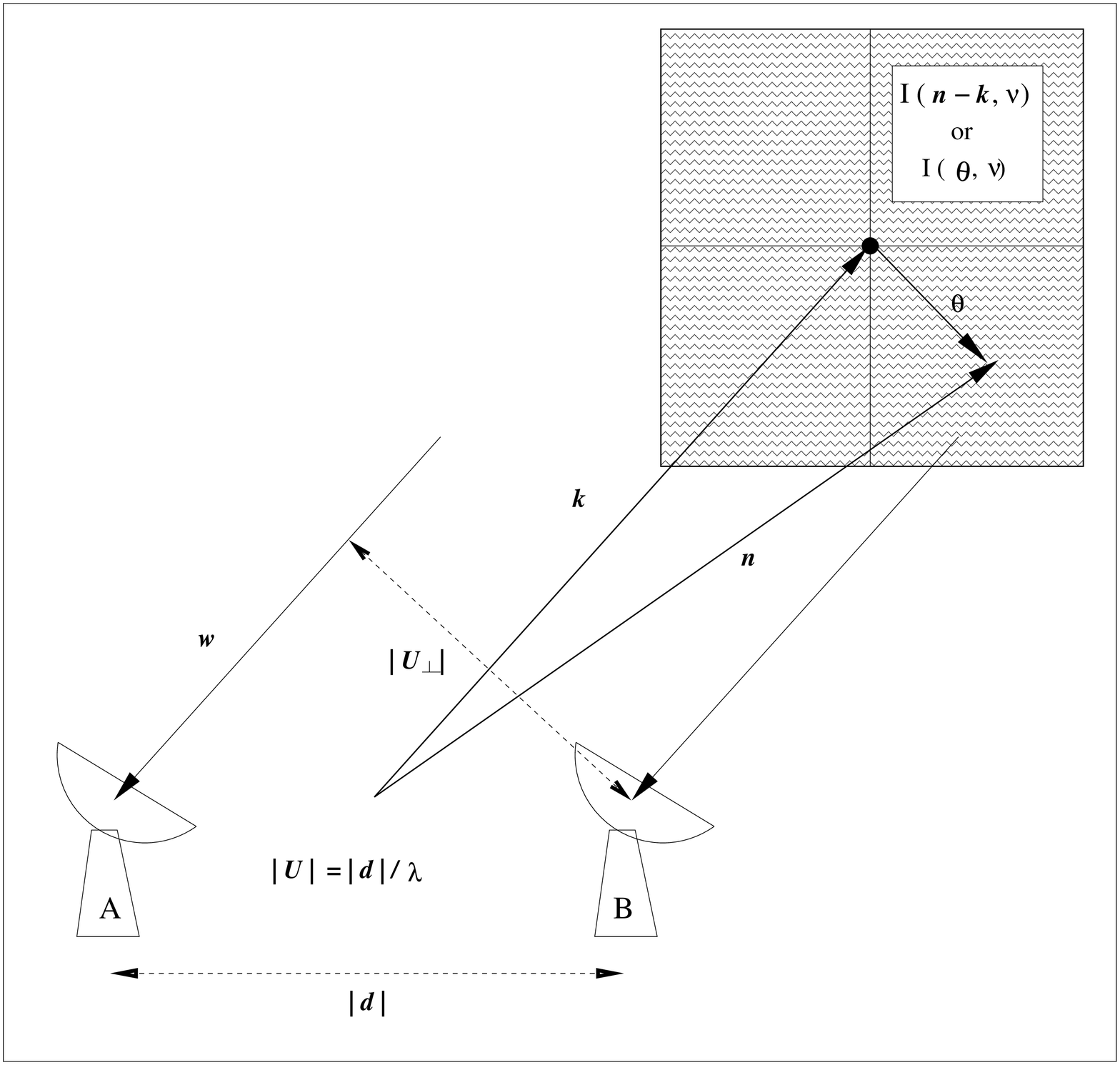,width=4.6in, angle=0}
\end{center}
\caption{Schematic representation of a particular visibility
measurement with a pair of antenna A and B separated by a distance
$\mid \d \mid$. The square box at the top right corner represents the
portion of sky observed or the field of view (FOV). The black dot at the
centre of the FOV corresponds to the direction to the phase centre
$\k$. Intensity fluctuations at a given direction $\n$ to the sky is
denoted as $I(\n - \k,\nu)$ or $I(\vt,\nu)$. Note that $\vt$ is a 2D
vector in the plane of the sky.} 
\label{fig:baseline}
\end{figure}
In the  $2D $ approximation we have
\begin{equation}
{\V} ^{2D}(\U_{\perp},\nu) = \int d \vt   \ e^{ 2 \pi i  \U_{\perp} \cdot
\vt} \ A(\vt,\nu) \  I(\vt,\nu)
\end{equation}
Writing  the specific intensity $I\tn$ as $I\tn = \bar{I}_{\nu} +
\delta I\tn $,  where the first term is a  constant background and the
second term is a fluctuation,  we have
\begin{equation}
{\V}^{2D}(\U_{\perp},\nu) =  \bar{I}_{\nu} \tilde{A}(\U_{\perp},\nu) + \tilde{A} (\U_{\perp},\nu)\otimes \tilde{\delta I}(\U_{\perp},\nu)
\end{equation}
where tilde represents a Fourier transform and $\otimes$ denotes a
convolution. 

The aperture function $\tilde{A}(\U_{\perp},\nu)$, peaks at
$\U_{\perp} = 0$ and has a finite width. Hence,
we shall retain the second term in all subsequent discussions. 

We note that, ignoring the
$w$-term leads to a simplification of the expression for visibility
and in the 2D approximation, $ \V^{2D}(\U_{\perp},\nu)$ is the
Fourier transform of $\ A\tn  \delta I\tn$.
Hence, we have,
\begin{equation}
A\tn \  \delta I\tn = \int  d \U_{\perp}'\  \V^{2D}(\U_{\perp}',\nu)
\ e^{-2 \pi i \, \U_{\perp}' \cdot \vt}   
\end{equation}
Substituting in {\bf Eqn. (\ref{eq:v3d})} we obtain
\begin{equation}
\V^{3D}\un = \int   d \U_{\perp}'  \ K(\U, \U_{\perp}', \nu) \  \V^{2D}(\U_{\perp}',\nu)
\end{equation}
Where the kernel $ K(\U, \U_{\perp}', \nu)$ is given as
\begin{eqnarray}
K(\U, \U_{\perp}', \nu)&=& e^{- 2 \pi i w} \ \int d \Omega_{\n}\ e^{-2 \pi i \ 
  (\U_{\perp}' - \U) \cdot (\n -\k)} \nonumber \\
&=& 4 \pi j_{0} \left ( 2 \pi \mid \U_{\perp}' - \U \mid \right ),
\end{eqnarray}
with $j_{0}$ denoting the Spherical Bessel function.

Defining a  quantity $\tilde{\mathcal A}\un $ as
\begin{equation}
\tilde{{\mathcal A}} \un = 4 \pi \int d \U_{\perp}'
\   j_{0} \left ( 2\pi \mid \U_{\perp}' - \U \mid \right) 
\tilde{A}(\U_{\perp}',\nu). 
\label{eq:modap}
\end{equation}
The scalar 3D visibility takes the form
\begin{equation}
\V^{3D}\un = \int d \U_{\perp}' \ \tilde{{\mathcal A}} (\U -
\U_{\perp}', \nu)  \ \tilde{\delta I}(\U_{\perp}')
\label{eq:V23D}
\end{equation}
It is to be noted that the `$w$' dependence of $\V^{3D}\un$ is
translated to the function $\tilde{{\mathcal A}} (\U -
\U_{\perp}, \nu)$. This can be regarded as a modified aperture
function. Note that this function is a real valued function, unlike
the complex Gaussian as discussed in \citet{2002MNRAS.334..569H}.  
 We investigate the nature of the modified aperture $\tilde{{\mathcal
     A}}\un$ as a function of $U_{\perp}$ for different values of $w$. 
We have assumed the primary aperture $ \tilde{A}( \U_{\perp}, \nu) $   to be a
Gaussian, $\exp \left [ -\frac{U_{\perp}^{2}}{2 U_{0}^{2}} \right ]$,
of width $ U _{0}$, and evaluated the  integral in
 {\bf Eqn.~(\ref{eq:modap})} numerically. We shall not explicitly
 write the $\nu$ 
 dependence of $\mathcal{A}$ and $\tilde{\mathcal{A}}$ henceforth. It
 can be shown that the modified aperture preserves the  circular
 symmetry of the primary aperture. The
 implicit $\nu$ dependence is present in these functions through
 $U_{0}$.  For an antenna of diameter $D$, 
 $U_{0}$ can be approximately written as $U_{0} \sim D /\lambda$ for
 observing wavelength $\lambda$. We use $U_{0} = D /\lambda$ with
 $D=45$ m   (specifications of the GMRT) for the subsequent
 discussion. This corresponds to a FOV of $4^{\circ}$ at
 $150$ MHz. We shall discuss the effect of larger FOV later.
\begin{figure*}
\begin{center}
\mbox{\epsfig{file=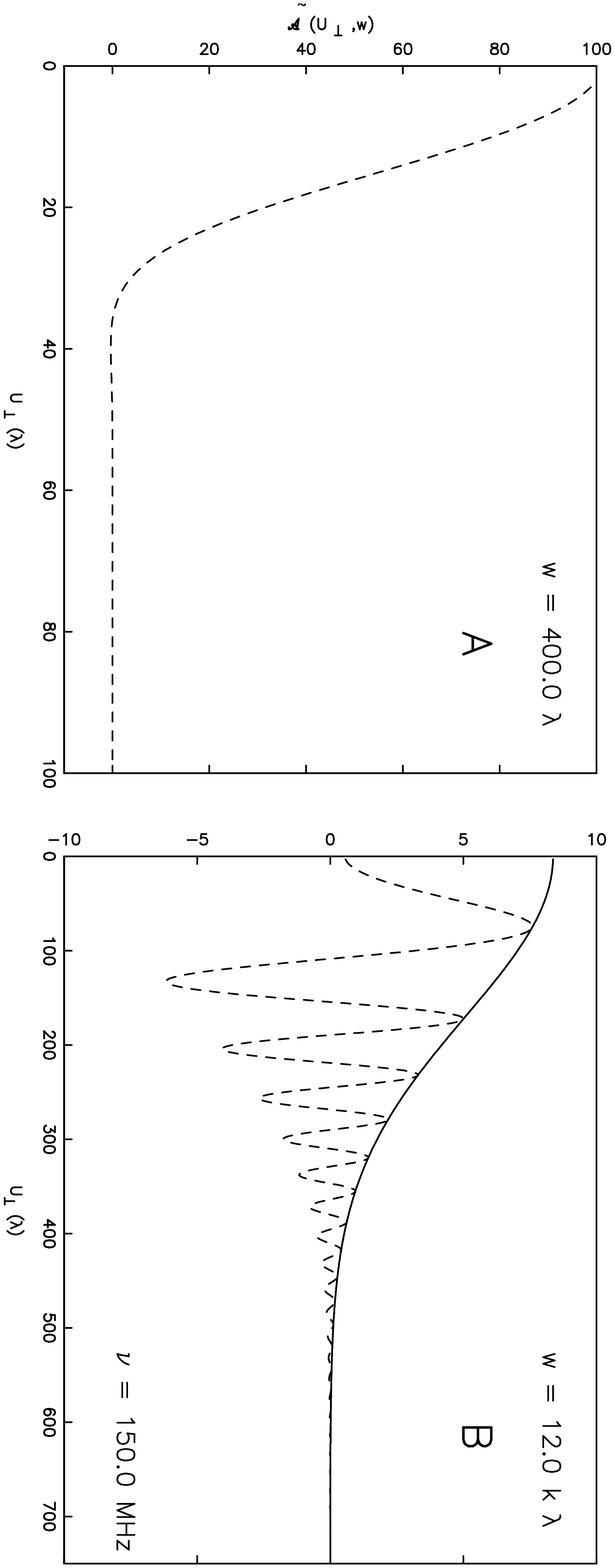,width=2.2in,angle=90}}
\end{center}
\caption{Modified aperture $\tilde {\mathcal A}(\U_{\perp},w )$
 plotted as a function of $\U_{\perp}$ for two different values of
  $w$, (A) $w=400. \lambda$ and (B) $w = 12.$ k$\lambda$ at $\nu
= 150$ MHz. Solid line in B shows the Gaussian envelope.}
\label{fig:1502p}
\end{figure*}

{\bf Figure~\ref{fig:1502p}} shows the numerically evaluated modified 
aperture $ \tilde{{\mathcal A}} (\U_{\perp}, w) $ as a function of $ 
U_{\perp}$ for two representative values of $ w $ (A: $ w = 
400\ \lambda$, B: $ w = 12\ {\rm k}\lambda$) at frequency $ \nu = 150$
MHz. 
\begin{figure}
\begin{center}
\mbox{\epsfig{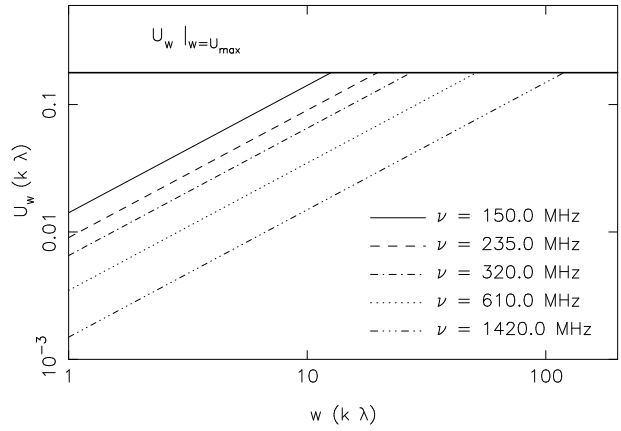}}
\end{center}
\caption{$U_{w}$ is plotted as a function of $w$ for different central
  frequencies of GMRT. Horizontal solid line corresponds to the
  maximum possible baseline.
}
\label{fig:Uw}
\end{figure}

\begin{figure}
\begin{center}
\mbox{\epsfig{file=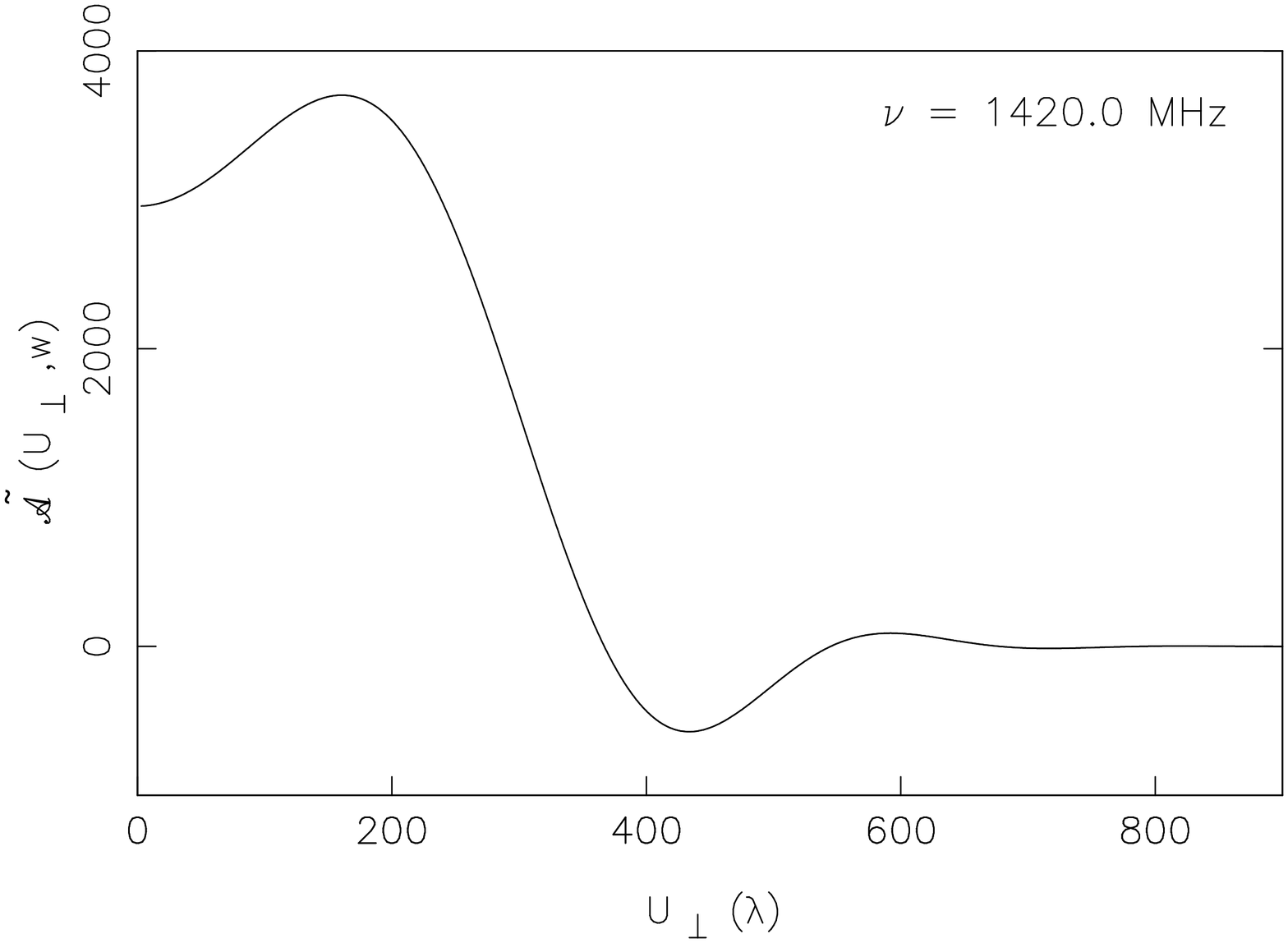,width=2.2in,angle=90}}
\end{center}
\caption{Modified aperture $\tilde {\mathcal A}(\U_{\perp},w )$
 plotted as a function of $\U_{\perp}$ for  $w=120.$ k $\lambda$ at
 $\nu=1420.$ MHz.
}
\label{fig:14201p}
\end{figure}
For $ w = 0 $ one reproduces the primary Gaussian aperture
trivially. We also note that for $ w \ll U_{\perp}  $ the Gaussian
profile is still maintained ({\bf Figure~\ref{fig:1502p}:A}). However for
large values of $w$ the 
aperture function manifests oscillations ({\bf Figure~\ref{fig:1502p}:B}).
The  period of these oscillations is found to be 
sensitive to $w$ (decreasing as $w$  increases).
The envelope of the modified aperture is maximum at $U_{\perp} = 0$
and falls of gradually as $U_{\perp}$ increases. We found that  a
Gaussian,   $ C  \exp \left [ - \frac{U^2}{2 U_w^2} \right ] $,
provides a good fit to the envelope.
Ignoring the effect of oscillations in $\tilde{{\mathcal A}}$, we note
that the 3D formalism can be recast in the same form as its 2D   
counterpart with $U_{w}$ taking the role of $U_{0}$. 
\citet{2002MNRAS.334..569H} have obtained a similar result 
 assuming small FOV, where, they have shown that the
 effect of $w$-distortion can be considered as turning the primary
 beam into a complex Gaussian.  

We next investigate the dependence of $U_{w}$ on $w$ and frequency. 
For a given frequency at large $w$, $U_{w}$ is found to increase
linearly with $w$, i.e, $U_{w} \sim m(\nu) w$.  This implies
that the $w$-term effectively broadens the aperture of the
instrument. {\bf Figure~\ref{fig:Uw}} shows the variation of $U_{w}$ with
$w$ for different frequencies in log-log scale.  We note that, the
slope $m(\nu)$ (as represented by the y-intercept in the
{\bf Figure~\ref{fig:Uw}}) determines the  effect of $w$-term for increasing 
values of $w$. A small value of $m(\nu)$ implies  a slow increase of
$U_{w}$ with $w$ and the effect of the $w$-term is
less. $m(\nu)$ is found to fall off as $\sim 1/\nu$ with
frequency. This indicates that the departure from the 
flat-sky approximation is more pronounced at the lower
frequencies. Redshifted 21-cm line observed at frequency $\nu$ probes
the redshift $z =\left [ \frac{1420 ({\rm 
  MHz})}{\nu} -1 \right]$. Hence, one may expect the $w$-term to have
a greater  effect while probing higher red-shifts. 

In this thesis we use 21-cm line  to study the ISM dynamics of the
nearby galaxies $(z \sim 0)$. {\bf Figure~\ref{fig:14201p}} shows the
modified aperture 
function  for frequency $1420$ MHz and $w = 120\ {\rm k}\lambda$
(this being the maximum $U$ for GMRT like arrays). At this frequency,
for GMRT, $U_{0} = 0.1  {\rm k}\lambda$, whereas $U_{w}\mid _{w =
  U_{max}} = 0.18  {\rm 
  k}\lambda$. It follows that the flat-sky approximation can be
safely used if the largest length scale probed, corresponds to a
$U_{\perp} \gg U_{w}\mid _{w =U_{max}}$. 

Till now, we have investigated the effect of $w$-term using $D=45$ m,
which corresponds to an FOV of $4 ^{\circ}$ at 150 MHz. We
estimated $U_{w}\mid _{w =U_{max}}$ assuming $D = 4$ m to $D=45$ m at
$\nu = 150$ MHz. At $D=4$ m,  (which corresponds to the largest
proposed FOV, $45^{\circ}$, of SKA), $U_{w}\mid _{w =U_{max}}
= 4$ k$\lambda$ for $U_{max} = 120$ k$\lambda$.  For all the cases
mentioned above, a Gaussian of the form $C  \exp \left [ -
  \frac{U^2}{2 U_w^2} \right ]$ provides a good fit to the 
envelope of the modified aperture function and we could set a value
for $U_{w}$. Further, we also observe that for these cases $U_{w}\mid
_{w =U_{max}} \simeq  \frac{d_{max}}{\pi D}$, where $d_{max}$ is the
largest baseline of the telescope array. This implies that  $U_{w}\mid
_{w =U_{max}}$ depends only on the geometry of the array
configuration.

\subsection {Visibility correlation and power spectrum estimation}
For 3D visibilities, power spectrum $P(U_{\perp})$ can be estimated by
performing the visibility correlation as
\begin{equation}
V_2^{3D}(\U) = \langle V^{3D}(\U)V^{3D}(\U)^{*} \rangle = \int  d^2 \U_{\perp}^{'} \ {|\tilde{\mathcal{A}} (\U_{} - \U_{\perp}^{'})|}^2  \ 
 P(U_{\perp}^{'})
\label{eq:est}
\end{equation}
Noting that the effect of the $w$-term is contained in the modified
aperture $\tilde {\mathcal A}$. We can retrieve the 2D estimator $V_2^{2D}$
defined in {\bf Section~\ref{sec:visicorr}} (consider same baseline
correlation) by replacing  $ \tilde{\mathcal A}$ with 
$\tilde{A}$. Hence, 
\begin{equation}
V_2^{2D}(\U_{\perp})= \int  d^2 \U_{\perp}^{'} \ {|\tilde{A} (\U_{\perp} -
    \U_{\perp}^{'})|}^2  \  
 P(U_{\perp}^{'}).
\end{equation}
We note that $V_2^{3D}(\U)$ and $V_2^{2D}(\U_{\perp})$ become
$V_2^{3D}(U)$ and $V_2^{2D}(U_{\perp})$ for circularly symmetric
primary aperture. We shall now discuss the effect of the  $w$-term on
the estimator defined in {\bf Eqn.~(\ref{eq:est})}.

\begin{figure}
\begin{center}
\mbox{\epsfig{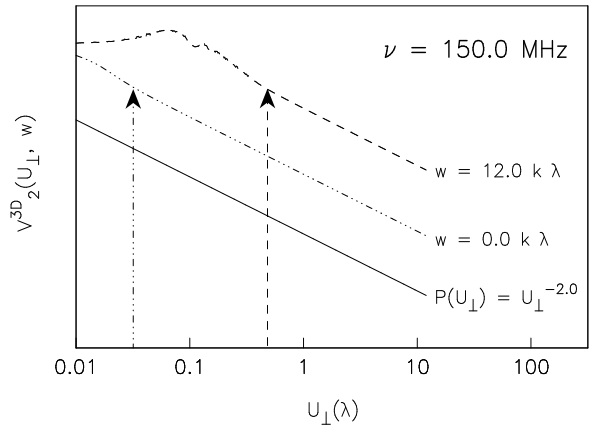}}
\end{center}
\caption{$V^{3D}_2$  as a function of $U_{\perp}$ for  $w=0$
  (dot-dash) and $w=U_{max}$ (dash) at $\nu = 150$ MHz, assuming $P(U_{\perp})
={U_{\perp}}^{-2}$. We also plot $P(U_{\perp})={U_{\perp}}^{-2}$
(solid line) for reference. The vertical arrows show the $U_{\perp}$
value above which the power law is recovered at $1 \%$. Note
that the plots are given arbitrary offset for clarity. 
}
\label{fig:pspec}
\end{figure}

{\bf Figure~\ref{fig:pspec}} shows $V^{3D}_2$  plotted
as a function of $U_{\perp}$ for two values of $w$, ($ w = 0 $ and $ w
= U_{max}$) at $ \nu = 150$ MHz, assuming $ P(U_{\perp})
={U_{\perp}}^{-2}  $. We have chosen $ U_{max}= 12 k\lambda$ (this
being  the largest baseline for the GMRT at $150 $ MHz). We  have
shown the  power 
spectrum $  P(U_{\perp})={U_{\perp}}^{-2}$ for comparison.
For  large values of  $w$, $V^{3D}_2$ show oscillations for $
U_{\perp} < U_w$,  which arises due to  the oscillatory nature of 
$ |\tilde{\mathcal{A}}(U)|^2$.

We find that  $ V^{3D}_{2}$
faithfully recovers the power law  $ {U_{\perp}}^{-2}$ (at $1\% $) for
$U_{\perp}$ greater than a certain value. This  value  is found to be
$3 \times(\sqrt{2}U_{0})$ for $ w = 0 $
and $ 3 \times(\sqrt{2}U_{w}) $ at $ w = 12\ {\rm k} \lambda$. Hence, a
non-zero $w$-term  changes the $U_{\perp}$
value beyond which the power spectrum estimation would be valid. 
\begin{table}
\centering
\begin{tabular}{c|cc}
\hline 
& $w = 0$ & $w \sim U_{max}$ \\
\hline
\hline
Aperture & $\tilde {A}(U_{\perp})$ & $\tilde{\mathcal A} (U_{\perp}, w)$ \\
width & $U_{0}$ & $U_{w}$ \\
\hline
Visibility & & \\
correlation & $V_{2}^{2D}(U_{\perp})$ & $V_{2}^{3D}(U_{\perp}, w)$ \\ 
\hline 
\end{tabular}
\caption{The effect of $w$-term, comparison between various quantities.} 
\label{table:t1}
\end{table}

\begin{table}
\centering
\begin{tabular}{c|ccc}
\hline 
& & $150$ (M Hz) & $1420$ (M Hz)\\
\hline
\hline
$U_{0}$ & & 0.01 &  0.1 \\
$U_{max}$& & 12.0 & 120.0 \\
$U_{w}\mid _{w = U_{max}}$& & 0.18 & 0.18 \\
\hline 
\end{tabular}
\caption{Relevant $U_{\perp}$ (k $\lambda$) values at different frequencies.} 
\label{table:t2}
\end{table}

The quantity of interest in power spectrum estimation using the
radio-interferometric observations used in this thesis is actually the quantity
\begin{equation}
{\mathcal E}(U_{\perp}) = \int_{0}^{U_{max}} dw \ \rho(w)\ V^{3D}_{2}(U_{\perp}, w),
\end{equation}
where, $\rho(w)$ is a normalized probability distribution of
$w$. The function $\rho(w)$ is specific to an observation as well
as to the array configuration of the interferometer. Hence, it is
difficult to make a general quantitative statement regarding the
effect of $w$-term in $\mathcal{E}$.
%We note that $\rho(w)\ V^{3D}_{2}(U_{\perp}, w) \sim 0$
%at large $w$ (since ${\mathcal A} \sim 0$ beyond a certain
%$U_{\perp}$).
 Since, for a given $w$, the largest baseline above which
$V^{3D}_{2}(U_{\perp}, w) \sim P(U_{\perp})$ is $U_{w}$,
 we can qualitatively state that ${\mathcal E}$ gives a good estimation
of the power spectrum for $U \ge 3\ U_{w} \mid_{ w  = U_{max}}$. It is
important to note that, for a specific array configuration, $U_{w}
\mid _{w = U_{max}}$ is independent of the frequency $\nu$, whereas
$U_{max} \propto
\nu$ ({\bf Table~\ref{table:t1}}). Hence, the $U_{\perp}$ range
amenable for power spectrum 
estimation is larger for large observing frequencies and  the
use of the 2D visibility for the analysis presented in this thesis is
justified. 

\newpage

\section{Error in the  visibility correlation estimator}
\label{app:visig}
The  real and imaginary parts  of the estimator $\P(U)$ have
uncertainties  arising from
 (1.) the sample variance and (2.) the system
noise. In this section we provide a method of estimating the error in
the visibility correlation estimator.

\subsection{The noise covariance matrix}
In radio astronomy the total power available at a radio telescope
terminal   is conveniently stated in terms of the system
temperature $T^{sys} = (\mathrm{Total\, power})/k_{B}$, where $k_{B}$ is
the Boltzmann's constant. The system temperature obtained for the antenna 
looking at a blank sky gives a measure of the total random noise in the
system. This noise is assumed to be Gaussian random with zero
mean and  its variance for the $j^{th}$ antenna can be given by
$\sigma^{2}_{j} = T^{sys}_{j}$.

 It can be shown that for most practical purposes, a pair of baselines
 $i$ and $j$ observing a  point source with flux 
density $S$, with an integration time of $T$, at a
bandwidth of $\Delta \nu$ the noise covariance matrix is  \citep{2007LFRA...NT} 
\begin{equation}
\sigma_{ij}^{2} = \frac{T^{sys}_{i}\ T^{sys}_{j}}{2\, T\, \Delta \nu}.
\end{equation}
It is important to remember that usually the interferometers do not
record the self correlations and hence all $i=j$ are not considered
from the above expressions. Now if the antenna pairs $i-j$ corresponds
to a baseline of $\U$, then variance of both the 
 real and imaginary part of  $\N(\U, \nu)$ is  $\sigma_{ij}$. For
a typical observation, the noise is also 
 uncorrelated and constant across the frequency channels.

An important property of the noise is  that at two baselines $\U_{A}$
and $\U_{B}$ the noise is uncorrelated, i.e, 
\begin{equation}
\langle \N(\U_{A}) \N(\U_{B}) \rangle = \delta_{A B}\, \Nt(\U_{A})
\label{eq:noiscorr}
\end{equation} 
where $\Nt(\U_{A})$ is noise covariance matrix for a given
baseline $\U_{A}$. Assuming Poisson system noise we can write,
\begin{equation}
\Nt(\U) = \frac{\left ( T^{sys}\right )^{4}}{4 T^{2}
  (\Delta \nu)^{2}} = N_{2},
\end{equation}
since $\Nt(\U)$ is independent of $\U$.

\subsection{Variance in the real and imaginary part of $\P(U)$}
\begin{figure}
\begin{center}
\epsfig{file=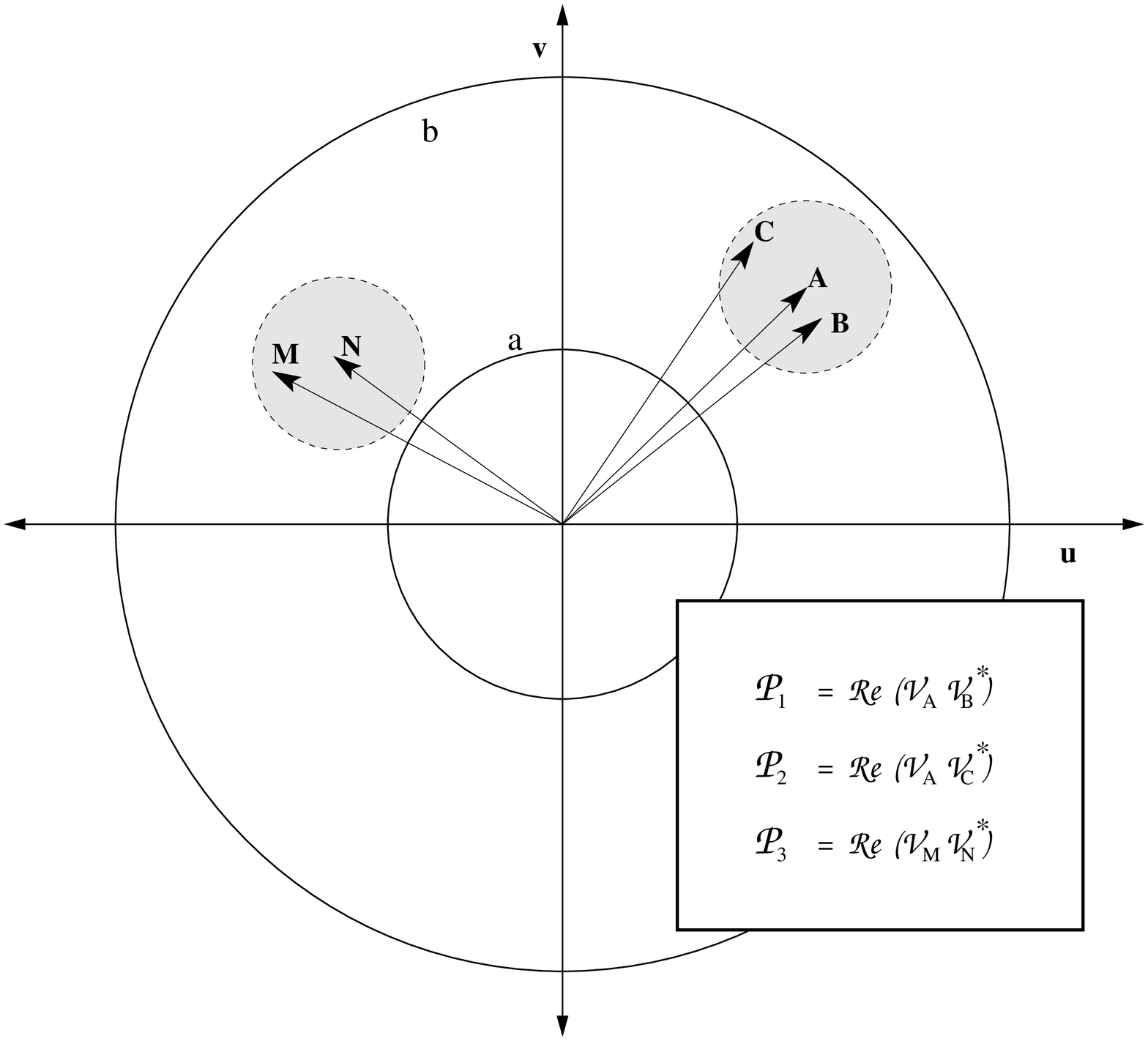, angle=0, width=4.8in}
\label{fig:est}
\end{center}
\caption{A particular annular region between $U_{A}$ and $U_{B}$ is
shown. Note that the visibility $\V(\U_{A})$ and $\V(\U_{N})$ are
correlated with all the visibilities in the corresponding shaded
circular regions. The estimates $\P_{1}$ and $\P_{2}$ are not
mutually independent estimates where as the estimates $\P_{1}$ and
$\P_{3}$ or $\P_{2}$ and $\P_{3}$ are mutually independent estimates.} 
\label{fig:estimates}
\end{figure}

We assume that the density fluctuations produced due to the ISM
turbulence is homogeneous and isotropic. In such cases we can replace
the ensemble average in the {\bf Eqn.~(\ref{eq:corrn})} by an angular
average. The real part of  each multiplication of the visibilities
$\V(\U_{i})$ and $\V(\U_{i} + \Delta \U)$ with $\Delta \U <
1/\pi \theta_{0}$, provide us an independent estimate,
\begin{equation}
\P(\U_{i}) = { \it Re} \ \left[  \V(\U_{i}) \V^{*}(\U_{i}
+ \Delta \U) \right]  = \P_{i}.
\end{equation}   
We consider annular regions in the $uv$ plane as shown 
in {\bf Figure~\ref{fig:est}} (the region between $\U_{a}$ and $\U_{b}$)
and average over all the individual estimates $\P_{i}$ in the annulus
to get 
\begin{equation}
\P = \frac{1}{N_b}\sum_{i=1}^{N_{b}} \P_{i},
\label{eq:est1}
\end{equation}
where, $N_{b}$ is the number of individual estimates $\P_{i}$ in the
given $\U$ bin. This value of $\P$ is then proportional to the power
spectrum $P_{HI}(U)$ for $U = (U_{a}+U_{b})/2$.

We will now  calculate the variance of the power spectrum estimator we
have just discussed. It is to be noted that, since visibilities at two
different baselines $\U_{i}$ and $\U_{i} + \Delta \U$ remains
correlated for  $\Delta \U \le 1/ \pi \theta_{0}$, all the estimates
$\P_{i}$ of {\bf Eqn.~(\ref{eq:est1})} are not independent.
{\bf Figure~\ref{fig:estimates}} shows three different
estimates of $\P$, namely $\P_{1}, \, \P_{2}$ and
$\P_{3}$, among which $\P_{1}, \, \P_{3}$ and $\P_{2}, \, \P_{3}$
are mutually independent estimates, but, $\P_{1}, \, \P_{2}$ are
not independent. This suggests that we can break the $\P_{i}$'s in say
$N_{g}$ groups such that, inside a group, the $\P_{i}$'s are not
independent, however, the estimates from different groups are
independent. Hence, we can write,
\begin{equation}
\P = \frac{1}{N_{g}} \sum_{i=1}^{N_{g}} \P_{N_{i}}
\label{eq:group}
\end{equation}
where, $N_{i}$ are the number of $\P$'s in the $i^{th}$ group and 
\begin{equation}
 \P_{N_{i}}=\frac{1}{N_{i}}\sum_{j=1}^{N_{i}} \P_{j}
\end{equation}
with $\langle \P_{N_{i}} \P_{N_{j}} \rangle = 0$ for $i \neq j$ and
$\sum_{i=1}^{N_{g}} N_{i} = N_{b}$. 
 
Variance of the estimator is defined as 
\begin{eqnarray}
\sigma_{\P}^{2} &=& \langle \P^{2} \rangle - \langle \P \rangle^{2}
\nonumber \\
&=& \frac{1}{N_{g}^{2}} \sum_{i=1}^{N_{g}} \ \sigma_{\P_{N_{i}}}^{2},
\label{eq:defs}
\end{eqnarray} 
where $\sigma_{\P_{N_{i}}}^{2} = \langle \P_{N_{i}}^{2} \rangle -
\langle \P_{N_{i}} \rangle ^{2}$. Assuming 
$\V_{2}(\Delta U) = \langle \V(\U_{i}) \V(\U_{i} + \delta \U)
\rangle$, we can write,
\begin{equation}
\sigma_{\P_{N_{i}}}^{2} = \frac{\V_{2}(\Delta U)^{2} +
  \V_{2}(0)^{2}}{2} + \frac{N_{2} ^{2} +
  \V_{2}(0) N_{2}}{2 N_{i}}
\label{eq:sigen}
\end{equation} 
Hence, using {\bf Eqn.~(\ref{eq:defs})}, we get
 the variance of the estimator given by 
\begin{eqnarray}
\sigma_{\P}^{2}&=&\langle \P^{2} \rangle - \langle \P \rangle^{2} \nonumber \\
&=&\frac{\V_{2}(\Delta \U)^{2} + \V_{2}(0)^{2}}{2 N_g} +
\frac{N_{2}^{2} + 2 \V_{2}(0) N_{2}} {2 N_{g}^{2}} \sum_{i=1}^{N_{g}}\frac{1}{N_{i}}
\label{eq:sigmag}
\end{eqnarray} 
In general, all $N_{i}$'s are different. If we assume $N_{1} = N_{2} =
\dots = N_{i}$, then above {\bf Eqn.~(\ref{eq:sigmag})} becomes
\begin{equation}
\sigma_{\P}^{2} = \frac{\V_{2}(\Delta \U)^{2} + \V_{2}(0)^{2}}{2 N_{g}} +
\frac {2 \V_{2} (0) N_{2} + N_{2}^{2} }{N_{b}}
\end{equation}
For an observation, when $N_{2} \gg \V_{2}(0)$, we can 
approximate the above equation as
\begin{equation}
\sigma_{\P}^{2} = \frac{\V_{2}(\Delta \U)^{2} + \V_{2}(0)^{2}}{2 N_{g}} +
\frac{N_{2}^{2}}{2 N_{b}}
\label{eq:sigsa}
\end{equation}
Since, we expect the value of the correlation $\V_{2}(\Delta \U)$ at two
different baselines separated by $\mid \delta \U \mid <<
\theta_{0}^{-1}$ is approximately same as  the correlation
$\V_{2}(0)$ at the same baselines, we can write,
\begin{equation}
\sigma_{\P}^{2} = \frac{\P^{2}}{N_{g}} +
\frac{N_{2}^{2}}{2 N_{b}}
\label{eq:sigsa}
\end{equation}
This expression is used by \citet{2008MNRAS.385.2166A} in calculating the
variance in the power spectrum of the foreground signal in EOR (Epoch
of Reionization)
observation. We can identify the first term in  {\bf
Eqn.~(\ref{eq:sigsa})} 
as a contribution from the  sample variance and the second term
arising due to the statistical fluctuation. 

For the sake of completeness we mention here that the visibility
correlation defined in {\bf Eqn.~(\ref{eq:corra})} also have an imaginary
part defined as  
\begin{equation}
\P^{I}(\U_{i}) = \it{Im} \ \left[  \V(\U_{i}) \V^{*}(\U_{i}
+ \Delta \U) \right]  = \P^{I}_{i}.
\end{equation}
However, when averaged over all the realizations, it can be shown that,
\begin{equation}
\langle \P^{I} \rangle = \langle \frac{1}{N_{b}} \sum_{i=1}^{N_{b}}
\P_{i}^{I} \rangle = 0
\end{equation} 
with
\begin{equation}
\sigma_{\P^{I}}^{2} = \frac{\V_{2}(0)^{2} - \V_{2}(\Delta \U)^{2}}{2 N_g} +
\frac{N_{2}^{2} + 2 \V_{2}(0) N_{2}} {2 N_{g}^{2}} \sum_{i=1}^{N_{g}}\frac{1}{N_{i}}
\label{eq:sigmagi}
\end{equation} 
Since, we expect the value of the correlation $\V_{2}(\Delta \U)$ at two
different baselines separated by $\mid \delta \U \mid <<
\theta_{0}^{-1}$ is approximately same as  the correlation
$\V_{2}(0)$ at the same baselines, we can safely neglect the first term
compared to the second term. Further, if $N_{1} = N_{2} = \dots =
N_{i}$, and for $N_{2} \gg \S_{2}(0)$, {\bf Eqn.~(\ref{eq:sigmagi})}
becomes 
\begin{equation}
\sigma_{\E^{I}}^{2} = \frac{N_{2}^{2}} {2 N_{g}}
\end{equation}

\newpage

\thispagestyle{empty}
\addcontentsline{toc}{chapter}{Curriculum Vitae} 
\chaptermark{Curriculum Vitae}
\setcounter{section}{0}
\setcounter{subsection}{0}
\setcounter{subsubsection}{2}
\setcounter{equation}{0}
\begin{singlespacing}
\newcommand{\entry}[2]{ {{\bf #1}} & &  {#2} \\}
\newcommand{\eskip}{& & \\}

\chapter*{Curriculum Vitae}
\vspace{2.5cm}
%\centering
\begin{tabular}{rcl}
\entry{\large {Name : }}{\large {Prasun Dutta}} 
\eskip
\entry{Affiliation : }{Department of Physics and Meteorology,}
\entry{}{Indian Institute of Technology, Kharagpur}
\entry{}{Kharagpur 721302, India}
\eskip
\entry{Date of Birth : }{1$^{st}$ August, 1981}    
\eskip
\entry{Email : }{prasundutta151@gmail.com}
\entry{}{prasun@cts.iitkgp.ernet.in}
\entry{}{prasun@phy.iitkgp.ernet.in}
%\eskip
%\entry{Contact no : }{+91-9433404193}
\eskip
\entry{Educational    }{{\bf Master of Science, Physics,} May (2005)}
\entry{Qualifications : }{Indian Institute of Technology, Kharagpur.}
\entry{}{{\bf Bachelor of Science, Physics (Hons.),} June (2003)}
\entry{}{University of Calcutta.}
\eskip
\entry{Research     }{Interstellar Medium Turbulence, Radio-interferometric }
\entry{Interests : }{observation techniques, Supernovae remnants.}
\end{tabular}
\end{singlespacing}
\newpage

%\clearpage{\pagestyle{empty}\cleardoublepage} %%%%%%%%%%%%%%%%%%%%

%\addcontentsline{toc}{chapter}{List of Publications} 
\renewcommand{\thefootnote}{\fnsymbol{footnote}}
\section*{List of Publications } 
\begin{singlespacing}
\subsection*{In Journals :}
\begin{itemize}
\item \footnotemark  {\bf Dutta, P.}, Begum, A., Bharadwaj, S., and Chengalur,
J.N. (2008). \\
 HI power spectrum  of the spiral galaxy NGC~628.\\
{\bf Monthly Notices of Royal Astronomical Society Letters}, 384, L34–L37. 

\item Roy, N., Bharadwaj, S., {\bf Dutta, P.}, and Chengalur,
 J.N. (2009). \\
Magnetohydrodynamic turbulence in supernova remnants.\\
{\bf Monthly Notices of Royal Astronomical Society Letters}, 393, L26–L30.

\item \footnotemark[\value{footnote}]  {\bf Dutta, P.}, Begum, A.,
  Bharadwaj, S., and Chengalur, J.N. (2009).\\
The scale-height of NGC~1058 measured from its HI power spectrum.\\
{\bf Monthly Notices of Royal Astronomical Society Letters}, 397, L60–L63. 

\item \footnotemark[\value{footnote}] {\bf Dutta, P.}, Begum, A.,
  Bharadwaj, S., and Chengalur, J.N. (2009). \\ 
A study of interstellar medium of dwarf galaxies using HI power
spectrum analysis. \\
{\bf Monthly Notices of Royal Astronomical Society}, 398, 887–897.

\item Roy, N., Chengalur, J.N., {\bf Dutta, P.}, and Bharadwaj,
S. (2010).\\
H I 21 cm opacity fluctuations power spectra towards Cassiopeia A.\\
{\bf Monthly Notices of Royal Astronomical Society Letters}, 404, L45–L49.

\item \footnotemark[\value{footnote}] {\bf Dutta, P.}, Begum, A.,
  Bharadwaj, S., and Chengalur, J.N. (2010).\\
Turbulence in the harassed galaxy NGC~4254. \\
{\bf Monthly Notices of Royal Astronomical Society Letters}, 405, L102–L106.

\item \footnotemark[\value{footnote}] {\bf Dutta, P.}, Guha Sarkar,
  T., and Khastgir, S.P. (2010).\\  
 The effect of the $w$-term on the visibility correlation and power
 spectrum estimation. \\
{\bf Monthly Notices of Royal Astronomical Society Letters}, 406, L32-34. 
%doi:10.1111/j.1745-3933.2010.00875.x.

\item {\bf Dutta, P.}, Khastgir, S.P., Roy, A. (2010).\\ 
Steiner trees and spanning trees in six-pin soap films. \\
{\bf American Journal of Physics}, 78, 215-221.
\end{itemize}
\footnotetext{\, \, \, Used for the present thesis}
\newpage
\subsection*{In Proceedings :}

\begin{itemize}
\item Roy, N., Bharadwaj, S., {\bf Dutta P.}, and Chengalur,
  J. N. (2009)\\ 
MHD turbulence in supernova remnants. \\
{\bf Proceeding of the 27th Meeting of the ASI in BASI, 2009}

\item \footnotemark[\value{footnote}] {\bf Dutta, P.}, Begum, A.,
  Bharadwaj, S., and Chengalur,   J.N. (2009).\\
Probing Turbulence in the Interstellar Medium of Galaxies.\\
{\bf Astronomical Society of the Pacific Conference Series}, volume 407, 83

\item Roy, N., Chengalur, J.N., Bharadwaj, S., and {\bf Dutta, P.}
  (2009).\\
Turbulence in Cold Neutral ISM and in Supernova Remnants.\\
{\bf Astronomical Society of the Pacific Conference Series}, volume 407, 272
\end{itemize}

\subsection*{In Preparation :}
\begin{itemize}
\item \footnotemark[\value{footnote}] {\bf Dutta, P.}, Begum, A.,
  Bharadwaj, S., and   Chengalur, J.N. (2009). \\ 
Probing Interstellar Turbulence in Spiral Galaxies Using HI Power
Spectrum Analysis.\\
To be submitted to {\bf Monthly Notices of Royal Astronomical
  Society}.

\item {\bf Dutta, P.}, Begum, A., Bharadwaj, S., and
  Chengalur, J.N. (2009). \\ 
Simulating HI spectral observation from a turbulent ISM.\\
To be submitted to {\bf Monthly Notices of Royal Astronomical
  Society}.

\end{itemize}
\footnotetext{\, \, \, Used for the present thesis}
\end{singlespacing}
\newpage

\newpage
\end{document}